\documentclass[useAMS,usenatbib]{mn2e}
\def\href#1#2{{#2}}
\usepackage{graphicx}
\usepackage{epsf}\LARGE
\usepackage{deluxetable}
\def\l{$\lambda$}
\def\mbh{$M_{\rm BH}$\/}
\def\lledd{$L/L_{\rm Edd}$}

\def\ne{$n_{\rm e}$\/}
\def\nc{$N_{\rm c}$\/}
\def\rfe{$R_{\rm FeII}$}

\def\feiiq{\rm Fe{\sc ii}$\lambda$4570\/}

\def\msol{M$_\odot$\/}

\def\ltsima{$\; \buildrel < \over \sim \;$}
\def\simlt{\lower.5ex\hbox{\ltsima}}            
\def\gtsima{$\; \buildrel > \over \sim \;$}

\def\simgt{\lower.5ex\hbox{\gtsima}}            

\def\ha{{\sc H}$\alpha$}
\def\lya{{Ly}$\alpha$}
\def\civ{{\sc{Civ}}$\lambda$1549\/}
\def\civnc{{\sc{Civ}}$\lambda$1549$_{\rm NC}$\/}

\def\cmq{cm$^{-2}$\/}
\def\cm3{cm$^{-3}$\/}
\def\hb{{\sc{H}}$\beta$\/}

\def\hbbc{{\sc{H}}$\beta_{\rm BC}$\/}
\def\hbvbc{{\sc{H}}$\beta_{\rm VBC}$\/}

\def\mgii{{Mg\sc{ii}}$\lambda$2800\/}

\def\ciii{{\sc{Ciii]}}$\lambda$1909\/}
\def\oiiiopt{{\sc{[Oiii]}}$\lambda\lambda$4959,5007\/}
\def\o4363{{\sc{[Oiii]}}$\lambda$4363\/}
\def\oiiiuv{{\sc{Oiii]}}$\lambda$1663\/}
\def\siiii{ Si{\sc iii]}$\lambda$1892\/}
\def\aliii{Al{\sc iii}$\lambda$1860\/}
\def\heiiuv{He{\sc{ii}}$\lambda$1640}
\def\nv{{N\sc{v}}$\lambda$1240}

\def\feiiopt{{Fe \sc{ii}}$_{\rm opt}$\/}
\def\feii{{Fe\sc{ii}}\/}

\def\feiii{{Fe\sc{iiiÄ}}\/}
\def\fe{{\sc{Fe}}\/}

\def\vr{{$v_{\mathrm r}$}}
\def\heii{{\sc{Heii}}$\lambda$4686\/}
\def\fe76087{{\sc [Fe vii]}$\lambda$6087\/}

\def\oii{{\sc [Oii]}$\lambda$3727}

\def\kms{km~s$^{-1}$}

\def\ergss{ergs s$^{-1}$\/}

\def\hi{H{\sc i}\/}

\def\heii{{{\sc H}e{\sc ii}}$\lambda$4686\/}

\def\mb{$m_{\mathrm B}$\/}
\def\lledd{$L_{\mathrm{bol}}/L_{\rm Edd}$\/}
\def\lbol{$L_{\rm bol}$\/}
\def\caii{Ca{\sc ii}\/}
\newcommand{\lsim}{{\, \lower2truept\hbox{
${< \atop\hbox{\raise4truept\hbox{$\sim$}}}$}\,}}
\newcommand{\gsim}{{\, \lower2truept\hbox{
${> \atop\hbox{\raise4truept\hbox{$\sim$}}}$}\,}}

\title[BLR Physical Conditions along the Quasar E1 Sequence]{Broad Line Region Physical Conditions\\along the Quasar Eigenvector 1 Sequence}
\author[P. Marziani et al.]{P. Marziani$^{1}$\thanks{E-mail:
paola.marziani@oapd.inaf.it}, J. W. Sulentic$^{2}\thanks{Professor Emeritus at the University of Alabama}$, C. A. Negrete$^3$, D. Dultzin$^3$,  S. Zamfir$^4$,  R. Bachev$^5$\\
$^{1}$INAF, Osservatorio Astronomico di Padova, Vicolo dell' Osservatorio 5, IT 35122, Padova, Italy\\
$^{2}$Instituto de Astrof\'\i sica de Andaluc\'\i a (CSIC), C/ Camino Bajo de Hu\'etor 50, 18008 Granada, Spain\\
$^{3}$Instituto de Astronom\'\i a, Universidad Nacional Aut\'onoma de M\'exico, DF 04510, Mexico\\
$^{4}$Department of Physics \& Astronomy, University of Alabama, Tuscaloosa, AL 35487,  USA\\
$^{5}$ Institute of Astronomy, Bulgarian Academy of Sciences, 72 Tsarighradsko Shousse Blvd., 1784 Sofia, Bulgaria}
\begin{document}

\date{Submitted: }

\pagerange{\pageref{firstpage}--\pageref{lastpage}} \pubyear{2002}

\maketitle

\label{firstpage}

\begin{abstract}
We compare broad emission line profiles and estimate line ratios for all major  emission lines between \lya\ and \hb\ in a sample of six quasars. The sources were chosen with two criteria in mind:  the existence of high quality optical and UV spectra as well as  the possibility to sample the spectroscopic diversity in the 4D Eigenvector 1 context (4DE1). In the latter sense each source occupies a region (bin) in the FWHM(\hb) vs. \feiiopt\ strength plane that is significantly different from the others. High S/N  \hb\ emission line profiles are  used as templates for modeling the other lines (Ly$\alpha$, \civ, \heiiuv, \aliii, \siiii,  and \mgii).  We can adequately model all broad lines assuming the existence of three components distinguished by blueshifted, unshifted and redshifted centroids (indicated as blue, broad and very broad component respectively). The broad component  { (electron density \ne $\sim 10^{12}$  \cm3; ionization parameter $U \sim 10^{-2}$; column density \nc $\ga  10^{23}$ \cmq)} is present in almost all type-1 quasars and therefore corresponds most closely to the classical broad line emitting region (the reverberating component). The bulk of \mgii\ and \feii\  emission also arises in this region. The blue component emission ($\log$ \ne $\sim 10$; $\log U\sim -1$; $\log$ \nc$<$ 23) arises in less optically thick gas; it is often thought to arise in an accretion disk wind. The least understood component involves the very broad component (high ionization and large column density) which is found in no more than half (but almost all radio-loud) type-1 quasars and luminous Seyfert nuclei.  It is perhaps the most distinguishing characteristic of  quasars with FWHM \hb $\ga$4000 \kms\ that belong to the so-called Population B of  our 4DE1 space. Population A quasars (FWHM \hb $\la$4000 \kms) are dominated by broad component emission in \hb\ and blue component emission in \civ\ and other high ionization lines.  4DE1 appears to be the most useful current context for revealing and unifying spectral diversity in type-1 quasars.
\end{abstract}


\section{Introduction}
\label{intro}
\subsection{Introduction}

Quasar spectra are not self-similar even if we restrict ourselves to active galactic  nuclei that show broad emission lines (type-1 AGN). Examination of the same broad emission  line, in many different sources reveals a striking diversity. This diversity begins with  the profile width but extends to every other measurable property (e.g. shape, line shift,  equivalent width). The \hi\ Balmer line H$\beta$\ is a good place to start a line profile  characterization because so much data exist. It can be observed from the ground out  to $z$$\sim$0.9 and through numerous IR windows to $z \ga $3. Thus one can compare the  line over a wide range of redshift and source luminosity. It is usually a strong line and  suffers only moderate contamination falling in a gap between strong optical \feii\ blends. At this time good S/N and moderate resolution spectra exist for more than 600 low $z$\ sources \citep[including the brightest $\approx$400 SDSS quasars]{marzianietal96,marzianietal03a,zamfiretal08,zamfiretal10}. Table 2 in \citet{sulenticetal00a} tabulates  earlier references with high S/N spectra. Comparable quality IR spectra of the \hb\ region now exist for more than 50 sources in the range $z$ = 0.9 -- 3.1 \citep{sulenticetal04,sulenticetal06,marzianietal09}.

The C{\sc iv} line at \l 1549 is the other current broad line-of-choice providing a high ionization (HIL) counterpoint to low ionization (LIL) \hb. High quality \civ\ spectra for low $z$\ sources  are more difficult to obtain with the HST archive providing UV spectra of the \civ\ region for  $\sim$130 sources \citep{bachevetal04}. Above  $z\sim$1.5 \civ\ is accessible at optical wavelengths out to  $z\sim$4.8. The largest compilation of high quality and high $z$ \civ\ spectra came from the Palomar surveys almost 20 years ago \citep{youngetal82,sargentetal89,bartheletal90}. A uniform study of the Palomar and other good quality HIL spectra can be found in \citet{tytlerfan92}.  Unfortunately most sources with good \hb\ spectra lack a corresponding \civ\ spectrum and vice versa. Many of the other broad lines have been studied but usually without the benefit of spectra for other broad lines in the same source. A fundamental desideratum involves strong narrow lines that provide  a reliable determination of the quasar rest frame (e.g. \oiiiopt, and especially \oii). Full exploitation of the Palomar spectra is impeded by the lack of rest frame determinations for the sources and the lack of matching \hb\ spectra.

Broad line profiles provide some of the most direct clues about the nature, structure and kinematics of the central regions in quasars. Inter-line profile comparisons for individual sources (Seyfert 1 nuclei as well as radio-quiet (RQ) and radio-loud (RL) quasars) have been carried out for a long time although there has been little agreement on the systematics \citep{gaskell82,wilkes86,espeyetal89,corbin90,carswelletal91,tytlerfan92,marzianietal96,richardsetal02}. Most of  these studies have sampled: (a) small numbers of sources, (b) only a few of the broad  lines in each source, and/or (c) less than ideal spectra { (see \cite*{shangetal07} for a notable exception).} If we consider the best studied HIL and LIL features (\civ\ and \hb) we find little evidence for profile (shape/shift/EW) similarity for sources with FWHM\hb $\la$ 4000 \kms\ while above that value the profiles appear to be more similar, after proper correction  for narrow line \civ\ emission is made \citep{marzianietal96,sulenticmarziani99,bachevetal04,sulenticetal07}. Ideally we would like to have spectral coverage for all of the principal broad lines in each of a sample of sources that reflect the impressive source-to-source diversity now known. It would also be ideal if we could achieve a consensus on how many distinct components each line contains. This paper will address both of these desiderata.

\subsection{Eigenvectors of Quasars}

We are never satisfied with the S/N and resolution of the available quasar spectra even though current best data exceed our ability to explain them theoretically. Minimum acceptable values for good spectra are arguably 20-30 in the continuum and $\lambda/\Delta\lambda \sim 1000$\ respectively. We are tempted to indiscriminately average line profiles in order to improve even more the S/N but a composite spectrum generated from a large sample obscures most of the  important astrophysical  information. Such composite spectra have heuristic value since they provide a wide spectral coverage and give a global view of the continuum shape \citep{vandenberketal01}. Different lines represent averages of different sources making comparison impossible  unless all quasar spectra are fundamentally similar. In the context of broad line physics  indiscriminate averaging obscures rather than reveals profile differences. It is the ``why'' of differences that yields physical insight. Not only do quasar spectra show significant  diversity but measured line parameters do not scatter symmetrically around an average.  Eigenvector analysis showed that the diversity of optical and UV spectra is not entirely  random, but that there are systematic trends \citep{borosongreen92,sulenticetal00b}. We  expanded the original optically based Eigenvector 1 parameters to include \civ\ and soft X-ray photon index (see also \citealt{wangetal96}) in an effort to enhance our ability to identify profile differences \citep[4D Eigenvector 1 $\equiv$ 4DE1][]{sulenticetal00a,sulenticetal00b,sulenticetal07}.  As examples of the lack of symmetric scatter we note that sources occupy well-defined non-random sequences in e.g. FWHM(\hb) vs. normalized \feiiq\ strength (see Figure \ref{fig:e1}) or vs. normalized  \civ\ line shift \citep[Figure 2d in][]{sulenticetal07}.

Profile diversity persists when composite spectra are generated in some fundamental context like 4DE1. For example, sources in a restricted region of the 2D parameter space involving FWHM \hb\ and \feiiopt\ strength measures  show considerable profile diversity (Figure \ref{fig:e1} which is the optical plane of 4DE1). The composite spectrum for sources in a restricted region (bin) presumably  reveals the underlying ``stable'' emission component(s). The source-to-source diversity within a  bin likely reflects minor changes in profile shape connected with profile variability. The bin-to-bin diversity  in a contextual space like 4DE1 have the greatest potential to  yield clues about  broad line region (BLR) structure and kinematics (convolved with source orientation). Contexts for  such binned profile composites can be empirical (e.g. FWHM, \feii\ strength) or physical (e.g. black hole mass \mbh, Eddington ratio \lledd). A comparison between e.g. radio-quiet and radio-loud quasars will yield any result that one wants unless the empirical context for profile binning is carefully chosen and is independent of radio properties.

\subsection{How Many Broad Line Components?}

We would like to compare all principal broad lines in a small sample of sources that
reflect the diversity found in the 4DE1 binning. At present however adequate multiwavelength data exist  for very few sources so studies encompassing all the principal broad lines are rare (e.g. Shang et al. 2007). The most basic question for facilitating such a study involves how many components one can identify in each broad line. This is particularly important for studies that use the widths of broad lines as virial estimators. Empirical studies  have made it clear that HIL (e.g. \civ) and LIL (e.g. Balmer series) show a wide range of strength, shape and shift properties. Our own work in the 4DE1  context has attempted to systematize these properties
\citep{sulenticetal00a,sulenticetal07,marzianietal01,marzianietal03a,marzianietal03b,marzianietal08}  with focus so far on comparisons of \civ\ and \hb\ assumed to be typical HIL and LIL features.

Several authors have considered the degree of stratification that might exist in the BLR including the idea that the innermost part of the BLR could be emitted in a quasi-distinct  very broad line region (VBLR). The VBLR concept may have originated with the  discovery that the core and wings of HIL \heii\ and LIL \hb\ in highly reverberated  NGC5548 show different responses to continuum change. \citep{petersonferland86}. The VBLR concept and/or terminology arose in several additional guises and this has led to some confusion. Is the VBC present in all quasars? Are there equal  numbers of blue- and red-shifted VBC \citep[see e.g.][]{corbin95}? Is there no  significant narrow emission line  component in \civ? If one rejects a significant \civ\ narrow component the ``peakiness'' of the \civ\ profile in many sources  suggests both narrow  intermediate line region (ILR) and very broad VBLR as the simplest \civ\ profile model \citep{brothertonetal94}. We think the answer to all of these questions is ``no'' \citep[see e.g.][]{marzianietal96,sulenticmarziani99,sulenticetal02,sulenticetal07}. Our studies show evidence for a VBC {\em only} in sources with FWHM(\hb) $\ga$ 4000 \kms\ \citep[Population B:][]{sulenticetal00a,sulenticetal00b,marzianietal03b,zamfiretal10}. The \hb\ profile in these sources requires both unshifted  gaussian (BC) and redshifted VBC components. Narrower (FWHM \hb $\la$ 4000 \kms) Pop. A sources can often be fitted with a simple Lorentz-like function although a blue  asymmetry is seen in the H$\beta$ profiles of  some pop. A sources. Whether FWHM \hb $\approx$\ 4000 \kms\ signifies a fundamental change in BLR physics (e.g. critical Eddington ratio?) or not, the highly significant changes in  profile shape above and below this value cannot be ignored.

\subsection{Defining an Empirical Approach}

This paper compares all of the principal broad lines (\lya, \civ, \ciii, \mgii, \hb) for a sample of sources where: (1) suitably high S/N spectra are available for all or most of the principal broad lines and (2) 4DE1 spectroscopic diversity is sampled. Our approach is empirical and uses the \hb\ profile fit for each chosen source to constrain the fits of all other broad lines in that source. High and low redshift Pop. B quasars show rather symmetric \lya\ and \civ\ profiles \citep{bartheletal90,bachevetal04} which is in sharp contrast to the appearance of \hb\ in Pop. B sources where a red asymmetry is common.  In Pop. A the situation is reversed with blue asymmetric \lya\ and \civ\ and symmetric Balmer line profiles.  Comparisons of \hb\ and \civ\ profiles suggest that three emission components are present in some or all sources.

\begin{enumerate}

\item A ``classical'' BC (FWHM = 600 --5000 \kms). The H$\beta$ profile is best modeled with a symmetric Lorentzian function for Pop. A sources (FWHM $<$ 4000 \kms)  and a Gaussian function for Pop. B (FWHM$>$4000 \kms). This component may be weak or, even, absent in some Pop. B sources. The bulk of the optical \feii\ emission appears to follow the BC \citep[e.g., ][]{marzianietal03a}.

\item A VBC (FWHM$\sim$10000 \kms). H$\beta$ is reasonably well modeled with a Gaussian  function. It is only observed in Pop. B sources where it shows a  redshift (FWHM in the range 1000-5000 \kms). This can be the dominant component in some  Pop. B sources.

\item A BLUE component (blue shifted or blue asymmetric profile) seen in the H$\beta$ profile of some Pop. A sources with \rfe = I(\feiiq)/I(\hb)  $>$0.5. It is more prominent in \civ\ where it can be defined as the residual emission after subtraction of an unshifted Lorentzian BC in Pop. A sources. In this paper we model this component with a Gaussian.
\end{enumerate}

If there is a semi-coherent structure that gives rise to all or most of the broad line emission then all broad lines in a source should show similar properties allowing  for some degree of stratification. \hb\ apparently does not show the same components in all sources motivating  the Pop. A-B nomenclature \citep{sulenticetal00a,sulenticetal00b,sulenticetal07} as an effective way to highlight this  difference. We assume that the distribution of sources in the optical plane of 4DE1 has physical meaning in the sense that source occupation is  driven by both physics and line-of-sight orientation. This assumption is reenforced by the discovery that sources in different regions of the optical plane (and especially Pop. A and B regions) show differences in almost all other multiwavelength measures. The sources chosen for this study allow us to test the hypothesis that the same three components  {\em observed in \hb\ and \civ}  can account  for the diverse line profiles of  \lya, \civ, \heiiuv, \siiii, \aliii, \ciii, \mgii, \hb\ and \ha. After testing this hypothesis (\S \ref{fits}) we measure  line intensities of the  components (\S \ref{fluxes}). Analysis of the properties of each line component yields insights about differences in physical conditions of the three identified emitting regions. (\S \ref{phys}).

\section{Source Selection and Observations}
\label{selec}

Sources were selected to be representative for the spectral diversity in the 4DE1 context. We selected six sources that include  one quasar in each of the six most populated  4DE1 spectral bins  \citep[A1, A2, A3, B1, B1$^{+}$\ and B1$^{++}$; ][]{sulenticetal02}. Figure \ref{fig:e1} marks  the location of these sources in the optical plane relative to the distribution of  the brightest 400+ SDSS DR5 quasars \citep{zamfiretal10}. { We remark that the validity of the analysis presented in this paper is not restricted to 6 objects only. Each object has been chosen as representative of each bin in the E1 plane of Fig. \ref{fig:e1}. These bins are occupied by  90\% of low-$z$ quasars \citep{zamfiretal10}. } While high S/N and moderate resolution spectral coverage from \lya\ to \hb\ exists for very few quasars we were fortunate to find representative sources for these six bins. UV coverage from \lya\ to \mgii\  comes from the HST archive for these low $z$ sources save for 3C 110 where \mgii\ is covered in our optical spectrum. A listing  of the available HST spectra can be obtained at the HST archive.  A list with the datasets actually employed to build the UV spectra  can be obtained from the authors.\footnote{It is available at the URL: http:// web.oapd.inaf.it/\textasciitilde marziani/ uvlog.pdf.} Optical spectra come from \citet{marzianietal03a}; a log of observation can be easily retrieved from their Table 2. UV and optical observations are generally not simultaneous, with a time mismatch of less than 3 years in all cases save Mark 478; in this last case, the UV observations were collected more than 5 years after the optical ones.

We make use of composite spectra for each of these bins \citep{sulenticetal02,bachevetal04,zamfiretal10} computed from the bright SDSS sample or from our spectroscopic atlas \citep{marzianietal03a}.   The \ciii\ blend  is unavailable for 3C110 and the region of \mgii\ has not  been observed in 3C 249. \ha\ spectra are only available for three sources. Our composite  spectra are used as surrogates for these gaps in our source data.

Table \ref{tab:obj}  identifies  the selected sources where: Col. 1 - IAU code name; Col. 2 - common name recognizable  by NED); Col. 3 - adopted redshift; Col. 4: B magnitude;  Col. 5 - Galactic B-band  absorption  from NED; Col. 6 - the spectral type following  \citet{sulenticetal02} and Col. 7 - additional information. Two sources have been reverberation mapped \citep[RM][]{petersonetal04}. Galactic absorption values $A_\mathrm{B}$\ are from \citet{schlegeletal98}; if $R_\mathrm{V} =3.1$, then $A_\mathrm{V}  \approx 0.756 A_\mathrm{B}$. This value has been used with the task {\tt deredden} of IRAF. No internal extinction correction was applied to our sources.


\section{Data analysis}

Data used in this paper are wavelength and flux calibrated. Spectra and spectral measures were shifted to rest frame wavelength and flux using the redshift listed in Tab. \ref{tab:obj} after correction for Galactic extinction.

The regions around \lya, \civ\ (including \heiiuv), the blend at $\approx$ 1900 \AA\ (including \aliii, \siiii, \ciii), \mgii, \hb\ (including the \feiiq\ blend), and \ha\ were analyzed using the {\tt specfit} task of {\tt IRAF} { \citep{kriss94}}. We used an \feii\ template based on an ESO spectrum of I Zw 1 to correct for \feii\ contamination. This template was already employed for data analysis in \citet{marzianietal03a,marzianietal03b} with improvements described in \citet{marzianietal09}. It is basically the \citet{borosongreen92} template with two major modifications: the \feii\ emission underlying  \hb\ was set according to a model simulation, due to the difficulty of distinguishing between \hb\ proper and underlying \feiiopt\ features \citep{marzianietal09}. In the red, our spectrum shows a significant feature at 6446 \AA\ 
In any case the \feiiopt\ correction is so
small that it would not affect any of our results concerning the \ha\ line shape.  In the UV, \feii\ emission around \mgii, \ciii\ and \civ\ is corrected applying a template obtained from a {\tt CLOUDY} \citep{ferlandetal98} simulation, assuming ionization parameter 10$^{-2.25}$, electron density 10$^{12.25}$ \cm3. \feiii\ emission, especially prominent around the 1900\AA\ line blend, was modeled using a template build following \citet{vestergaardwilkes01}. There is  considerable uncertainty regarding the intensity of the \feiii\ feature at 1914\AA; this feature is heavily blended with \ciii\ making it impossible a reliable estimate of the  intensity of both lines in Pop. A sources (i.e., in any case the iron spectrum is strong).

The three previously described line components were simultaneously fit along with a power-law continuum and appropriate  \feii\ (and \feiii\ around 1900 \AA) template solution employing $\chi^2 $\ minimization techniques. The relative intensity of the  components was left free to vary although we imposed the condition that velocity shifts and widths  roughly match the ones measured in \hb\ and in \civ\ (BLUE component only).

\section{Results}

\subsection{Multicomponent Fits}
\label{fits}

Fig. \ref{fig:profiles} shows the resultant fits to the line profiles of \lya, \civ+\heiiuv, 1900\AA\ blend, \mgii\ and  \hb\   for each source. \ha\ is not included since it does not add information to the line profile interpretation based on \hb. Residuals are shown below each panel. Table \ref{tab:comp} reports the total flux in the \civ\ line, and the normalized intensity ratios of BC, VBC and BLUE component; individual sources are discussed in Appendix \ref{individuals}.  The following results are consistent with trends visible in median composite spectra for the same bins \citep{sulenticetal02,bachevetal04,zamfiretal10}.


\paragraph*{Broad Component}
The BC dominates LIL (Balmer lines, \feii, \mgii) emission in Pop. A sources, while it becomes less prominent  in Pop. B.  The fraction of BC \hb\  flux decreases for Pop. B sources from bin B1 to B1$^+$\ and from B1$^+$\ to B1$^{++}$.  The 4DE1 sequence can be described as a sequence of  decreasing \hb\ BC prominence and may even involve BC demise. Evidence exists for  AGN with a very weak or absent BC, some showing no emission lines (e.g. BL Lacs) or only  narrow emission lines (type-2 AGN). Others more relevant to this study show lines with  strong VBC emission that can be mistaken for a typical type-1 AGN \citep[e.g. PG1416-129]{sulenticetal00c}.  We suspect that 3C 110 (also discussed in Appendix \ref{individuals}) may be an example of this situation. Rather than the two  component BC+VBC fits shown in Figure \ref{fig:profiles}  and \ref{fig:blue} the correct interpretation for 3C110 may  involve a single redshifted VBC component. FWHM $\sim$  10000 \kms\ would be unprecedented for the BC.

In Pop. A sources I Zw I and Mark 478 \ciii\ is closely blended with \feiii\ emission leaving open the possibility that BC \ciii\ may be very weak or absent (for this reason no \ciii\ ratios are reported in Tab. \ref{tab:bc}).   Nearby \aliii\ and \siiii\  are also very prominent in these two Pop. A sources.  Conversely, bin A1 source Mark 335  shows stronger  \ciii\ emission  and \aliii\ intensity more similar to Pop. B sources.

\paragraph*{Blueshifted Component (BLUE)}  Figure \ref{fig:blue}  provides the justification for including a blueshifted component  in the Figure \ref{fig:profiles}  fits by showing the  residual that arises from a two component fit to \civ\ (BC+VBC) for all six sources. Pop. A sources show a strong residual reflecting the well known blueshifted/asymmetric component.   BLUE is present in the \lya\ and \civ\ emission of all sources but 3C~110, with typical  shifts and widths  $\Delta v \sim -2000$ \kms  and $\sim$7000 \kms. Pop. A and B sources show similar shift and width values. Only bin A3 source I~Zw~1 may show a possible hint of this blue component in the  \hb\ line.  BLUE is not seen in \mgii\ while  the 1900 \AA\ blend is too complex to allow any conclusion to be made.


\paragraph*{Very Broad Component} A VBC component is one of the defining properties of the  broad line spectrum in Pop. B sources. A VBC is not observed in NLSy1 or even less extreme  Pop. A quasars \citep[see the cautionary note regarding very high luminosity quasars][]{marzianietal09}.   Using \hb\ as a template we therefore fit only two components (BLUE + BC) in Pop. A sources. The residuals of those fits show no excess flux  on the red wing of \hb\ that might be interpreted as a weak VBC signature. This reenforces the earlier proposed ideas that FWHM$\approx$4000 \kms\ reflects some fundamental change in the structure and/or kinematics of the line emitting region--perhaps  connected to a critical  value of the Eddington ratio.

Restricting attention to our  three Pop. B sources that show a VBC, shifts and widths are $\sim$2000 \kms\  and $\sim 10000$  respectively (3C 110 has two entries in  Table  \ref{tab:vbc}: the second one refers to the assumption of no BC presence in the emission lines; see Appendix \ref{individuals}). These values are consistent with the ones derived for median spectra of more than 150 Pop. B sources  in the SDSS bright sample \citep{zamfiretal10}.  Using \hb\  as the model we conclude that a VBC is  also present in the \civ\ profiles of Fairall 9 and 3C 249.1, giving rise to  the extended red wings visible in Figure \ref{fig:blue}.  These profiles look deceptively symmetric but both our empirical fits  argue that \civ\ counterparts to  the \hb\ VBC should be present. 



\subsection{The \heiiuv\  profile} \label{heii} The  equivalent width of He$^{+}$ lines is  an Eigenvector 2  parameter involving an apparent anti-correlation with source luminosity \citep{borosongreen92}. The \heiiuv\ profile  of Pop. B sources  is customarily flat-topped and rather extended on the blue side.  The profile can be interpreted  as due to the sum of the same components observed in \civ. An unambiguous decomposition of the \heiiuv+\oiiiuv\  blend for Pop. A sources is difficult. A fit of the \heiiuv + \civ\ profile reproduces the  shape of the entire blend   assuming that \heiiuv\ is due to blueshifted + broad component.   In Pop. B sources, the  analysis is even more complex because \heiiuv\ is blended with the red wing (VBC) of \civ. The  presence of blueshifted and VBC \heiiuv\ emission creates the apparent flat extension on the  red side of \civ\  that has not been satisfactorily explained as yet. This extension is visible  in the spectra of several RL sources at both low- and high- redshift \citep{marzianietal96,bartheletal90}. This  ``plateau'' effect is especially well seen  on the  red side of \civ\ in  3C 249.1 (Fig. \ref{fig:heii})  where the best fits for \heiiuv\ are obtained assuming the  blueshifted and VBC components as for \civ\ and \lya. The same  interpretation  seems to apply to \heii: a good supporting case involves  the  low-luminosity quasar PG 1138+22 \citep{marzianisulentic93}.



\subsection{Intensity Ratios}
\label{fluxes}

Tables  \ref{tab:bc}, \ref{tab:blue}, \ref{tab:vbc}  provide flux ratios for the three line components in each of the six sources. The rest-frame equivalent width for \lya\ is reported for each emission component in the last column of each table.  Ratios are reported with 2 significant digits to avoid roundoff errors.  { Uncertainties on ratios are estimated from the errors in the individual component  intensity which are typical $\pm$20\% for the strongest components (the ones of \lya\ and \civ, \mgii\ BC, \hb BC in Pop. A) or for individual or unblended features (\feii)  and $\pm$50\% \ for the weakest or most blended features (\heiiuv,\siiii, \mgii\ VBC). 
Applying simple error propagation yields, for the intensity ratio of two components, labeled 1 and 2: $\Delta (I_{1}/I_{2})= (I_{2}/I_{1}) \cdot \sqrt{ (\Delta I_{1}/I_{1})^{2}+ (\Delta I_{1}/I_{2})^{2}}$. For a typical error $\approx 40\%$ on line intensity, the error on ratio is   $\Delta (I_{2}/I_{2})= 0.56 \cdot (I_{2}/I_{1})$.}


\paragraph*{BC} The ``classical'' and  almost ubiquitous broad line component shows some line ratios that  yield  important constraints on the physical conditions:  \lya/\hb\ $\sim$ 10 \citep[see also ][]{netzeretal95},  compared to an expectation of $\sim$35 for pure recombination; \mgii/\lya  $\approx$ 0.1 -- 0.3; relatively large \rfe ($\ga  0.3$).  The \feii/\hbbc\ values assume that all or most \feii\ arises in the same region as H$\beta$ BC. Removal of the VBC  part of \hb\ in the \rfe\ estimation increases values over the previously published ones where no \hb\ VBC correction was made, yielding    \rfe $\sim 0.5$ for  Fairall 9 and 3C 249.1.  \civ/\lya\  estimates range from $\approx$ 0.15 for I Zw 1 and Mark 478 increasing to  $\approx$ 0.5 for Mark 335 and all Pop.  B sources. An apparent anti-correlation between \civ/\lya\ and \rfe\ as well as a positive correlation between \siiii/\civ\ and \rfe\ for Pop. A sources are visible in median composite spectra of \citet{bachevetal04}. Our data in Table \ref{tab:bc} confirm  a  positive trend  between \siiii/\civ\ and \rfe\ for the 3 Pop. A sources, although low \siiii/\civ\ values are observed for \rfe$\approx$0.5. The largest \siiii/\civ\ (and lowest \civ/\hb)  ratios are associated with the largest \rfe\ ($\ga$ 0.5) estimates, probably implying metal enrichment (\S \ref{metals}).

\paragraph*{BLUE} The blueshifted component is less constrained due to its weakness in many lines/sources. The only ratios that can be measured involve  \lya, \civ, and,  with greater  difficulty \heiiuv.  Given the importance of the \lya/\hb\ ratio we made a special effort to derive an upper limit for it. They are estimated assuming that any \hb\ emission is peaking at less than 3$\sigma$ the noise level, meaning that the upper limit to the line flux can be written as $I_\mathrm{p} \cdot \sqrt{2 \pi} \sigma \approx 1.067 I_\mathrm{p} \cdot \mathrm{FWHM}$, where $I_\mathrm{p}$ is the peak line intensity.  The lower limits indicate  that BLUE values of  \lya/\hb\  are much  higher than those of BC and  VBC. This has important  physical consequences (\S \ref{phys}).

\paragraph*{VBC}   Visually, the observed \feiiopt\ emission is fully consistent with an origin in the BC (i.e., no obvious evidence  for any any \feii\  VBC).  Even if we measure a FWHM of \feii\ consistent with the one of \hbbc\ for our 3  Pop. B sources, the VBC large width $\ga$ 10000 \kms\ could create a  pseudo-continuum underlying \hb.   To settle the issue of a possible \feii\ VBC  we attempted to fit  a VBC of  \feiiopt\   to the B1 and B1$^+$\ composites  from the bright SDSS sample of \citet[][S/N $\ga$200]{zamfiretal10}. The assumption of  \feii\ VBC with the same shift and width of \hbvbc\ leads  to   implausible results with very large $\chi^{2}$\ unless \feii\ VBC is negligible. We conclude that we have no strong evidence of a significant \feii\ VBC emission.

The red wing of the $\lambda$1900  blend is ascribed to \ciii\ VBC emission, as the iron spectrum of Pop. B sources is always  weak.


\subsubsection{The \mgii/\hb\ ratio}
\label{mg2}

With the goal of verifying whether the values of \mgii/\hb\ reported in Table \ref{tab:bc} and   \ref{tab:vbc} are  typical, we selected a sample of quasars brighter than $g$ =19.0 with SDSS spectra that included both \hb\ and  \mgii\ (redshift range $z $ = 0.40 -- 0.75). While most of these spectra show moderate-low S/N they are good  enough to allow one to assign the sources to 4DE1 bins in which context high S/N composite spectra can be computed (normalized to specific flux at 5100 \AA). We obtained 160 sources in bin B1 and 58 in B1$^+$.  The VBC \mgii/\hb\ ratios we derive from the bin B1 and B1$^+$ composite spectra are 0.7 and 0.5, respectively. Given Pop. A  values in Table \ref{tab:vbc}  and the consistent value \mgii/\hb $\approx 1.5$ computed on the median spectra,  we conclude that the VBC shows \mgii/\hb\ ratios 2--3 times lower than the BC. Due to the lesser prominence of the VBC in Pop. B sources, the FWHM of \mgii\ becomes significantly lower than that of \hb\ \citep{wangetal09}.

\subsection{Physical Conditions }\label{phys}
We generated a multidimensional grid of {\tt CLOUDY} \citep{ferlandetal98} simulations in order to infer ionization  parameter $U$\ and electron density \ne\ values from our spectral measurements. Simulations span the density range  $7.00 \le \log$ \ne $\le 14.00$, and $-4.50 \le \log U \le 00.00$  with an interval of 0.25 in both $\log U $\  and $\log$\ne\  \citep[c.f. ][]{koristaetal97}. Each simulation was computed for a fixed ionization parameter  and density assuming  plane parallel geometry. Part or all of this 2D grid of simulations was repeated  assuming  \nc = $10^{21}, 10^{22}, 10^{23}, 10^{24}, 10^{25}$ \cmq. Metallicity was assumed to be either solar or five times solar.  Two alternative input continua were employed: (1) one assumed to be the standard AGN continuum by {\tt CLOUDY} which  is equivalent to the continuum described in \citet{mathewsferland87} and (2) the low-$z$\ quasar continuum of \citet{laoretal97b}. We exploit the simulations to deduce constraints on $U$, \nc, and \ne\ from the most  reliable intensity ratios we derived in this paper.

\paragraph*{BC} The BC is the better constrained component  with relatively large \rfe\ and ratios \aliii/\siiii, \siiii/\civ\ requiring high density ($\log$ \ne\ $\sim$ 12), low ionization ($-3 \la \log U \la -2 $) and large column density (\nc $\ga 10^{23} $\cmq).  These conditions yield values of the important \lya/\hb\ and \mgii/\lya\ ratios that are consistent with the ones observed, and explain \rfe\ $\la 0.5 - 1.0 $ as well. Larger \rfe\ values suggest super-solar metallicity (\S \ref{metals}).  Hereafter we will refer to the BC-emitting region as the LIL-BLR while we will retain the term BLR in its more general meaning i.e., as the region where all of the broad emission is produced.


\paragraph*{BLUE} The very large \lya/\hb\ ratio suggests radically different conditions including lower optical depth and consistency  with \nc $<$ 10$^{23}$\cmq\ (see \S \ref{thin}) .
The \lya/\hb\ and the \heii/ \civ\ ratios suggest high ionization $\log U \sim -1$ and densities in the range \ne $\sim 10^{9.5} - 10^{10.5}$ \cm3. There is no detection of a \ciii\ blueshifted  component in the $\lambda$1900 blend, but this is consistent with high ionization gas at density $\log $\ne $\sim$ 10: the expected \ciii/\civ\ ratio for $\log U \sim -1$ is $\approx$0.1, and any blueshifted \ciii\ emission would be lost in the 1900\AA\ blend. We included  a component with the same shift and width derived for \civ\ in the fits of F9 and 3C 249.1, but {\tt specfit} yielded 0 or negative values in the final fit.

\paragraph*{VBC} If \rfe$\sim$0 in the VBC and given the low \mgii/\lya\ ratio, we infer a high ionization parameter $\log U \simgt -1$.   However, Results for VBC are contradictory with all ratios consistent with high ionization but some ratios indicating high and others low density.

We  observe \ciii/\civ $\sim $ 0.1 -- 0.2 which indicates a relatively low density emitting region with \ne $\sim 10^9 - 10^{10}$ \cm3. The observed   \heii/\civ\ ratio (0.1 -- 0.2) is also consistent with this picture. Much higher density would collisionally quench \ciii\ to undetectable levels. Not all VBC line ratios are consistent with moderate density picture. If the Balmer  decrement is very flat,  and the \lya/\hb\ ratio in between 5 and 10, then very high density gas (\ne $\sim 10^{13}$\cm3) is required \citep{netzeretal95} as well as a very  large optical depth if photoionization is assumed. A large column density \nc $ > 10^{23}$cm$^{-2}$\ yields conditions consistent with these   line ratios measured for the VBC. \citet{sneddengaskell07} computed line  profile ratios  \lya/\ha\ and \ha/\hb\  for several luminous Seyfert 1 nuclei (mainly Pop. B) and reached similar conclusions in favor of large \ne\ and \nc.

\subsubsection{Chemical Abundances and \feiii\ Contribution}
\label{metals}

The LIL-BLR is the only region allowing for  tentative abundance considerations;  results should  obviously be applicable to the other emitting regions. Ratios involving metal lines should be considered the most robust for deriving physical inferences. Intermediate-ionization lines (like \aliii\ and \siiii) are  produced in a fully ionized region which makes them less dependent on column density than the LILs \citep{negreteetal10}. From the simulations, we find that the ratio \aliii/\siiii\ is not strongly dependent on $Z$: the ratio increases by about 40\%\ passing from $Z = 1 Z_{\odot}$\ to $Z = 5 Z_{\odot}$, for $\log $\ne $\approx 12$ and  $\log U \approx$--2.  However,   \civ/\siiii\ (or, equivalently \civ/\aliii) is much more affected. The low   \siiii/\civ\ and the other line ratios we measure  for  Mark 478 can be all accounted for  with $Z = 5 Z_{\odot}$, $\log$ \ne  $\approx 11.6$, and $\log U \approx$ -2.4. No solution with a well defined value of \ne\ and $U$\ is found if $Z=Z_{\odot}$\ is assumed. Even higher metallicity,  $\sim 10 Z_{\odot}$,   is strongly suggested for I Zw 1.  The assumption of metallicity above solar is consistent with  observations of high-$z$\ quasars  \citep[e.g. ][]{hamannferland99,juarezetal09}, and the two sources where the evidence of higher abundance is strong also show the most prominent \nv\ line in the \lya\ panels of Fig. \ref{fig:profiles}, as expected. 

In the cases of high metallicity we expect that  \feiii\ should be especially prominent in the \l 1900 spectral region. The physical conditions derived for the LIL-BLR  remind to the  ÒWeigelt blobsÓ of $\eta$\ Carin\ae, located in the equatorial plane of the system, perpendicular to the symmetry axis of the bipolar lobes forming the ``Homunculus'' nebula. The Weigelt blobs are believed to be dense gas photoionized by the radiation associated to the central, massive star and to a possible companion \citep[e.g., ][]{johanssonetal00,davidson05}. The spectrum of the Weigelt blobs shows very weak  \ciii\  along with a prominent line at \l1914, ascribed to the $z^7 P_3^0 \rightarrow a^7 S_3$ \feiii\  transition. Although this is not a resonant line (the lower level is $\approx$ 3eV above ground), the line appears very strong because the upper level is populated by \lya\ fluorescence.  Indeed, in I Zw 1 where lines are narrow, the peak emission at \l1914 is actually visible, and the \feiii\ emission around 1900 \AA\ may resemble the one of Model 3 of \citet{vestergaardwilkes01}.

\section{Discussion}

\label{disc}

{ In the following we will discuss measurements in fixed interpercentile velocity ranges (\S \ref{iv}) as an approach complementary to our multicomponent analysis. } After considering properties of the BLUE component (\S \ref{thin}) we will show how our results on the physical properties of the different emitting regions do not favor a predominance of line emission from an accretion disk (\S \ref{disk}) or from a binary BLR (\S \ref{binary}).    We suggest that the properties of the BLUE, BC and VBC emitting regions as well as their trends along the E1 sequence can be explained by the interplay of gravitational and radiation forces (\S \ref{ledd} and \S \ref{gravrad}). { Last, we will discuss our interpretation of the BLR and compare it to past work (\S \ref{ps}). } 

\subsection{Interpercentile Velocity Measurements}\label{iv}

{ The true  profile shapes of the various line components are uncertain. The Lorentz-like shape of Pop. A \hb\ profiles can be considered the most accurate assuming that the peak and the wings do not represent independent emission components \citep[e.g., ][]{lamuraetal09}. The lack  of inflections in even the highest S/N composite profiles \citep{zamfiretal10} motivated us to favor the single component Lorentz fits. Uncertainties in our decompositions force us to restrict the analysis to estimates  of the relative component intensities, average line shifts and widths. A double Gaussian function is a good  approximation but the individual Gaussian  fits may provide uncertain estimates of the relative BC and VBC fluxes in the core of the line where the two components overlap in radial velocity. It is difficult to say whether a stronger core component and a less broad, fully redshifted VBC provide a worse fit, although the SDSS  composite spectra of \citet{zamfiretal10}  and earlier work on PG~1416-129 \citep{sulenticetal00c} favor our BC-VBC decompositions. Also, if our interpretation of some sources  showing \hb\ that is almost pure VBC \hb\ (3C110 in this paper) is correct, then the Gaussian functions that we employ represent a reasonable description of the VBC profile shape. Measuring line ratios at fixed inter-percentile velocities is an alternative approach that might have  a more direct physical meaning if the BLUE and VBC matter only in the highest radial velocity ranges.  We therefore report emission line fluxes measured in three radial velocity ranges of $\Delta$\vr $= 2000$\kms\ centered at 0 \kms\ (``0'' range), $\pm$4000 \kms\ (B and R ranges for Pop. A sources) and 0 \kms, $\pm$6000 \kms\ (Pop. B sources)  in Table \ref{tab:iv}. The reported measurements were carried out after subtraction of \feii, \feiii\ and  of contaminating narrow and broad lines. Errors are estimated at a $2\sigma$\ confidence level as twice the rms scatter in the radial velocity range times the $\Delta \lambda$ in \AA. Two different centerings of the radial velocity ranges  are needed to account for the line width difference of Pop. A and B sources. The overall interpretation as deduced from the multicomponent fit is confirmed. In particular, the data of Table  \ref{tab:iv} indicate that there is no evidence of an additional red component for Pop. A objects.  There is also no convincing evidence of a BLUE component in \hb\ of Pop.  B sources. The large error bars associated with the \lya/\hb\ ratios support the same conclusion motivated by the multicomponent fits for the BLUE component: the upper limits and the possible detection of I Zw 1 imply \lya/\hb $\gg$ 10.  Line ratios for the B, 0 and R ranges yield  the same qualitative conclusions as the multicomponent fits, with the limitation that we cannot obtain meaningful results for  the heavily blended features at $\lambda$1900. Also, the large error bars in the  \hb\ R range of Pop. B sources reflect the residuals of [{O {\sc iii}}]\l4959 subtraction.  A comparison of our \lya/\hb\ ratios with those of \citep{netzeretal95} and of our \ha/\hb\ ratios (from the multicomponent fits) with those of \citet{sneddengaskell07} show agreement especially in the redshifted line wings i.e., for the VBC. This suggests that our approach yields consistent  values although the true shape of the VBC component is uncertain especially in the range $-1000  $\kms $\le$\vr$\le 1000$\kms. }


\subsection{Optically Thin Gas?}
\label{thin}


The weakness of the response to continuum changes in the \hb\ wings and inter-line comparisons e.g.   have motivated arguments that the VBC is mainly produced in optically thin gas \citep{morrisward89,zheng92,shieldsetal95,sulenticetal00c,corbinsmith00}. On the other hand, several authors have interpreted the response of the blue wings in \civ\ and \hb\ as evidence for predominantly optically thick gas  \citep[e.g. ][and references therein]{turlercourvoisier97,koristagoad04,sneddengaskell07,shapavalovaetal10}.    \citet{koristagoad04} showed that an optically thick VBLR can explain the stronger core response of \hb\ to continuum changes. The small \lya/\hb\ and \ha/\hb\ ratios suggest a large column density \nc\ for both LIL-BLR and VBLR.  This interpretation may be correct for some transient sources but is unlikely to apply to more stable sources that show a strong red-ward asymmetry (i.e. Pop. B quasars). In an optically thick medium the intensity of a recombination line depends on the luminosity  of the ionizing continuum; if the medium is optically thin it depends on the amount of emitting gas. In practice this places a constraint on the column density \nc.  In high luminosity quasars optically thin gas may not be adequate, even with a covering factor approaching unity, to explain the equivalent width and luminosity of the observed VBC  \citep{marzianietal06,sneddengaskell07}.

Is the blueshifted component due to optically thin gas perhaps ejected from the central engine?  Fig. \ref{fig:thin}  shows the predicted equivalent width of \lya\ and \lya/\hb\ intensity ratio (thin lines) as a function of the ionization parameter $U$ for \nc $= 10^{21}$ cm$^{-2}$\ and \nc $= 10^{22}$ cm$^{-2}$. Computations assume total covering, solar metallicity and \ne$=10^{10}$\cm3. W(\lya) remains approximately constant until the optically thin domain is entered at a ``critical'' value of $U$. Beyond that value  W(\lya) is roughly proportional to the inverse of $U$: in the optically thin domain the  increase in continuum luminosity does not give rise to a corresponding increase in line luminosity. The most stringent observational boundary is set for Mark 478 and Fairall 9, with \lya/\hb $\ga$45. At the same time, the median \civ/\lya\ and \heiiuv/\civ\ values suggest $-1.5 \la U \sim -1$\ if $\log $\ne = 10. Considering that the observed W(\lya) is in the range  15 -- 60 \AA, low column density gas at \nc $= 10^{21}$ cm$^{-2}$\ can explain the largest observed equivalent widths only if there is substantial covering of the source, i.e., the source is almost completely covered by a { geometrically} thin shell ($\sim 10^{11}$ cm) of optically thin gas. A smaller covering fraction is required if \nc $= 10^{22}$ cm$^{-2}$ \ or larger.

\subsection{Alternative Interpretations}

\subsubsection{Accretion Disk Contribution}
\label{disk}
Some Pop. B sources show very broad double-peaked H$\beta$ profiles that have been interpreted as the signature of accretion disk line emission \citep[e.g., ][]{eracleoushalpern03,stratevaetal03}. { A  double peaked signature has been found for \mgii\ in Arp 102B \citep{halpernetal96}. } Others have argued that double-peaked profiles are too rare and disk model solutions too disparate to  justify the disk emission hypothesis \citep{sulentic89}. Attempts have been made to find larger samples of double-peaked sources \citep{stratevaetal03,bonetal09}; however, most (90\%) of these source are not double-peaked but rather red  or blue asymmetric. A large sample of the highest S/N SDSS spectra binned in the 4DE1 context \citep{marzianietal09,zamfiretal10} do not show any hint of a double-peak signature in high S/N ($\approx$250) composites. Broader pop. B sources unsurprisingly favor red asymmetries. Sources included in a given Pop. B  bin of Figure 1 show a diversity of red/blue bumps and asymmetries however bin composite spectra show a BC+VBC blend similar to the one observed in e.g. Fairall 9 and 3C 249.1 of this paper.  Where is the disk signature? If a significant part of the broad lines is produced in an accretion disk then it must be obscured or masked by non-disk emission components \citep[e.g.][]{zhangetal09}.  For Pop. B sources, one could argue that the {\em sum} of the BLUE and VBC  represents a double peaked component accounting for the broadest part of the line profile with  the BC emitted elsewhere.  A difficulty with this interpretation is that the gas in the BLUE and VBC appear to be arise in regions with different physical conditions. The BLUE component is not consistent with high density and  large column density as expected in standard accretion disk models \citep[e.g.,][]{dumontcollinsouffrin90b}.  In the case of Pop. A sources the narrower Lorentz-like profiles, interpreted as a disk signature, would require line emission from a disk with radius $\sim$10$^6$ gravitational radii which is theoretically and observationally disfavored \citep{collinzahn99}.  The evidence continues to grow weaker for a predominance of accretion disk emission in Pop. A and Pop. B line profiles. { Only a minority of very-broad sources whose prototype is Arp 102B remain consistent with accretion disk predictions \citep{stratevaetal03,eracleoushalpern03,gezarietal07}.}

\subsubsection{Binary BLR}
\label{binary}

If quasar activity is driven by merger and accretion processes then we might expect to find some/many sources with evidence for a binary black hole. At least some of these sources might involve two AGN with associated BLRs. It has been proposed that binary BH configurations might give rise to double-peaked line profiles \citep[][see \citealt{gaskell09} for a recent review]{gaskell85,halpernfilippenko88,shenloeb09,lauerboroson09}. Two co-orbiting BH each with a BLR producing (typical) single-peaked broad lines could give rise to double or single peaked composite line profiles depending on the line-of-sight orientation and radial separation of the BHs. A large peak shift   $v_\mathrm{r} > 2000 $ \kms\ is not unphysical since  $v_\mathrm{r} \approx 2000 $ \kms\ is less than the typical virial velocity at the reverberation estimated radius of the BLR. 

This idea is not easily applicable to Pop. A sources which show quite narrow and symmetric H$\beta$ profiles. The situation for Pop. B sources appears more favorable if one considers, for example, a binary BLR origin of the BLUE and redshifted VBC components. The {\it reductio ad extremum} would view all pop B sources as binary pop A sources. One problem with this view would involve the fact that virtually all pop. B sources are pop. B. However in some pop. B sources unshifted BC gas might be associated with a low-ionization region at larger radius and distributed around the center of mass of the two orbiting BH. The radial velocity  of the first component would be $v_\mathrm{r,1} = \sqrt{G M/d} \cdot\sin i \cdot \sin (2\pi t/ P) \cdot m_{2}/M$, while for the second   $v_\mathrm{r,2} = \sqrt{G M/d}\cdot\sin i \cdot \sin (2\pi t/ P + \pi) \cdot m_{1}/M$ at any given orbital time $t$. $P$\ is the orbital period,  $m_{1}$\ and $m_{2}$\ the BH masses, with $M = m_{1}+m_{2}$, $d$ the black hole separation. The inferred properties of the  BLUE and VBC components  lead one to infer that the masses of the two BHs should be roughly equal. The \mbh\ similarity again raises questions about  why we find very different \lya/\hb\ ratios and inferred physical conditions for the BLRs around the two BH. In addition, the broadest profiles should ``narrow'' dramatically on a time scale of $\approx \pi d / 2 v \approx 10 d_{16} v^{-1}_{1 000} $yr.  Note that the  profile presented in Figure 10 of the \citet{shenloeb09} paper is {\em not} double-peaked but a prototypical Pop. B sources showing the red asymmetry that we suggest is due to a distinct VBC emitting region. If one ascribes a (broader) core component to a more massive black hole and the redward asymmetry to a narrower component (less massive BH), then the signs of the two component radial velocities $v_\mathrm{r,1}$\ and $v_\mathrm{r,2}$\ should always be opposite, which is  not the case in the  majority of Pop. B sources. { The expected change in radial velocity is not seen  in Arp 102B-like double peaked sources  either \citep{eracleousetal97,gezarietal07}. }

\subsection{Role of Mass and Eddington Ratio}\label{ledd}

Table \ref{tab:mass} reports the \mbh\ and \lledd\ values for the sources following the prescription  of \citet{vestergaardpeterson06} where: Col. 1 - source name; Col. 2 - specific flux at 5100 \AA; Col. 3 - bolometric  luminosity assuming a correction factor of 10 from the observed luminosity at 5100 \AA; Cols. 4-5 - FWHM of the entire \hb\ profile and FWHM of \hbbc\ as  derived from the $\chi^{2}$ fits; Cols. 6-7 and 8-9 present the corresponding black hole masses and Eddington ratios. \mbh\ values are reduced by $\Delta \log $ \mbh$\approx$ 0.3 when the BC component is employed.  The effect is larger  (factor of 10) if \civ\ is used as a virial proxy for \hb\ in Pop. A sources \citep{sulenticetal07}. Sources follow a sequence of increasing mass and especially decreasing \lledd. Pop. A sources show the highest  Eddington ratios by an order of magnitude or more. A correction by radiation pressure  would leave the Eddington ratio trend unaffected \citep{marconietal08}.  It is interesting to note that the VBC occurs only in Pop. B sources where \lledd $< 0.1$, but BLUE is strong in both Mark 478 and 3C 249.1 whose \lledd\ is different by almost an order of magnitude.


\subsection{Gravitational and Radiative Acceleration}
\label{gravrad}
If line emitting gas is optically thick to the Lyman continuum then the radiation will exert an outward acceleration that  is inversely proportional to the column density and proportional to the ionizing luminosity. The ratio between the radiative and gravitational  accelerations is:

\begin{equation}
r_\mathrm{a} = \frac{a_{\mathrm{rad} } }{a_{\mathrm{grav} } } \approx 0.088 L_{44} M_\mathrm{BH,8}^{-1} N_\mathrm{c,23}^{-1}
\end{equation}

where \mbh\ is  in units of 10$^8$ solar masses, and $L_{44}$\ is the luminosity of the ionizing continuum ($\lambda < 912$\AA) in units of 10$^{44}$\ergss. The equation can be written in the convenient form

\begin{equation}
r_\mathrm{a} =  \frac{a_{\mathrm{rad} } }{a_{\mathrm{grav} } } \approx 7.2\, \frac{L_{\mathrm{ bol}}}{L_{\mathrm{Edd}}}  N_\mathrm{c,23}^{-1}.
\end{equation} If $r_\mathrm{a} \gg1$ then radiative acceleration dominates. This may be the case for the BLUE component if we interpret the blueshift as a Doppler shift. A condition of equilibrium may be reached at $r_\mathrm{a} \approx 1$. If \lledd$ = 1$, the corresponding \nc\ is $\simlt 10^{24}$ cm$^{-2}$,  a plausible value for the LIL-BLR. Finally if $r_\mathrm{a} \ll1$ the emitting gas may be unable to withstand the central black hole gravity and may fall toward the center giving rise to the observed redshifted VBC. This interpretation  is compatible with large \nc\ values for the VBLR and would naturally explain why the VBC is observed in objects with low \lledd $\simlt 0.1$. Of particular relevance to Pop. A sources are emitting clouds moving under the combined effect of gravitation and radiative acceleration \citep{mathews93,netzermarziani10} that  produce the observed Lorentz-like profiles. If \lledd$\la$0.1, as in Pop. B, gas of $\log $ \nc $\sim 22$\ would be pushed away  while gas even larger column density could flow out in Pop. A sources.

\subsection{Radio Loudness}

{ If we compare the \hb\ profile of RQ and RL {\em in Pop. B only} i.e., where
most radio loud objects are found, there is no major difference although RL objects show a preference toward more extreme redward asymmetries \citep{marzianietal03b}. If we look at high ionization lines of RL objects, the \civ\ profile can show a blueward asymmetry although a redward asymmetry is more frequent \citep{marzianietal96,sulenticetal07,punsly10}. Radio quiet objects {\em of Pop. B} show a somewhat different behavior: \civ\ BLUE is systematically stronger and the profile is more often symmetric or blueward asymmetric. The \citet{sulenticetal07} data suggest that radio loudness may modulate the relative intensity of BLUE and VBC, lowering BLUE, perhaps because radio ejections affect the radial flow seen in RQ quasars \citep{marzianietal96}. BLUE is however definitely observed in RL objects \citep{bartheletal90,willsetal95,smalletal97}.  } 

\subsection{Previous Studies and the Present Interpretation of the BLR}
\label{ps}

{ We are aware of only one other attempt to model all major line profiles in a sample of low $z$\ quasars. \citet{shangetal07} studied 22 sources and modeled all lines using two components--broad and very broad. Their sample is larger and allows for some correlation analysis. However, Pop. A sources are likely overrepresented, a result of their use of an X-ray selection that favors Pop. A sources. The results of \citet{shangetal07} should be therefore  related mainly to three Pop. A sources included in our investigation even if some Pop. B present in their data suggest a change in trends at FWHM(\hb)$\approx$ 4000 \kms\ (their Fig. 5). Their major empirical result, namely that  \civ\ and \lya\ are very different from \ha\ and \hb\ in terms of line width, asymmetry, and velocity shifts  is confirmed by the present study as far as \hb\ is concerned. Our results also suggest, along with \citet{shangetal07}, that a  simple radially stratified ionization structure of the BLR is unable to explain the different line profiles.  A wind component has been isolated through our E1-based approach  and three components appear to be a minimum set for the reasons summarized in the introduction.  The 4DE1 approach unifies the lines within specific sources and allows for the definition of  a trend among the sources.  BLUE and LIL-BLR are characterized by well defined properties. The VBLR is a region that could be defined by virtue of high ionization and, if our considerations are correct, large \nc.  Within these constraints a large range of parameters -- especially density -- is suggested. The VBLR is therefore the only region where a locally optimally emitting cloud scheme  -- where a range of properties occurs at every location \citep{baldwinetal95}-- could apply. 

Recent reverberation mapping studies attempted to consider response of lines to continuum changes in bins of radial velocity \citep{bentzetal08,denneyetal09a}. Infall is suggested by the observation of a faster response to continuum change in the red wing of \hb\ in NGC 3516 and Arp 151. Arp 151 belongs to a perturbed system, and infalling gas can give rise to a redward asymmetries  on spatial scales much larger than the one of the BLR \citep[e.g., ][]{rafanellimarziani92,marzianietal03c}. Even if it is too early to generalize, these results are consistent with the interpretation of  the \hb\ redward asymmetry/redshift  in Pop. B sources \citep[e.g., ][]{gaskell09a}. As far as the VBC is concerned, a possible problem with an infall interpretation is related to the line anisotropy. Very high \nc\ clouds should prevalently emit lines facing the central photoionizing continuum source; this is true also for \hb\ \citep{ferlandetal92}. Therefore, if line emission is anisotropic, gas should be visible from the far side of the continuum source and infalling gas should yield a net blueshift. However, if emission comes from the non-illuminated side of the cloud \citep{ferlandetal09}, and the more distant part of the infalling gas is obscured a net shift to the red is expected as frequently observed in the line profiles. 

 }


\subsubsection{On the Origin of the Broad Component}
{ 
The  BC may be distinct from the other emitting regions and in many sources (Pop. A) may be the dominant or only visible  component in all LILs. 
Two defining BC properties involve low ionization degree and symmetry of the line profiles. We find BC width FWHM$\leq$ 5000 \kms with fractional peak shift $\Delta v_{\mathrm r}$/FWHM $\la 0.1$. This leads to the conclusion that the LIL-BC may be the only H$\beta$ component that  is a valid virial estimator. The width of \feii\ and especially of \mgii\ likely offer the next best virial estimators \citep{sulenticetal06,wangetal09}. 

The necessity of a low-ionization emitting sub-region within the BLR has been stressed since long, especially by S. Collin and collaborators \citep{collinsouffrinetal88}. Observational support in favor of this idea came by the realization of the very different profiles of LILs and HILs, as reviewed in the introduction. More recently, the analysis of the \caii\ IR triplet  suggests that a rather dense gas with density $\log$ \ne\ 11.5 and $\log U \sim -2.5$ is present in the BLR and that is  primarily responsible for the observed O{\sc i}, Ca {\sc ii}, and Fe{\sc  ii} lines, based on the resemblance of their profiles \citep{matsuokaetal07,matsuokaetal08}.  There is a close agreement between the values derived by \citet{matsuokaetal07} and the values derived in this paper through the ratios \aliii/\siiii, \civ/\siiii\ \citep[see][for a more detailed report]{negreteetal10}. 

LIL emission from AGN has been the subject of a controversy related to the ionization mechanism: \citet{collinsouffrinetal88} and \citet{collinsouffrindumont89} suggested mechanical heating to explain the LIL prominence and especially the \feii\ intensity.   Explaining strong LIL emission under the assumption of solar abundance has been a challenge for photoionization models. The  \civ/\heiiuv\ and \siiii/\civ\ intensity ratios lead us to conclude that the assumption of metal enrichment is appropriate. Among Pop. A objects there could be an important role of metallicity in farther defining the properties of the LIL-BLR.We see indeed a trend in the prominence of the \nv: the most prominent \nv\ is observed in I Zw 1 and Mark 478. If the \nv/\civ\ ratio can be used as a metallicity indicator, then our six objects might follow the accretion rate-metallicity relationship pointed out by \citet{shemmeretal04}.  }  An increase in iron abundance from a few to several  times solar \citep[in agreement with much previous work, e.g.][]{hamannferland99} is suggested for sources with \rfe $\ga 1$\ (A2+ bins). 

The low-ionization BC seems to be preferentially associated with young, or rejuvenated, systems like NLSy1 galaxies \citep{mathur00,sulenticetal00a,sulenticetal08} whose \lledd\ is relatively large ($0.2 \la$\lledd$\la 1$).  Enhanced star formation in NLSy1s has been recently demonstrated from analysis of Spitzer data \citep{sanietal10} while clues have existed for a long time \citep[e.g. ][and references therein]{sandersetal88,krongoldetal01}. Higher metallicity is likely a consequence of top-heavy starbursts that may be frequent in galaxy nuclei \citep[e.g.][]{bonnellrice08}. Such starbursts are expected to  sustain the high  accretion rate typical of Pop. A sources  and to provide the enriched LIL-BLR material in extreme Pop. A sources like I Zw 1.

\section{Conclusion}

Ignoring the composite nature of broad emission lines in quasars will likely result   in a poor or erroneous understanding of their emitting regions \citep[as known for 30 years; e.g.,][]{gordonetal81}. The contextualization of 4DE1  not only shows that quasars are  spectroscopically diverse but it provides a way to organize the differences in order to  facilitate interpretation of the emitting components in  different classes of sources.  Analysis of  6 low-$z$ quasars that explores 4DE1 diversity { and are believed to be representative of the wide majority of low-$z$ quasars} suggests that the strongest  optical/UV emission lines involve one or more of three different components that can be  defined by their shifts with respect to the rest frame of each quasars: a redshifted VBC,  an unshifted BC component, and a blueshifted (BLUE) component.   The three components have  been tentatively isolated through models of the strongest  lines from \lya\ to \mgii\ and \feiiopt\ using \hb\ as a template, and characterized as follows.

\begin{enumerate}

\item The BC is the broad line component present in the overwhelming majority of type-1 AGN.  \feii\ and \mgii\ emission likely arise in the same region. Other broad lines usually show a  BC component although there may be exceptions:  \ciii\ may be very weak  in extreme Pop. A sources where metallicity enhancement is most in evidence.  At the other extreme of the 4DE1 sequence some Pop. B sources may show only weak or even zero  BC emission.

\item Pop. A sources show dominant BC emission in LILs and strong BLUE component  emission in HILs.  The BLUE component is most prominent in NLSy1s, but is detected in most quasars, including Pop. B sources. The  prominence of the BLUE  component in extreme Pop. A NLSy1s (almost always radio-quiet) suggests that it is not closely  related to any relativistic outflow due to radio jets. It has been interpreted as the signature of an accretion disk wind where the receding part of the flow is obscured by the disk \citep{marzianietal96}. If the interpretation of \S \ref{gravrad} is correct, then BLUE  may be due to gas upon which radiative forces dominate. { We are able to constrain ionization parameter, density and column density of this outflowing region. These quantities are important to assess feedback effects on the interstellar medium of the quasar host galaxy.}

\item  The redshifted VBC component can be described as  a defining LIL property in Pop. B sources.  The BC is distinguished empirically from the VBC  by its much larger FWHM and significant  redshift. The low Balmer decrement and \lya/\hb\ ratio suggest, if we assume photoionization as the heating source of the gas, high ionization and large column density. The physical conditions in the VBLR are not clear, and probably involve a stratification of properties  that we are unable to resolve at this time. On the one hand, the \ciii/\civ\ ratio suggests high  ionization and moderate density. On the other  hand, the strong \heiiuv\ emission, the low \lya/\hb\ ratio as well as the flat Balmer decrement (\ha/\hb$\approx1$) are better explained by extreme density at very high ionization. In this region a locally optimally emitting cloud scheme \citep{baldwinetal95} could be appropriated. \end{enumerate}

\thanks\newpage
\hoffset=-0cm\null
\begin{table*}
\begin{minipage}{150mm}
\caption{Selected Sources\label{tab:obj}}
\begin{tabular}{@{}lllrclll}
\hline
{IAU Code} &   NED Name    & $z$  & \mb & $A_\mathrm{B}$ & Sp.T. & Comments\footnote{RM: Reverberation Mapping data; listed in \citet{petersonetal04}.}
\\ \hline
J00535+1241 & UGC 00545 & 0.0605&14.4 &0.28 & A3 &  $\equiv$  I Zw 1, \ha\\
J14421+3526& MRK 0478& 0.0771&15.0& 0.06 &A2&\\
J00063+2012& MRK 0335& 0.0258& 14.0&0.15 & A1 & RM, \ha \\
J01237$-$5848 &Fairall 9& 0.0462&13.5 & 0.12& B1 &  RM. \ha \\
J11042+7658 & PG 1100+772 & 0.3116 &15.7& 0.15 & B1$^+$ &$\equiv$ 3C 249.1, no MgII \\
J04172$-$0553 &3C 110&  0.7744&15.8& 0.19 & B1$^{++}$& no CIII \\
\hline
\end{tabular}\end{minipage}
\end{table*}

 \begin{table*}
 \begin{minipage}{150mm}
  \caption{Relative Intensity of BLUE, BC, and VBC in the \civ\ Line \label{tab:comp}}
  \begin{tabular}{@{}llcrrrrr@{}}
  \hline
   Sp. T.   &  Name   & F(\civ)\footnote{F(C{\sc iv}) = BLUE + BC + VBC}& BC(C{\sc iv})/  & VBC(C{\sc iv})/  & BLUE(C{\sc iv})/ \\
                 &               &                                                                                       & F(C{\sc iv})      & F(C{\sc iv})          & F(C{\sc iv})       \\
                 &               &  [\ergss cm$^{-2} $]  &       &           &       \\
\hline
A3 &I Zw 1        & 0.74E-12 & 0.36  &       0.00 & 0.74\\
A2& Mark 478  & 1.17E-12 & 0.61 &  0.00 & 0.39  \\
A1 & Mark 335  & 6.31E-12 & 0.87& 0.00 &  0.13 \\
B1& F 9 &  5.84E-12   &  0.43  & 0.27  &   0.30\\
B1$^+$ & 3C 249.1  &   2.11E-12   & 0.17      &  0.31   &   0.38  \\
B1$^{++}$&  3C 110 &  1.60E-12  & 0.00 & 1.00 & 0.00 \\
\hline
 \end{tabular}
\end{minipage}
\end{table*}

\begin{table*}\
 \begin{minipage}{150mm}
  \caption{Broad Component  (BC): 1000 \kms $\la$ FWHM $\la 5000$ \kms, $\Delta v_\mathrm{r}\sim$ 0   \kms  \label{tab:bc}}
  \begin{tabular}{@{}llrrrrrrrrrr@{}}
  \hline
   Sp. T.   &  Name   & \multicolumn{8}{c}{Intensity Ratio}& W\footnote{Equivalent width in \AA.} \\ \cline{3-10}
   && C{\sc iv}/& Al{\sc iii}/ & Si{\sc iii}/  &Mg{\sc ii}/ & Fe{\sc ii}/ & \ha/ & \lya/&Mg{\sc ii}/ & \lya \\
   & & \lya &  Si{\sc iii}  & C{\sc iv} &\lya & \hb& \hb & \hb& \hb & \\
\hline
A3 &I Zw 1                 &   0.14 & 0.61 & 0.97  & 0.31   & 1.80  &  4.2              &  5.5  & 1.70 &  61 \\
A2& Mark 478           & 0.14 &0.41    & 0.47  & 0.09   &  0.95 &\ldots             & 10.3   &  0.94 & 87 \\
A1 & Mark 335          & 0.45 & 0.33   & 0.07  & 0.07   & 0.47  & 4.6               & 15.8  & 1.03 & 102 \\
B1& Fairall 9             &  0.53 &  0.32 & 0.18  & 0.21    & 1.04 & $\simlt$6.1 & 12.3  & 2.54 & 158\\
B1$^+$ & 3C 249.1  & 0.44 &  0.68  & 0.17  & \ldots & 0.54 & \ldots            & 10.7  &\ldots & 32 \\
B1$^{++}$&  3C 110 & 0.55 & \ldots & \ldots &0.08  &$\simlt$0.36\footnote{Assigned following the criterion that related FWHM -- S/N and minimum detectable intensity, as in \citet{marzianietal03a}. }& \ldots  & 17.0& 1.3 & 71 \\
   \hline
 \end{tabular}
\end{minipage}
\end{table*}

\begin{table*}
 \begin{minipage}{150mm}
  \caption{Blueshifted component (BLUE): FWHM $\sim 7000$ \kms, $\Delta v_\mathrm{r}\sim$ -3000 \kms\label{tab:blue}}
  \begin{tabular}{@{}llrrrrrrr@{}}
  \hline
   Sp. T.   &  Name   & \multicolumn{4}{c}{Intensity Ratio}& W\footnote{Equivalent width in \AA.} \\ \cline{3-6}
   &    & C{\sc iv}/ & He{\sc ii}/ & \ha/ & \lya/ &  \lya\\
   & & \lya & C{\sc iv} & \hb & \hb\footnote{Lower limits to \lya/\hb\ are estimated by the maximum contribution expected by a component of the same shift and width if peaking at 3$\sigma$ the noise level. See text for a detailed explanation.}  & \\
\hline
A3 &I Zw 1 & 0.25 & 0.41 & 4.2 &  $\sim$18 &  60\\
A2& Mark 478 & 0.66   & 0.17   &   \ldots &$\simgt$46 & 15 \\
A1 & Mark 335 &   0.45 & 0.67  &  \ldots &$\simgt$32 &16 \\
B1& F 9 &  1.05 & 0.14 & \ldots & $\simgt$46 & 30\\
B1$^+$ & 3C 249.1& 0.59  & 0.17 &\ldots &   $\simgt$32 & 53 \\
B1$^{++}$&  3C 110\footnote{Consistent with 0 intensity in all lines.}  & \ldots & \ldots & \ldots &\ldots &0 \\
   \hline
 \end{tabular}
\end{minipage}
\end{table*}

 \begin{table*}
 \begin{minipage}{150mm}
  \caption{Very Broad Component (VBC): FWHM $\sim 10000$ \kms, $\Delta v_\mathrm{r}\sim$ +2000 \kms  \label{tab:vbc}}
  \begin{tabular}{@{}llrrrrrrrr@{}}
  \hline
   Sp. T.   &  Name  & \multicolumn{7}{c}{Intensity Ratio}& W\footnote{Equivalent width in \AA.} \\ \cline{3-9}
   && C{\sc iv}/&He{\sc ii}/ & C{\sc iii}/  & Mg{\sc ii}/& \ha/ & \lya/ &Mg{\sc ii}/ &\lya\\
   & & \lya & C{\sc iv}    &C{\sc iv}/ & \lya  & \hb & \hb &\hb  &  \\
\hline
A3 &I Zw 1 &  \ldots & \ldots &\ldots & \ldots & \ldots & \ldots & \ldots &0\\
A2& Mark 478 & \ldots & \ldots &\ldots & \ldots & \ldots & \ldots & \ldots &0  \\
A1 & Mark 335 & \ldots & \ldots &\ldots & \ldots & \ldots & \ldots & \ldots &0 \\
B1& F 9 & 0.36 &  0.13 & 0.18 & 0.08  & 1.26& 9.0 & 0.68  & 88 \\
B1$^+$ & 3C 249.1 & 0.67 &   0.17 & 0.17 &  \ldots  & \ldots & 4.5 & \ldots & 41 \\
B1$^{++}$&  3C 110  & 0.37  & 0.10 & \ldots & 0.04 & \ldots  & 5.0 & 0.2 & 27\\
   &     3C110\footnote{VBC only; see text for details.}        &          0.50   &  0.10 & \ldots & 0.07 & \ldots & 12.6 & 0.83 & 97 \\
\hline
 \end{tabular}
\end{minipage}
\end{table*}
\clearpage

\rotate
\voffset=6cm\pagestyle{empty}
\begin{table*}\setlength{\tabcolsep}{1.5pt}
 \begin{minipage}{190mm}\small
  \caption{{ Interpercentile Velocity Measurements \label{tab:iv}}}
  \begin{tabular}{@{}lcccccccccccccccccccccc@{}}
  \hline
Name		&&\multicolumn{3}{c}{\lya\tablenotemark{a}} && \multicolumn{3}{c}{\civ\tablenotemark{a}} &&	 \multicolumn{3}{c}{$\lambda$1900\tablenotemark{a,d}} && \multicolumn{3}{c}{\mgii\tablenotemark{a}} && \multicolumn{3}{c}{\hb\tablenotemark{a}} \\
\cline{3-5}\cline{7-9}\cline{11-13}\cline{15-17}\cline{19-21}
& & B & 0 & R && B & 0 & R && B & 0 & R && B & 0 & R && B & 0 & R &&\\
\hline
I Zw 1\tablenotemark{b} 	          && 4.2$\pm$0.7& 15.5$\pm$3.4   & 1.1$\pm$0.9  && 1.7$\pm$0.8& 2.4$\pm$1.3 & $\simlt$0.4 && $\simlt$0.2  & 0.8$\pm$0.3 & $\simlt 0.4$ && $\simlt$0.3 & 4.2$\pm$0.8 & 0.4$\pm$0.3\tablenotemark{e} && 0.15$\pm$0.10 &2.4$\pm0.3$ &  $\simlt$0.1	 \\
Mark 478\tablenotemark{b}	&& 3.4$\pm$1.1 & 33.0$\pm$2.4 & 1.3$\pm$1.2 && 1.6$\pm$0.6 & 5.1$\pm$1.0 & $\simlt$0.4 && $\simlt 0.3$ & 0.6$\pm$0.4 & $\simlt 0.4$ && $\simlt 0.8$ & 2.8$\pm$0.5 & $\simlt 0.4$ && $\simlt 0.1$ & 3.2$\pm$0.2 & $\simlt 0.2$    \\
Mark 335\tablenotemark{b}	&&	7.5$\pm$2.2 & 75.9$\pm$4.9 & 3.9$\pm$1.8 && 3.8$\pm$1.2 & 30.8$\pm$2.5 & $\la 1.2$\tablenotemark{g} && $\simlt$0.9 & 6.1$\pm$1.8 & $\simlt$0.8 && $\simlt$0.7 & 4.5$\pm$0.7 & $\simlt$0.5 &&  0.15$\pm$0.10 & 4.0$\pm$0.2 & $0.2\pm0.1$ \\
								
Fairall 9\tablenotemark{c}	&&	4.0$\pm$1.5 & 28.5$\pm$3.2 & 7.2$\pm$2.6 && 2.2$\pm0.7$ & 16.3$\pm$2.4  & 2.7$\pm$1.0 && $\simlt$0.8 & 2.5$\pm$0.9 & $\simlt$0.5 &&$\simlt 0.4$ & 4.6$\pm$0.6 & 0.5$\pm$0.4 && $\simlt 0.1$ &2.2$\pm$0.2 & 0.6$\pm$0.3	 \\
3C 249.1\tablenotemark{c}	&&2.4$\pm$0.5 & 6.0$\pm$1.2 & 1.5$\pm$0.7 && 1.4$\pm$0.2  & 3.6$\pm0.6$ & 1.1$\pm$0.3 && $\simlt$0.2 & 0.8$\pm$0.2 & 0.2$\pm$0.1 && \ldots & \ldots & \ldots && 0.15$\pm$0.05 & 0.6$\pm$0.1 & 0.3$\pm$0.1\\
3C 110\tablenotemark{c}	&&1.4$\pm$0.3 & 4.2$\pm$0.8 & 3.0$\pm$0.2 && 0.7$\pm$0.2 & 2.4$\pm$0.4 & 1.5$\pm$0.2 && \ldots & \ldots & \ldots &&$\simlt$0.1 & 0.5$\pm$0.1 & 0.15$\pm$0.01 && 0.1$\pm$0.1& 0.4 $\pm$0.1& 0.3$\pm$0.2 \\
\hline
 \end{tabular}

\tablenotetext{a}{Fluxes in units of 10$^{-12}$\ergss\ \cmq.}
 \tablenotetext{b}{B: $-5000  $ \kms\ $\le$ \vr\ $\le -3000$ \kms; 0: $1000  $ \kms\ $\le$ \vr\ $\le 1000$ \kms; R: $3000  $ \kms\ $\le$ \vr\ $\le 5000$ \kms.}
\tablenotetext{c}{B: $-7000  $ \kms\ $\le$ \vr\ $\le -5000$ \kms; 0: $1000  $ \kms\ $\le$ \vr\ $\le 1000$ \kms; R: $5000  $ \kms\ $\le$ \vr\ $\le 7000$ \kms.}
\tablenotetext{d}{\aliii\ is measured for I Zw1 and Mark 478; \ciii\ for the remaining objects.}
\tablenotetext{e}{Strong \feii\ residual.} \tablenotetext{g}{R: $3500  $ \kms\ $\le$ \vr\ $\le 5500$ \kms\ due to a blemish in the spectrum. } 
\end{minipage}
\end{table*}

\begin{table*}
 \begin{minipage}{150mm}
  \caption{Black Hole Mass and Eddington Ratio \label{tab:mass}}
  \begin{tabular}{@{}lcccccccccc@{}}
  \hline
Name		& $f_{\lambda}$ & $\log$ \lbol\	&\multicolumn{2}{c}{FWHM} && \multicolumn{2}{c}{$\log$\mbh} &&	 \multicolumn{2}{c}{$\log$\lledd} \\
\cline{4-5}\cline{7-8}\cline{10-11}
 & at 5100 \AA&& \hb\ &\hbbc && \hb\ &\hbbc && \hb\ &\hbbc \\
& [\ergss cm$^{-2}$ \AA$^{-1}$] &  [\ergss] & [\kms] & [\kms] && [\msol] & [\msol] \\

\hline
I Zw 1	         &8.40E-15&	45.5&	1100&	 1100     &&	7.3	&7.3	&&-0.3	&-0.3\\
Mark 478	&7.13E-15&	45.7&	1300	&       1300	&&7.5	&7.5	&&-0.4	&-0.4\\
Mark 335	&9.00E-15&	44.8&	1500 &	15000    &&	7.2 &7.2	&&-0.9&-0.9\\
								
Fairall 9	&7.52E-15&	45.3&	6500	& 4550&	&8.7	& 8.4 &&	-2.0&	-1.7\\
3C 249.1	&3.00E-15&	46.5&	8300	 &5050&&	9.5&	9.0	&&-1.6&	-1.2\\
3C 110	&1.50E-15&	46.8&	14000&	10700&&	10.1&	9.9&&	-1.9&	-1.6\\
\hline
 \end{tabular}
\end{minipage}
\end{table*}
\clearpage\eject\clearpage
\null
\voffset=0.0cm

\section*{Acknowledgments}
DD and AN acknowledge financial support from  PAPIIT UNAM, grant IN11161013. PM wishes to thank Prof. H. Netzer for discussions on topics closely related to the present paper.{  The authors also thank an anonymous referee for suggesting some necessary additions. }

\bibliography{biblio}{}
\bibliographystyle{apj}
\vfill \eject \clearpage
\clearpage

\newpage\pagestyle{empty}
\setcounter{figure}{0}
\begin{figure*}
\includegraphics[scale=0.78]{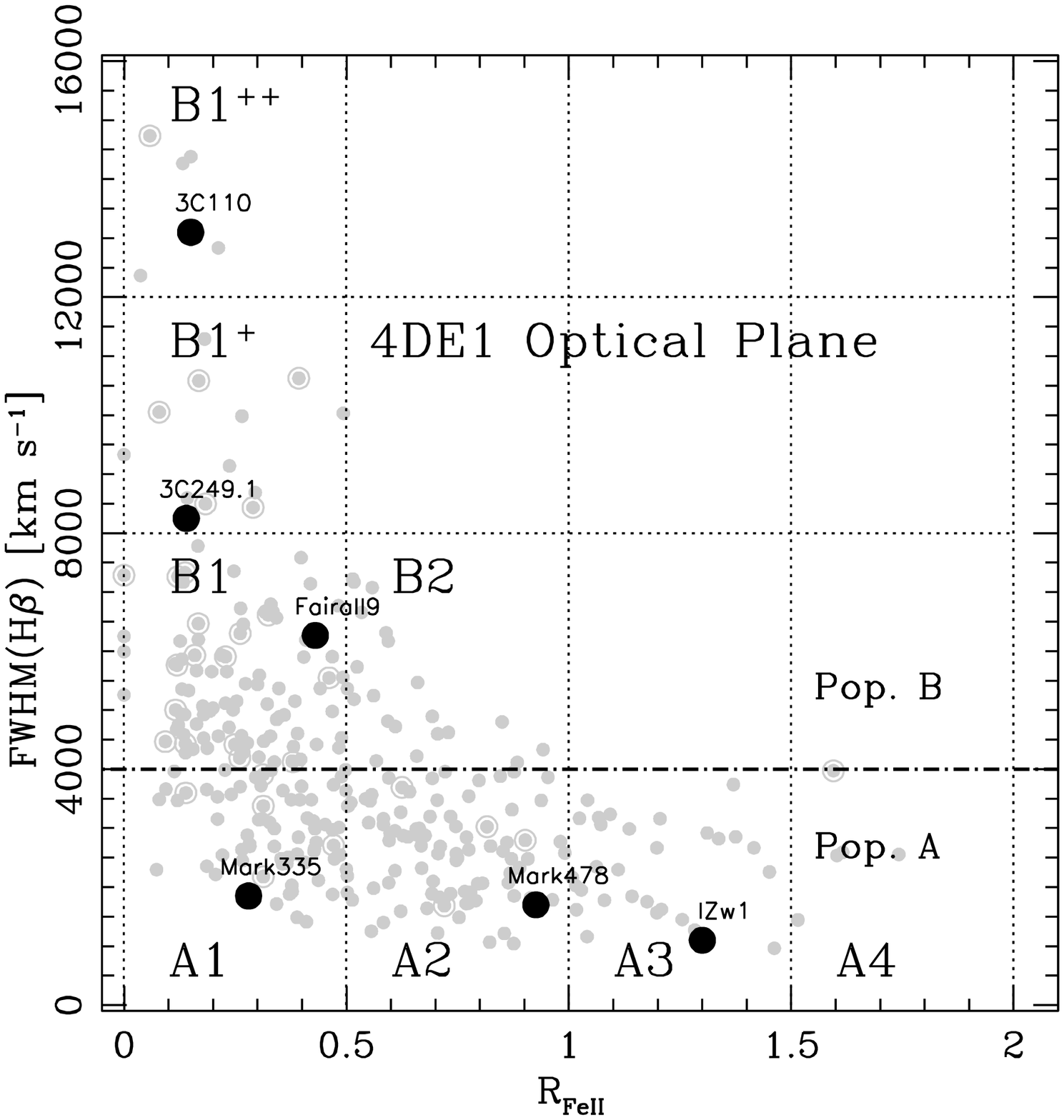}
\caption{Optical plane of the 4DE1 parameter space, i.e., FWHM(\hb) vs. \rfe. \rfe\ measures the prominence of \feiiq\ with respect to \hb\ (see text for definition). Occupied spectral bins are labeled. Grey dots are from the sample of \citet{zamfiretal08}. Large black spots mark the position of the six sources considered in this study.\label{fig:e1}}
\end{figure*}

\voffset=5cm
\rotate
\begin{figure*}%
\includegraphics[scale=0.222]{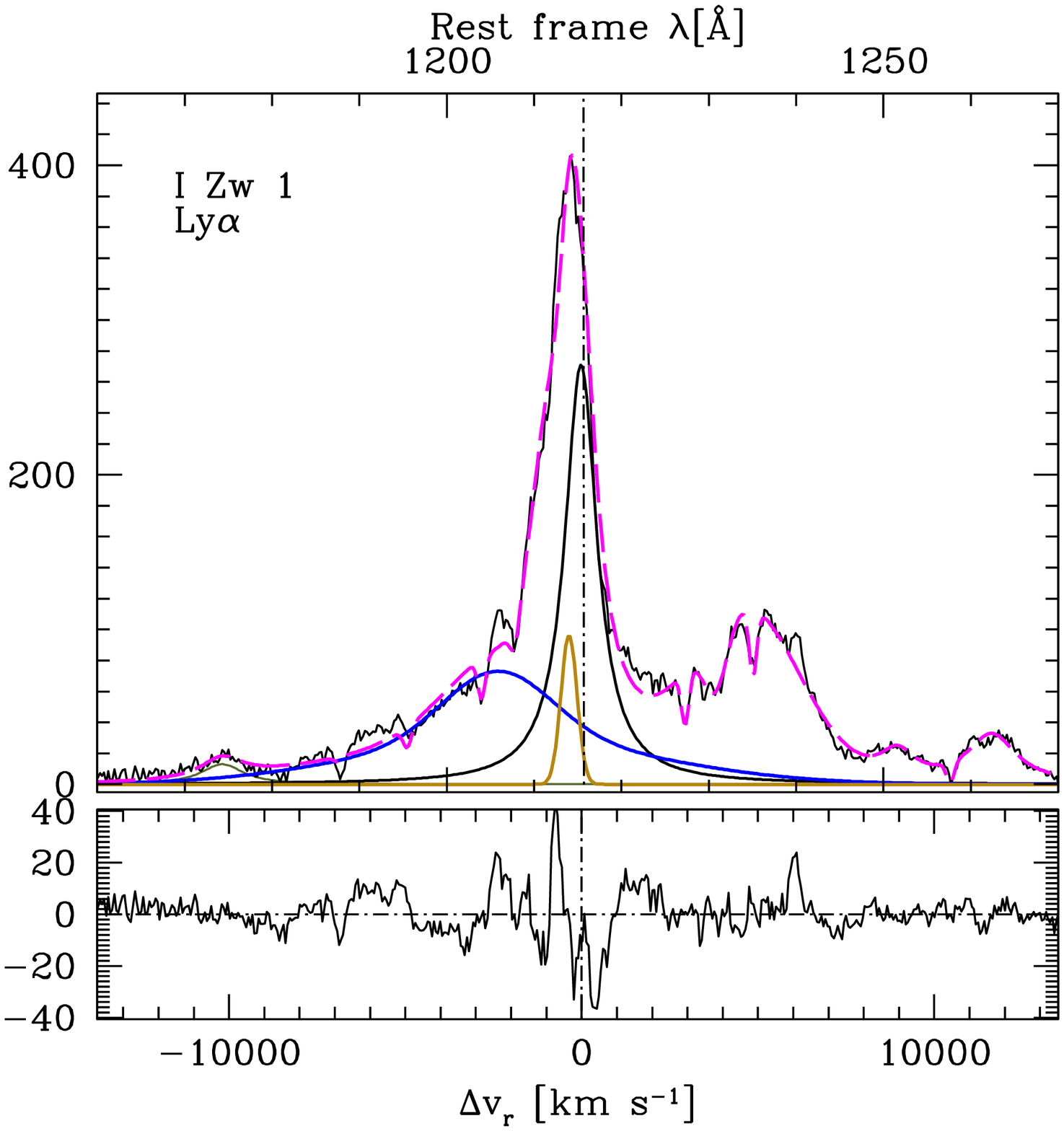}\includegraphics[scale=0.22]{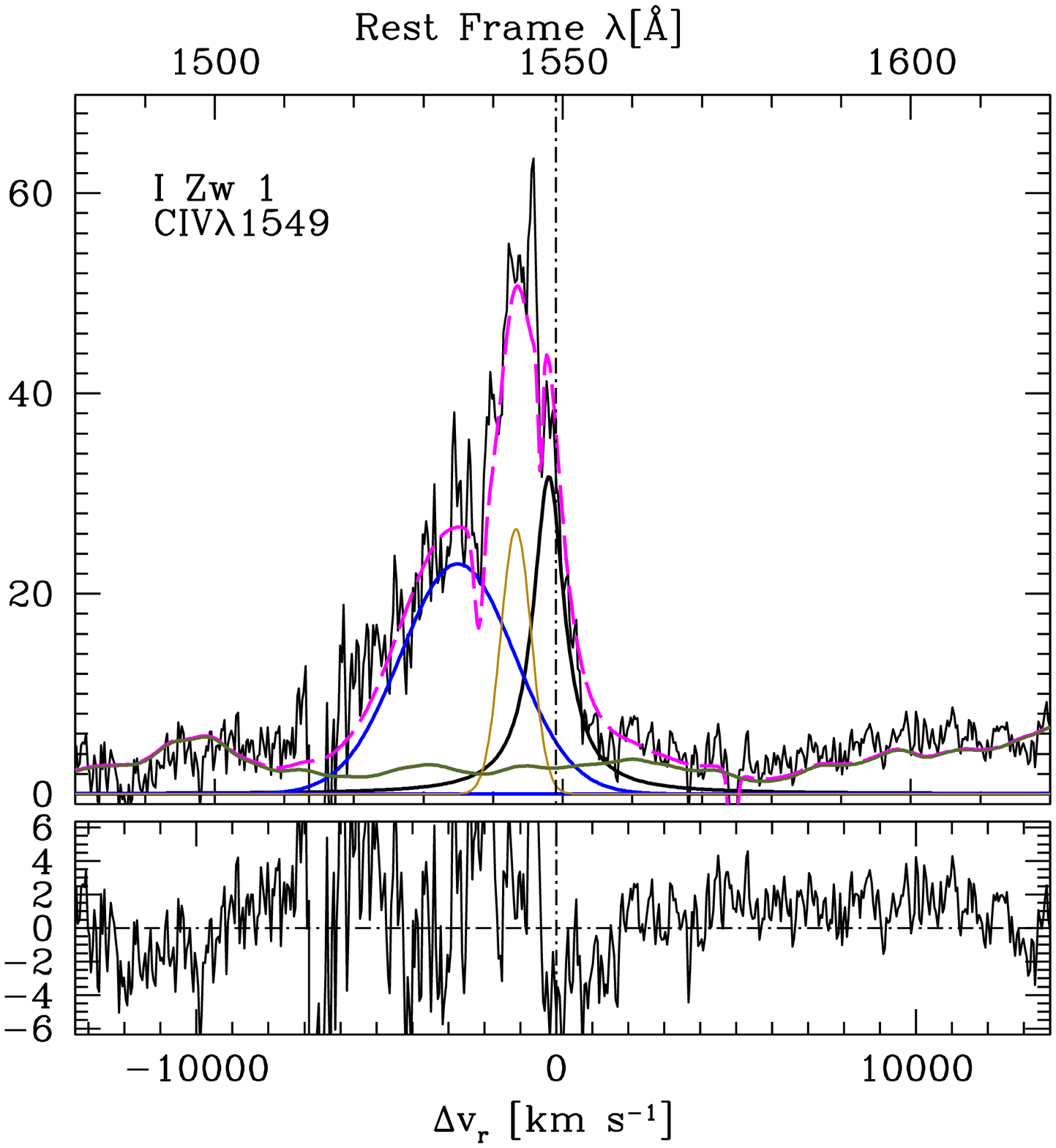}\includegraphics[scale=0.22]{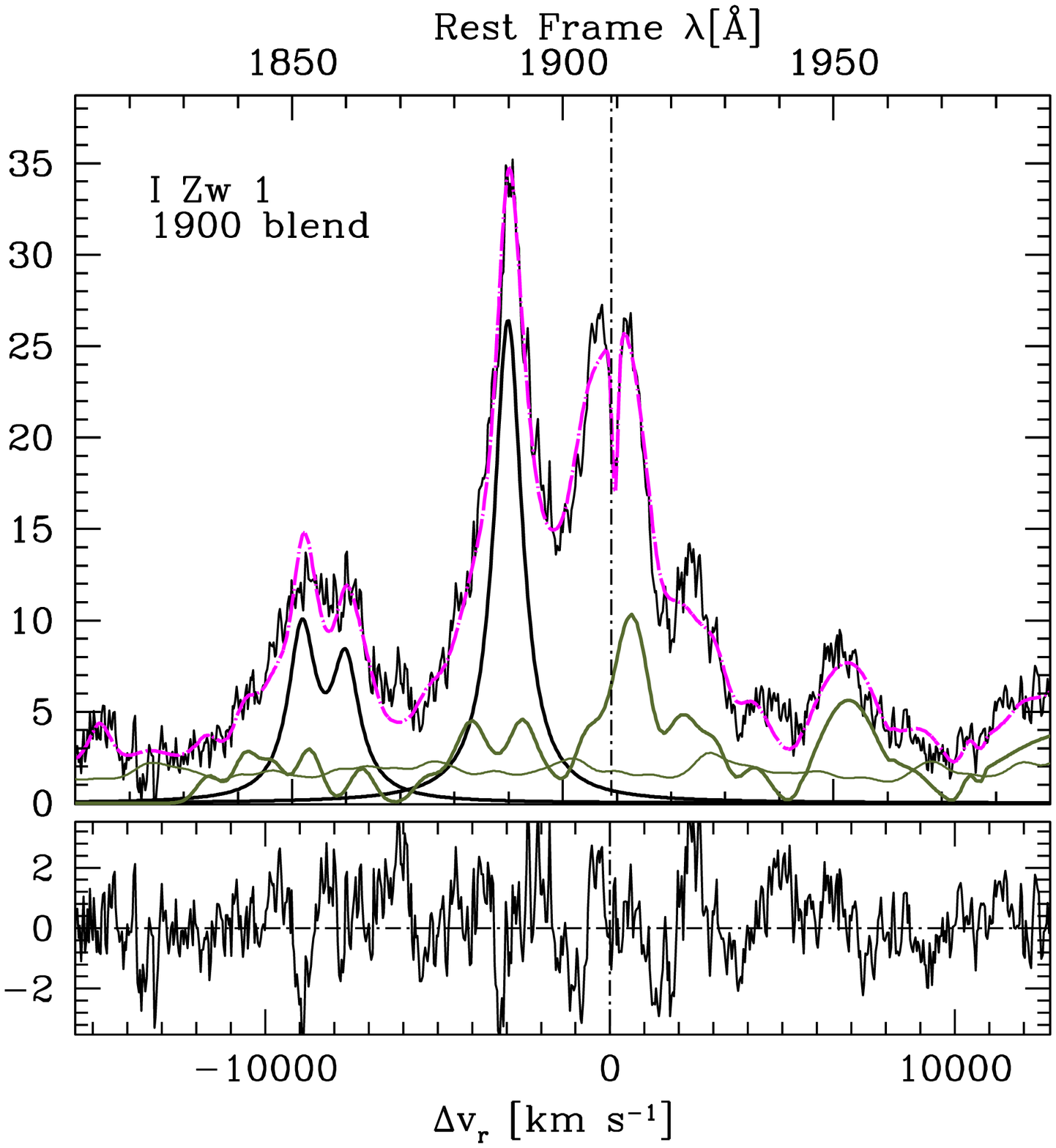}\includegraphics[scale=0.22]{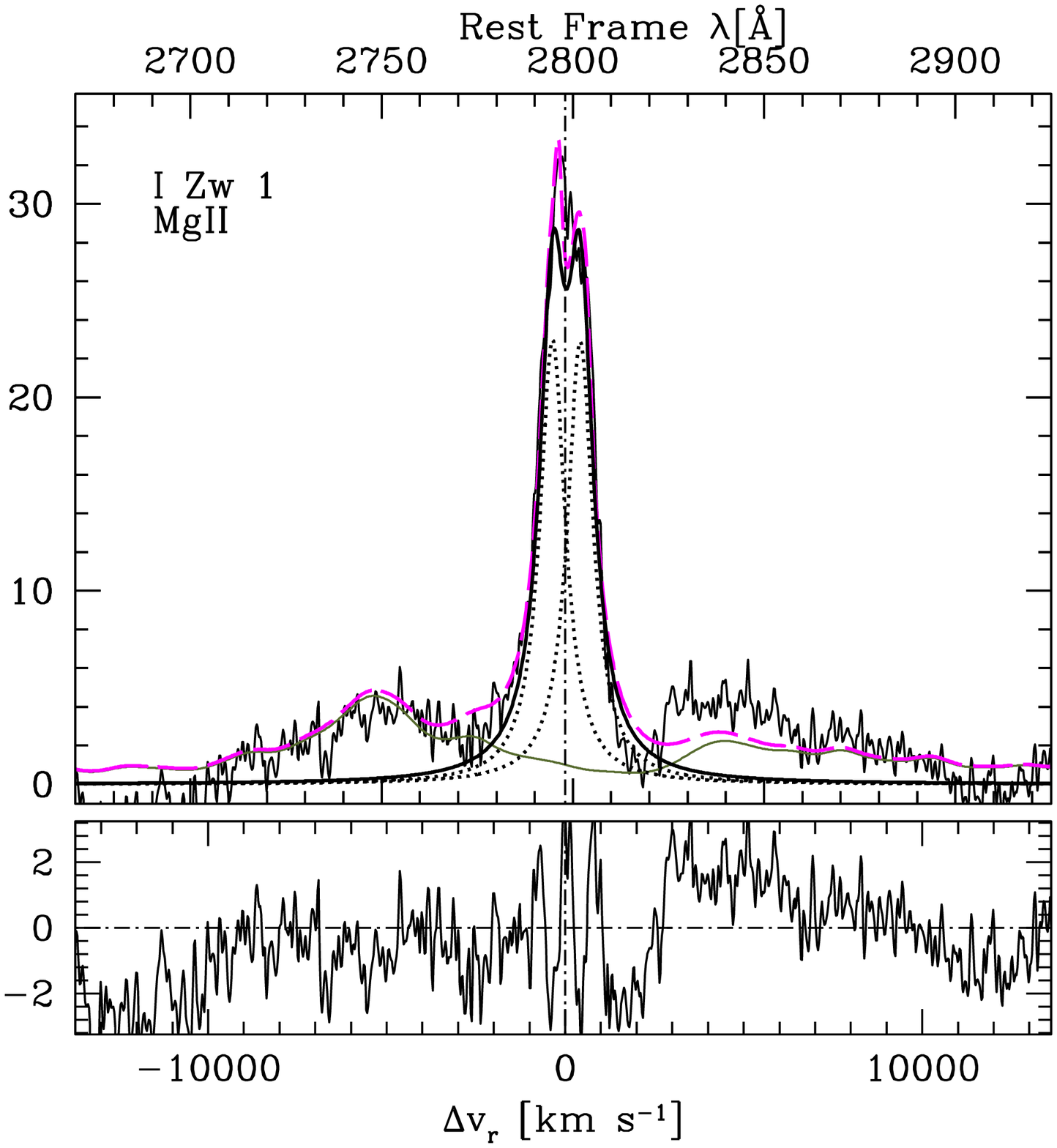}\includegraphics[scale=0.222]{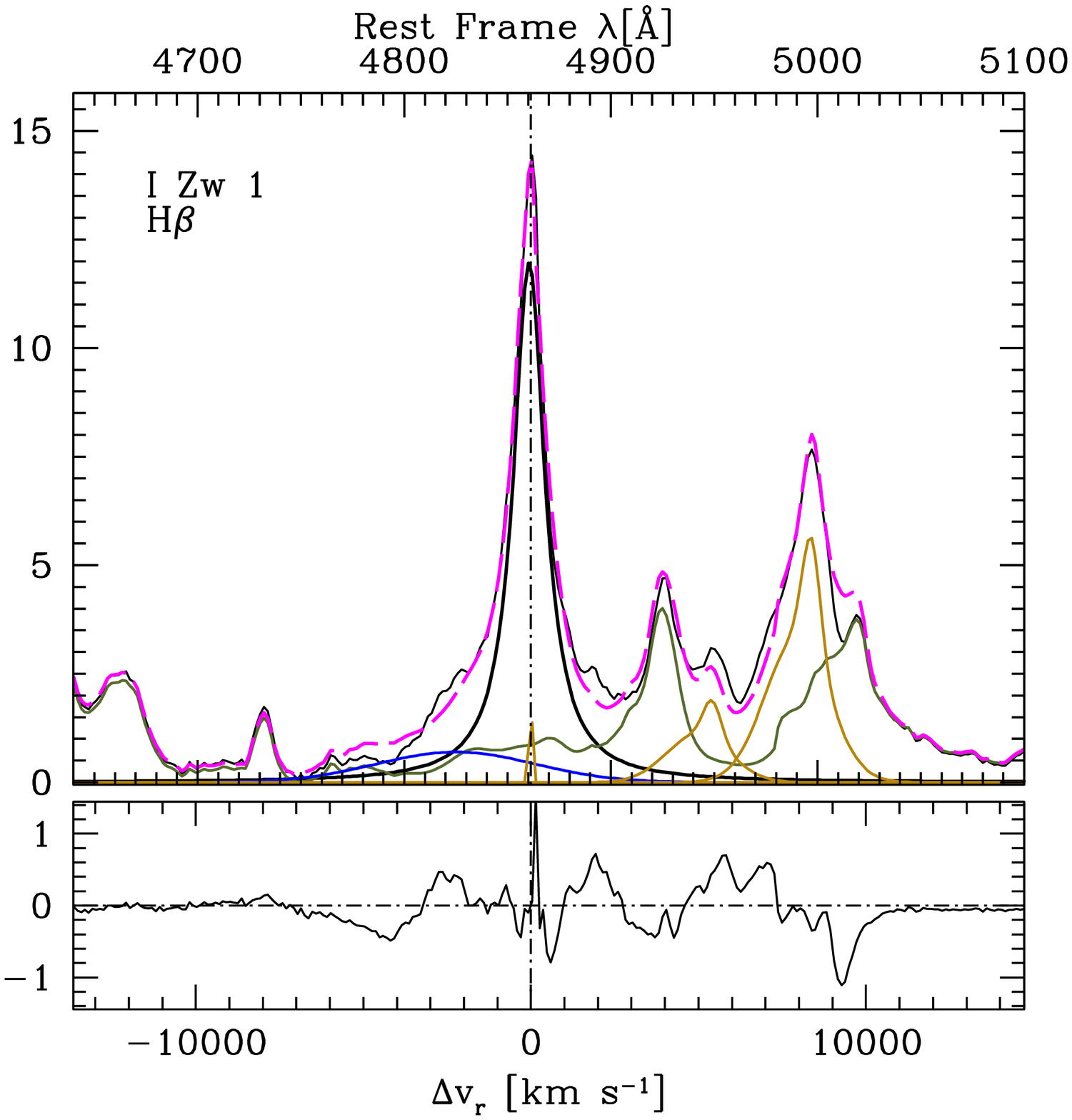}\\
\includegraphics[scale=0.22]{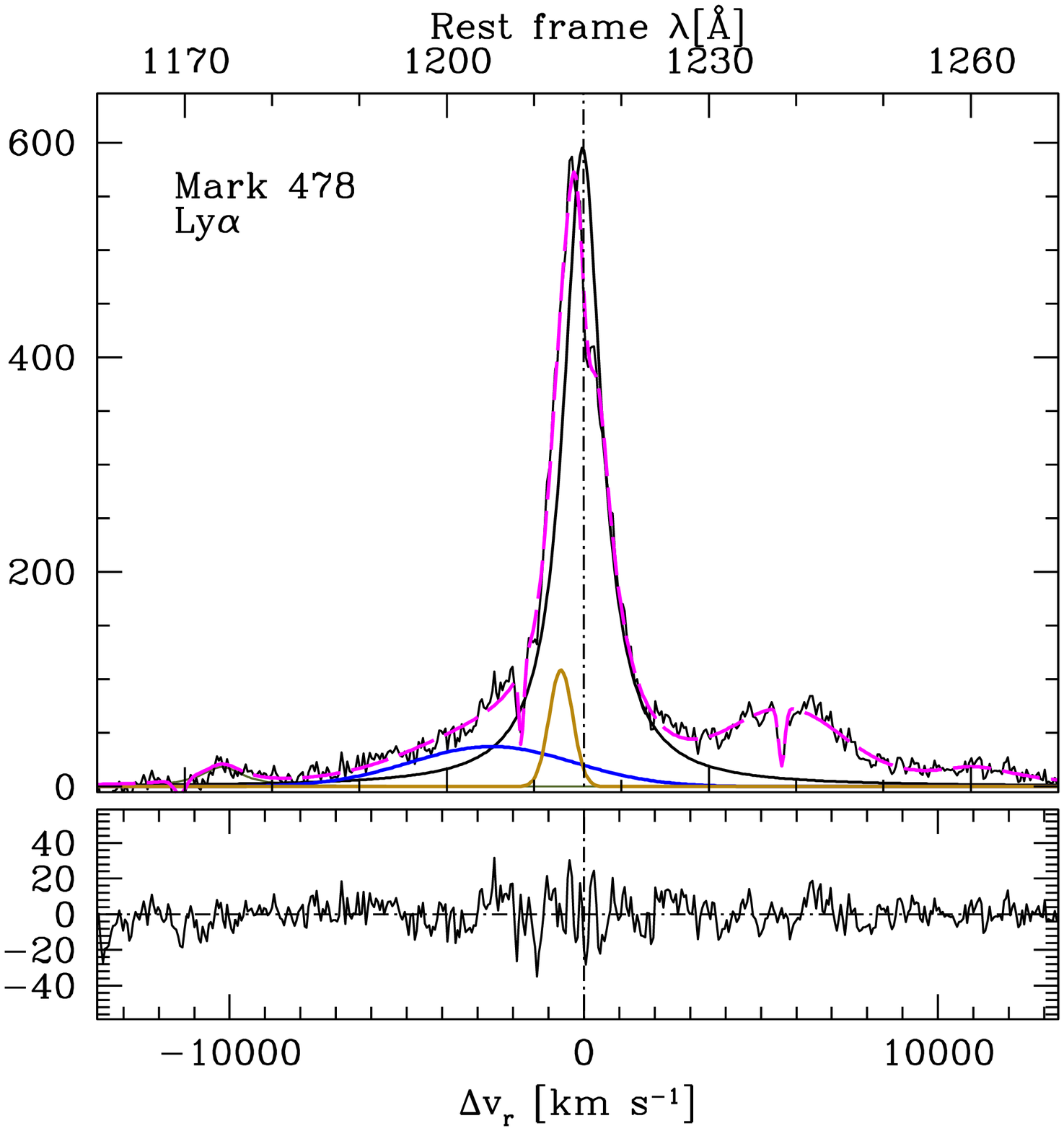}\includegraphics[scale=0.22]{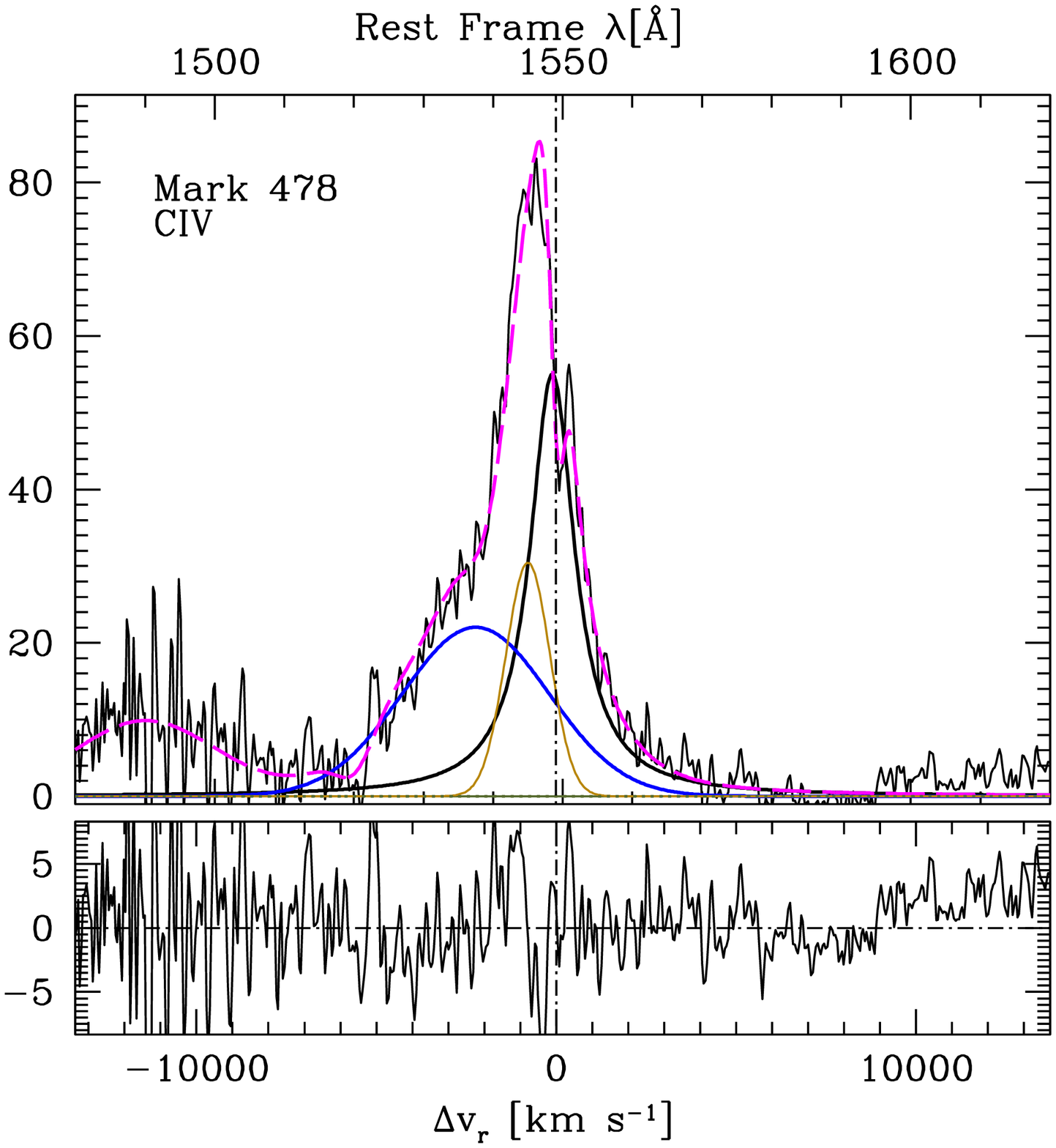}\includegraphics[scale=0.22]{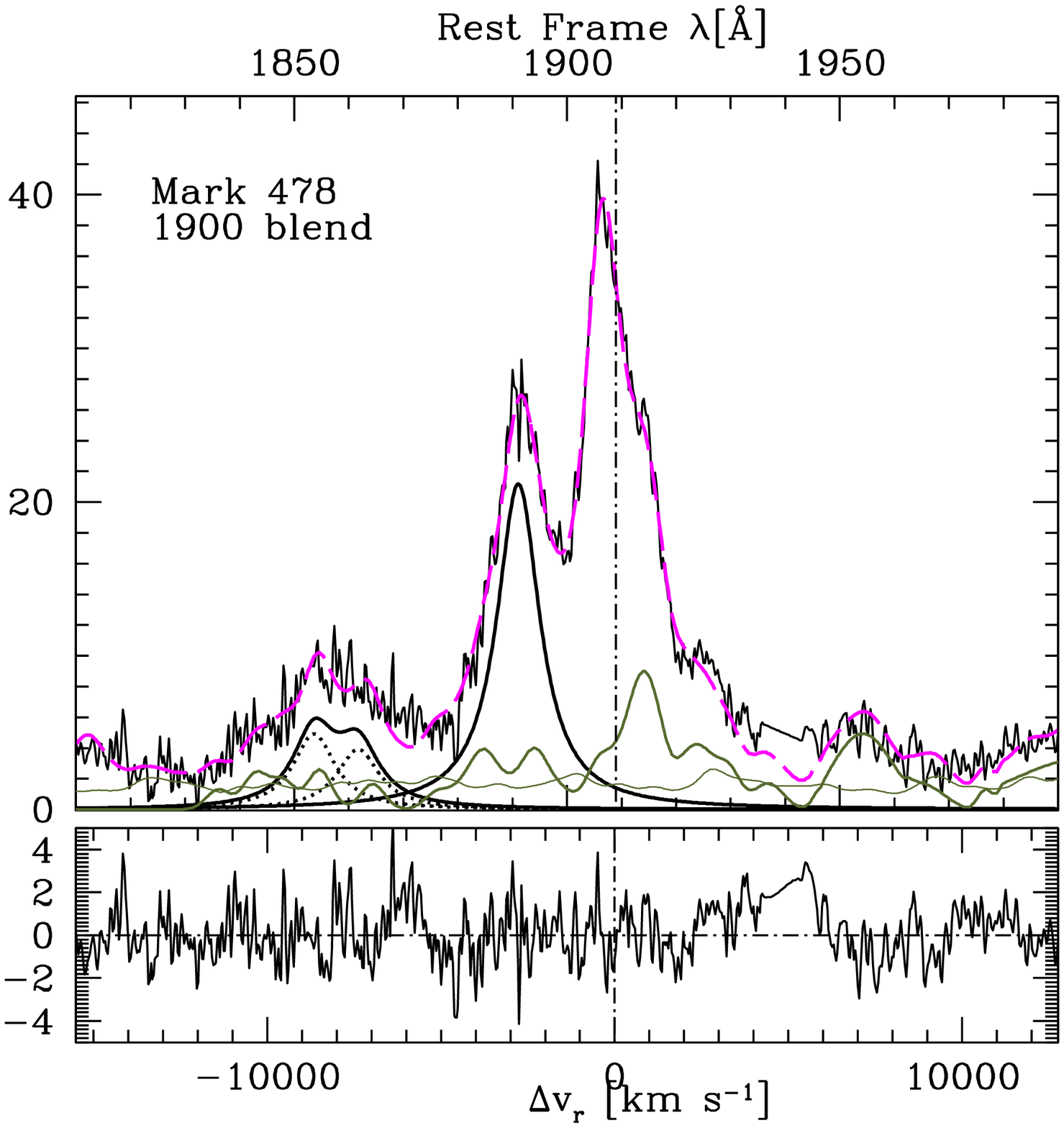}\includegraphics[scale=0.22]{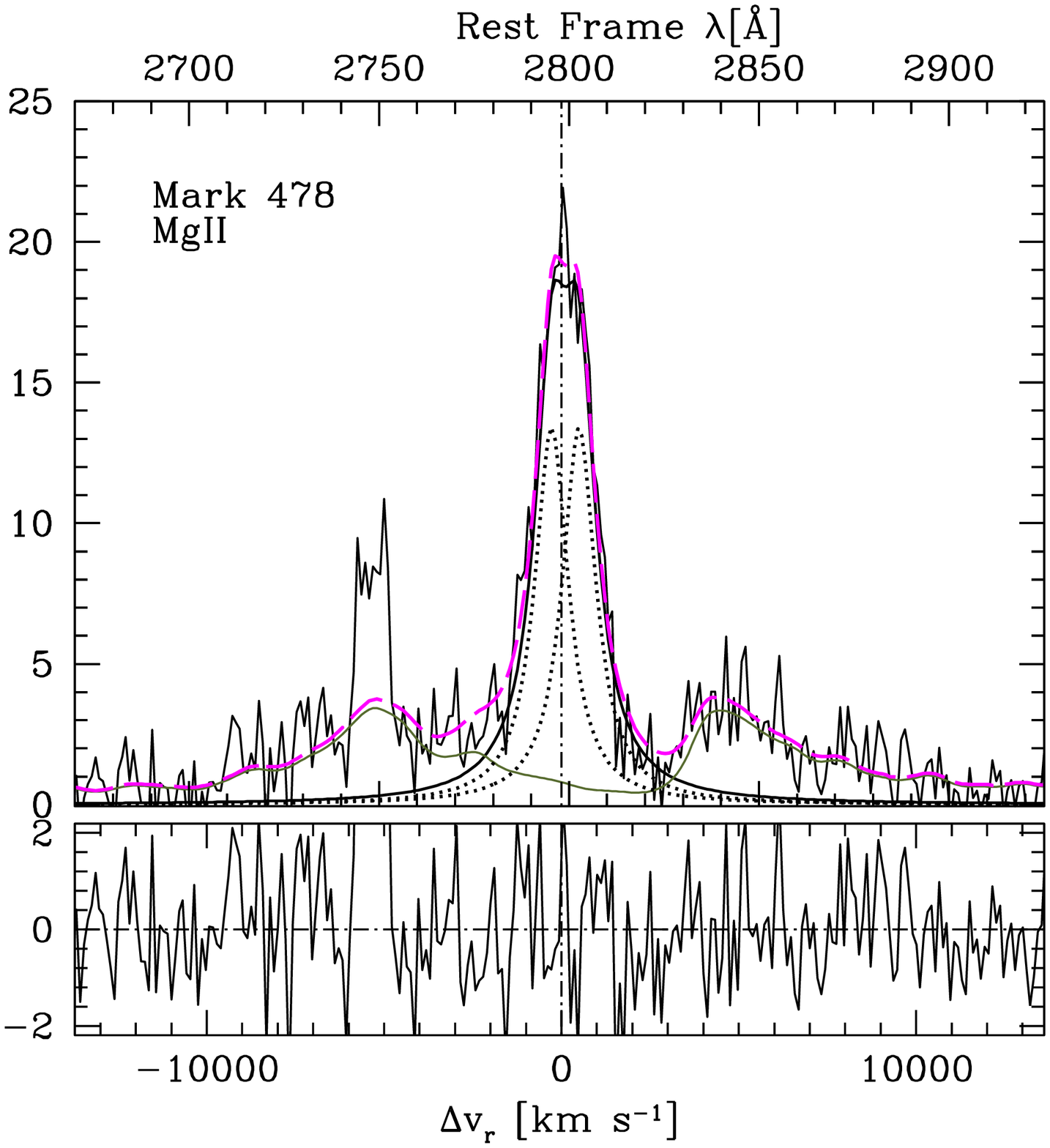}\includegraphics[scale=0.22]{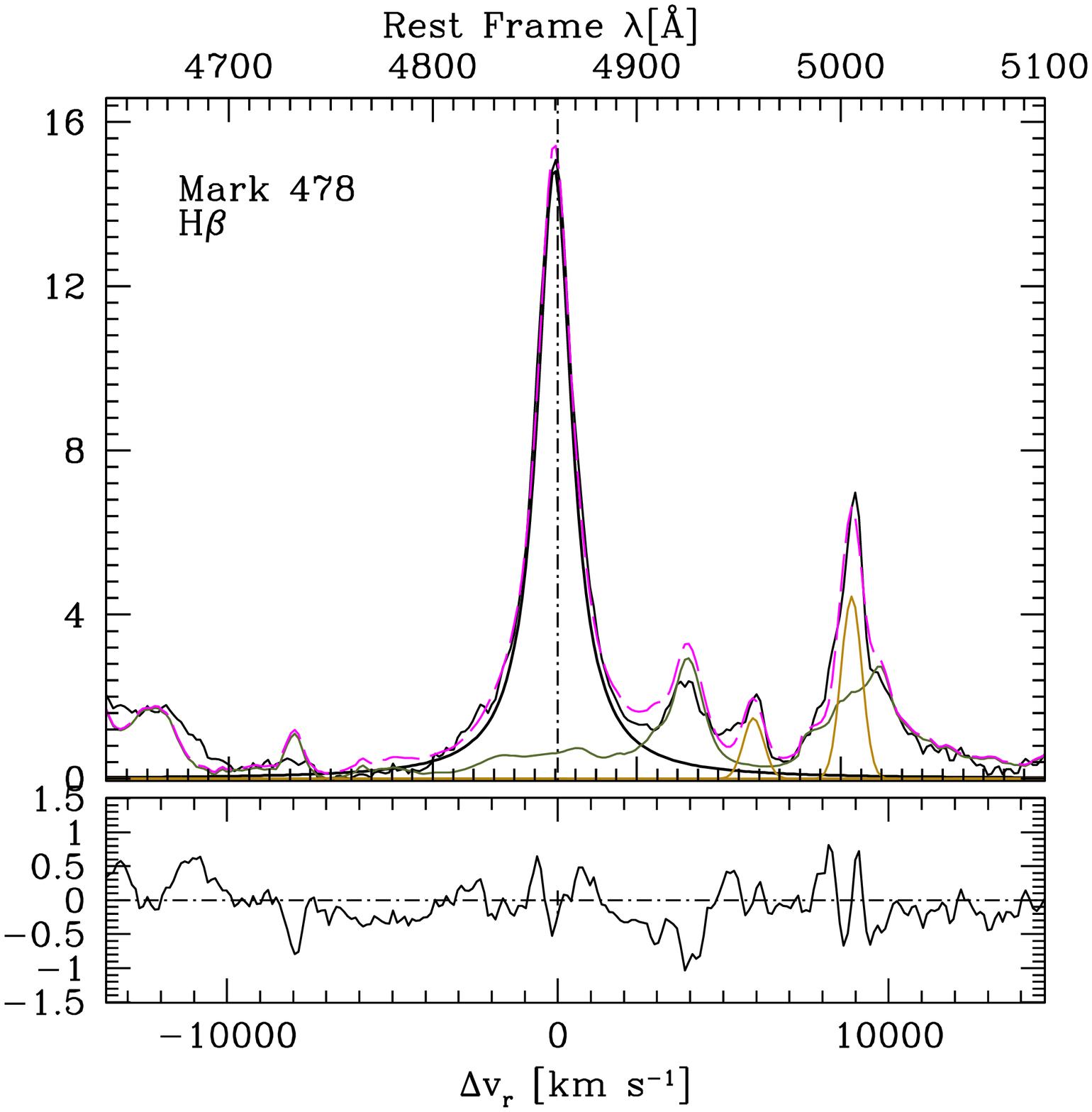}\\
 \includegraphics[scale=0.222]{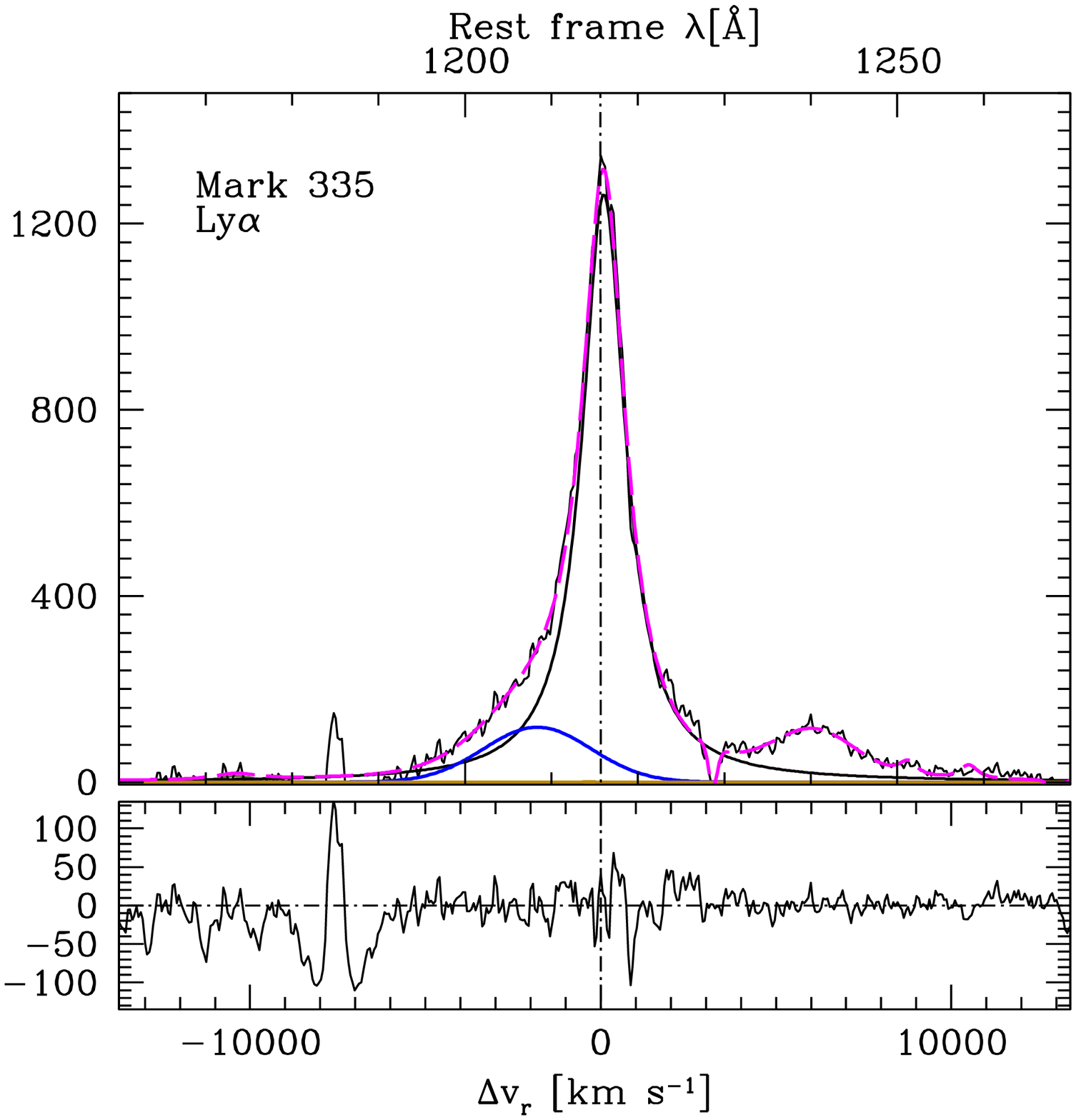}\includegraphics[scale=0.222]{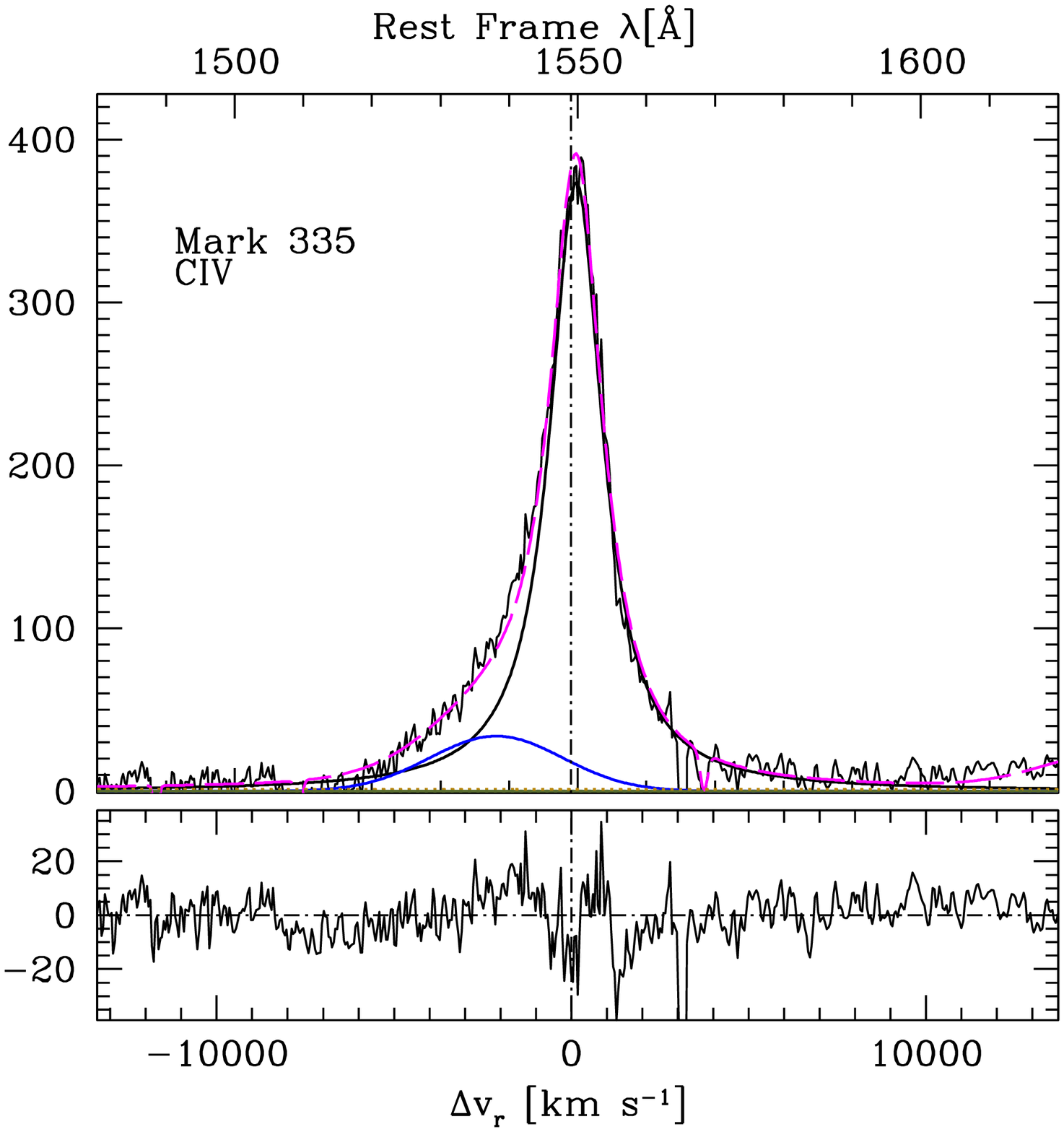}\includegraphics[scale=0.222]{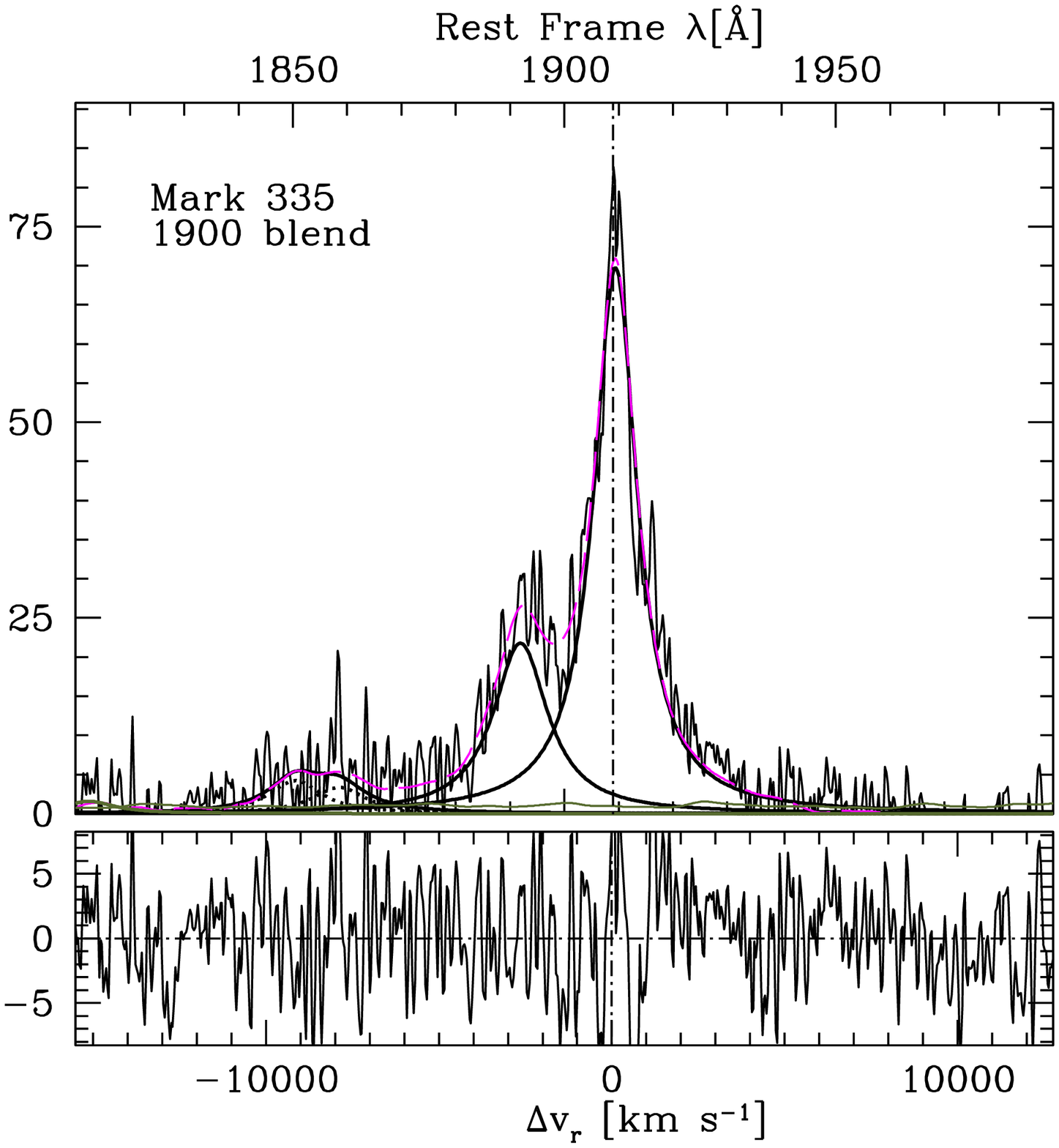}\includegraphics[scale=0.222]{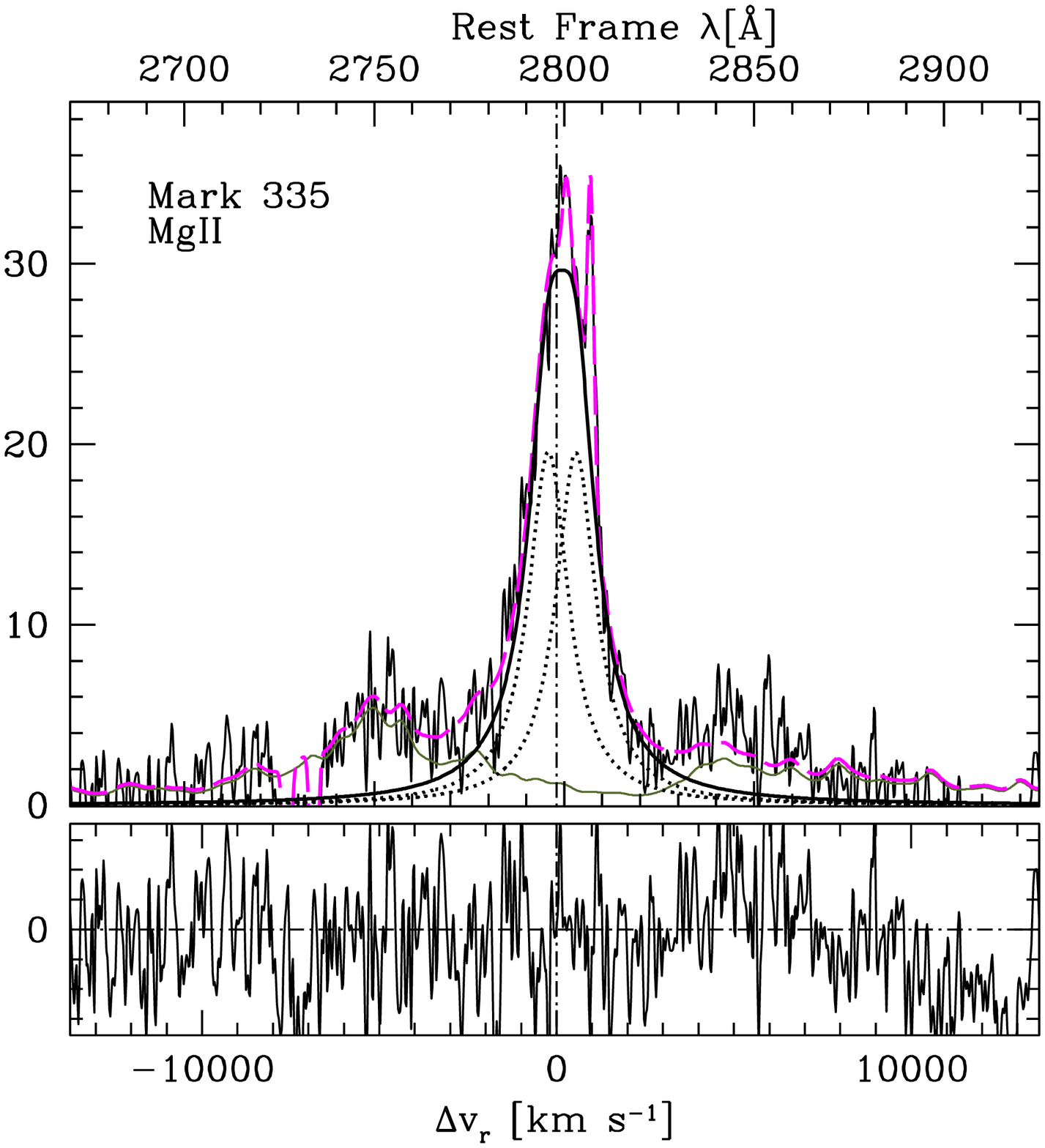}\includegraphics[scale=0.22]{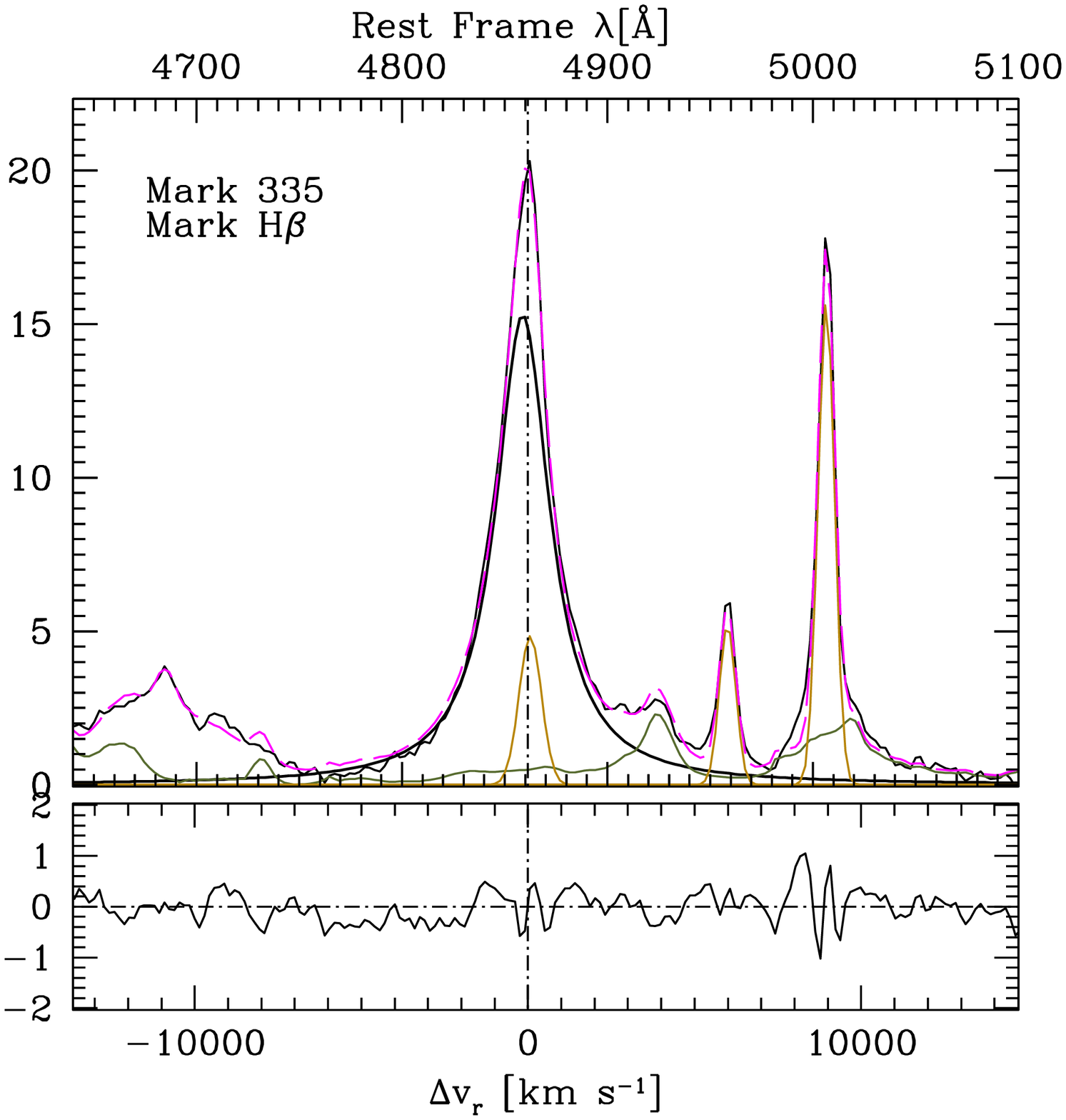}
\caption{Multi-component fits of the \lya, \civ, 1900 \AA, \mgii, \hb\ spectral regions for the six object of this work ordered along the E1 sequence. All spectra are reduced to rest-frame and continuum subtracted. Ordinate is specific flux in units of 10$^{-15}$ \ergss\ cm$^{-2}$ \AA$^{-1}$. The lower part of each panel shows residuals as a function of radial velocity. Thin black line: continuum subtracted rest-frame spectrum; thick dashed (magenta) line:  multicomponent fit result. Thick continuous lines: blueshifted (blue), broad (black) and very broad (red) components. Thin orange line: narrow line components; thin dark green line: \feii\ template emission. In the 1900 blend panel, the thick dark green line shows the adopted \feiii\ template. In case of severe \feiii\ contamination the \ciii\ line is not shown. \label{fig:profiles}}
\end{figure*}

\setcounter{figure}{1}
\begin{figure*}\rotate
\includegraphics[scale=0.222]{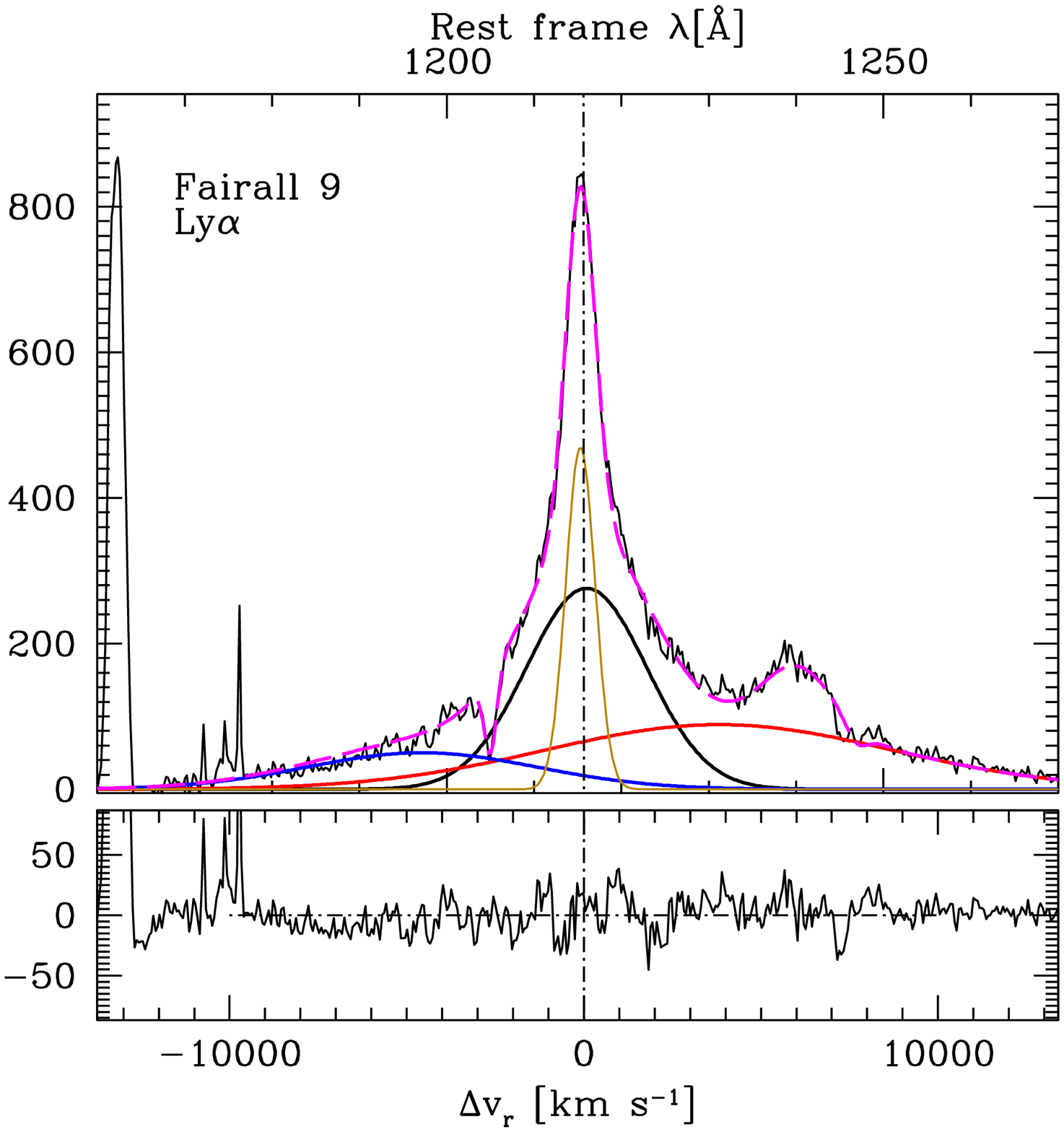}\includegraphics[scale=0.222]{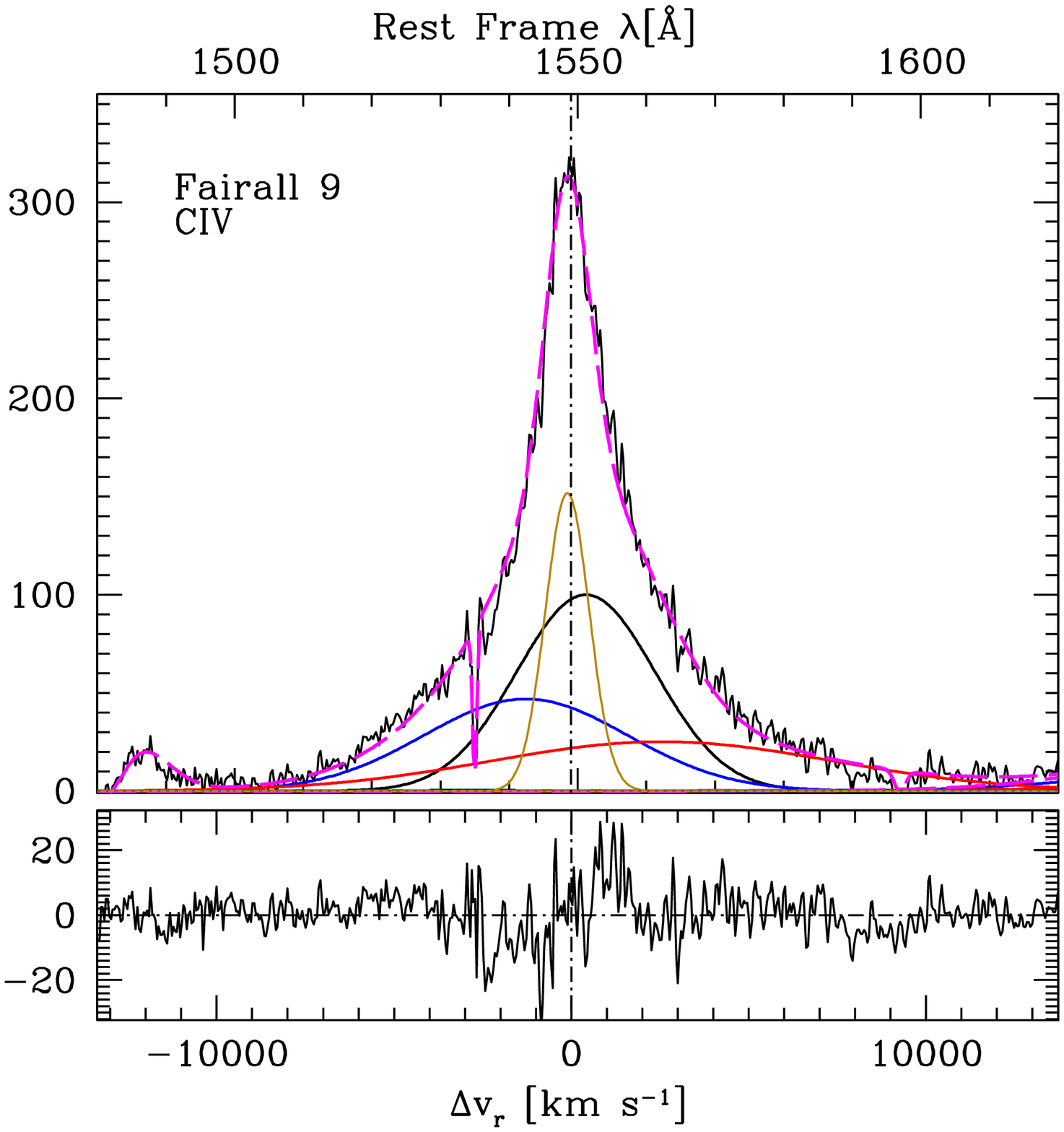}\includegraphics[scale=0.222]{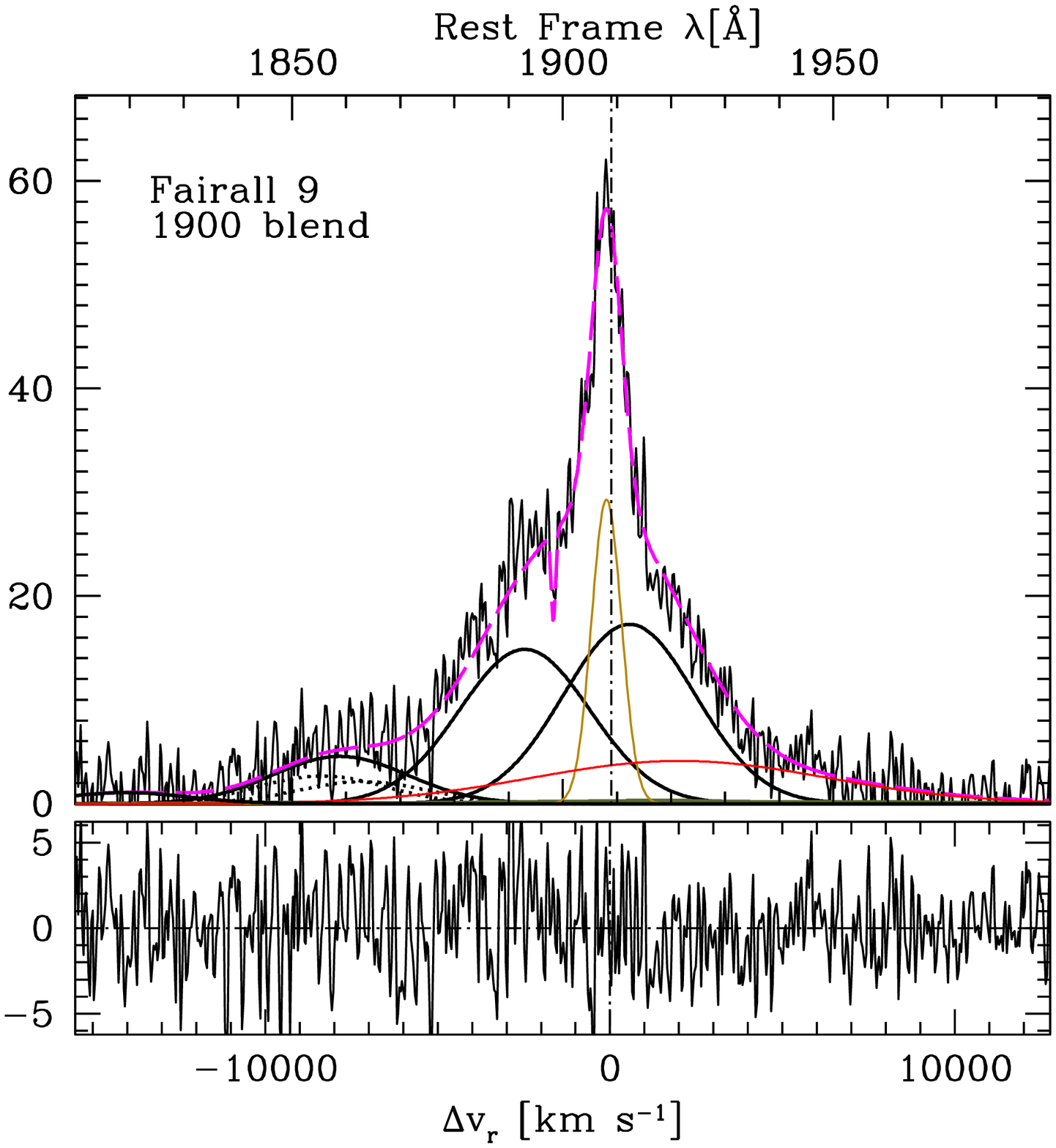}\includegraphics[scale=0.222]{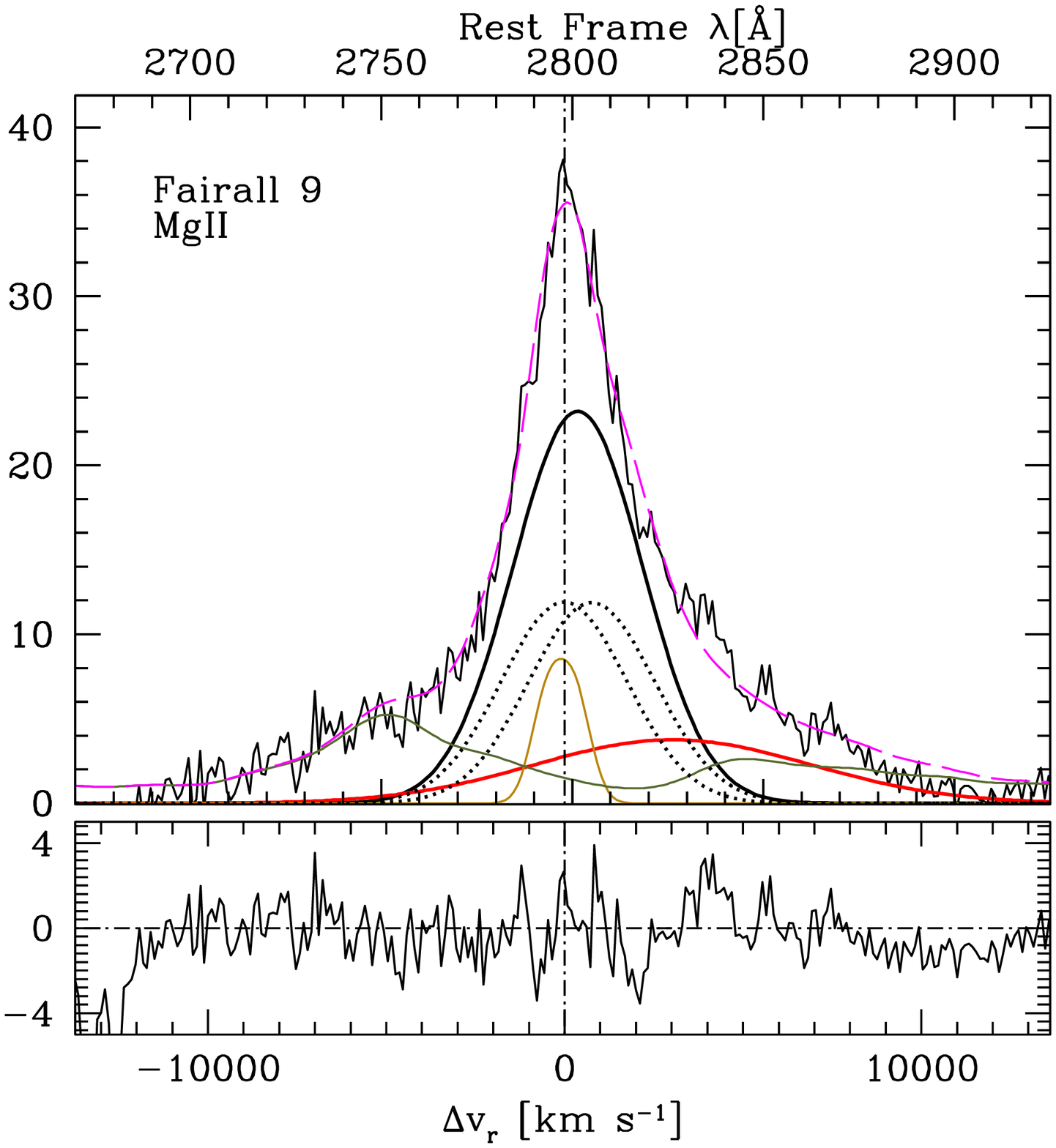}\includegraphics[scale=0.222]{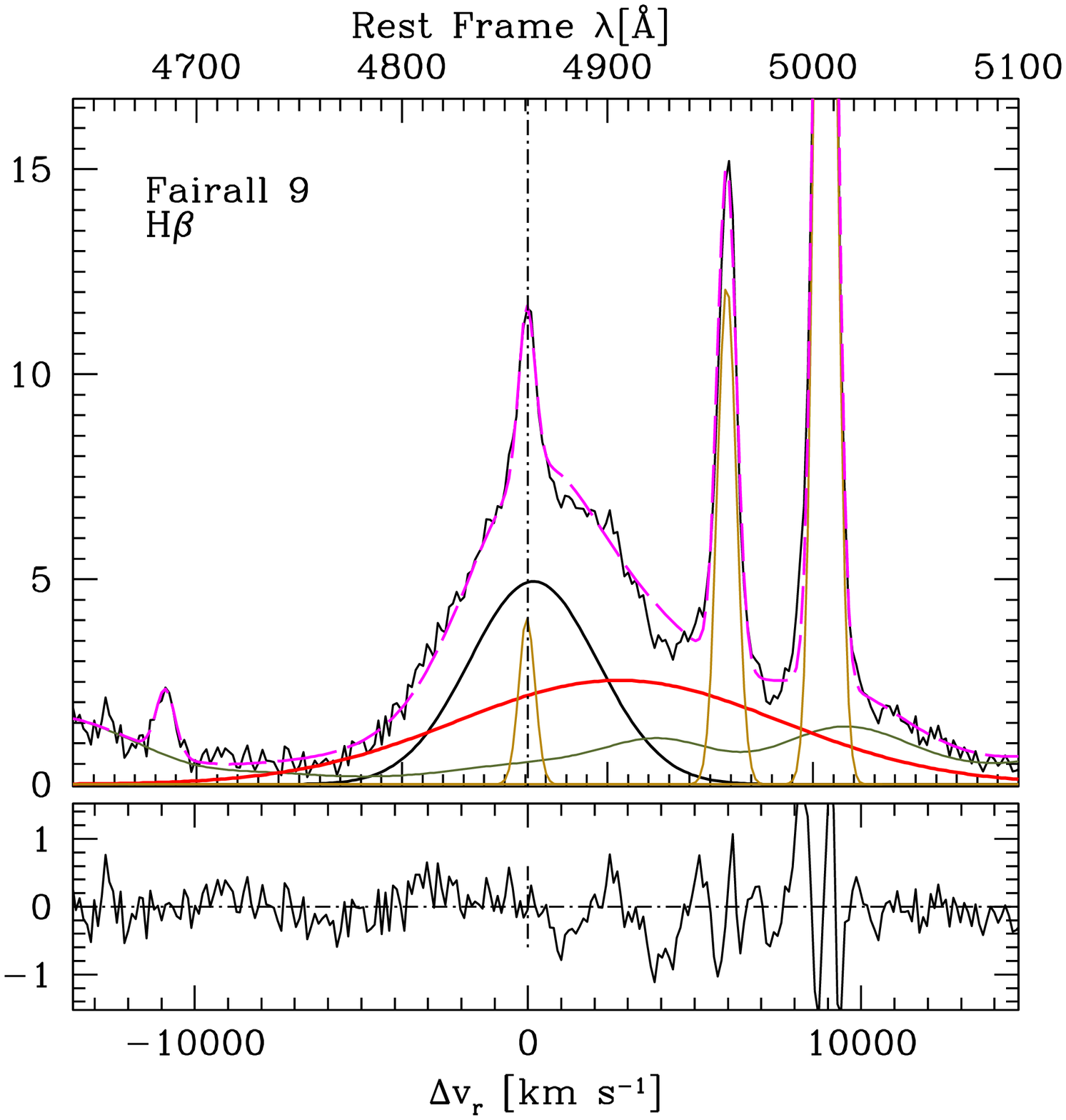}\\
\includegraphics[scale=0.22]{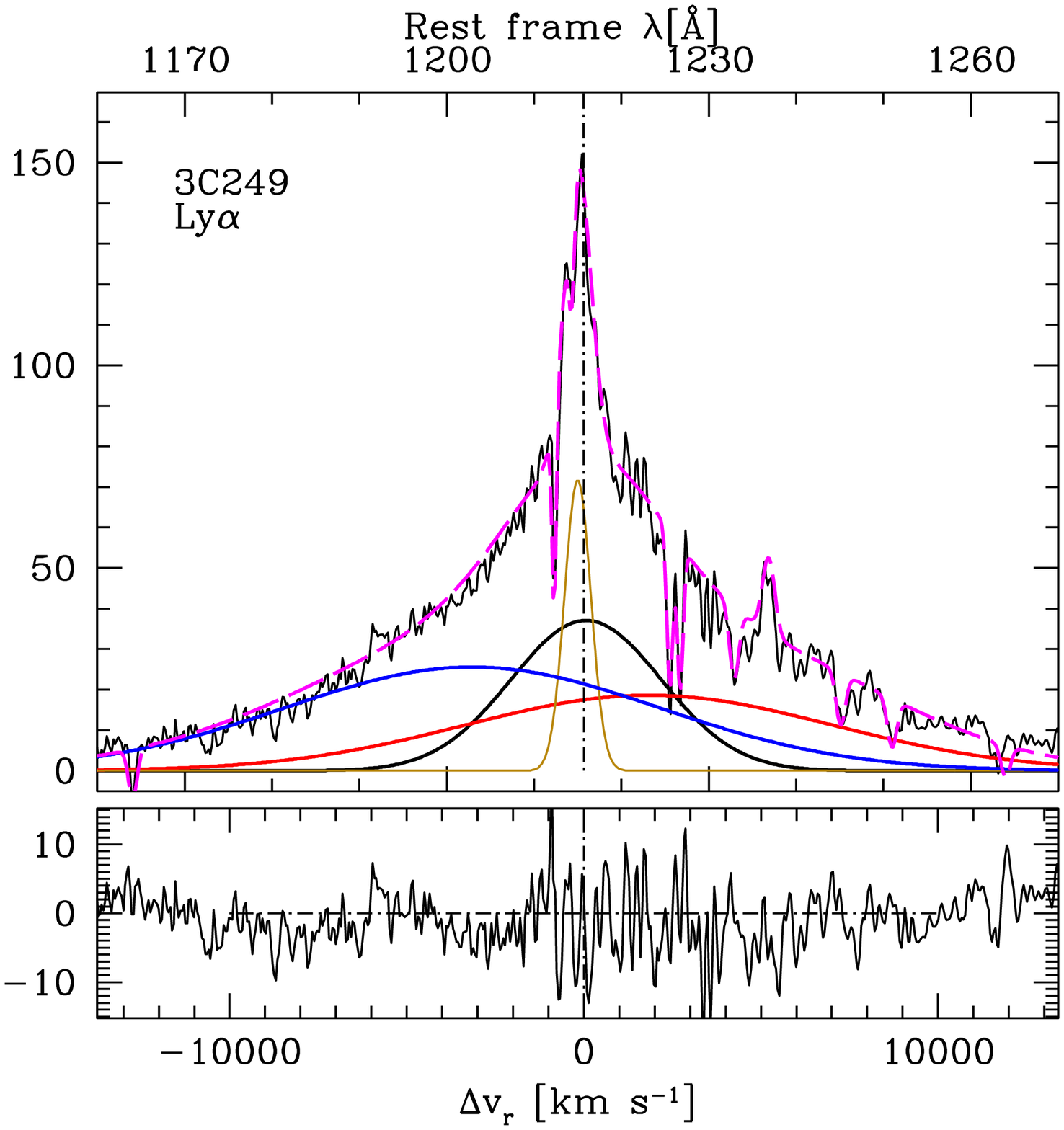}\includegraphics[scale=0.22]{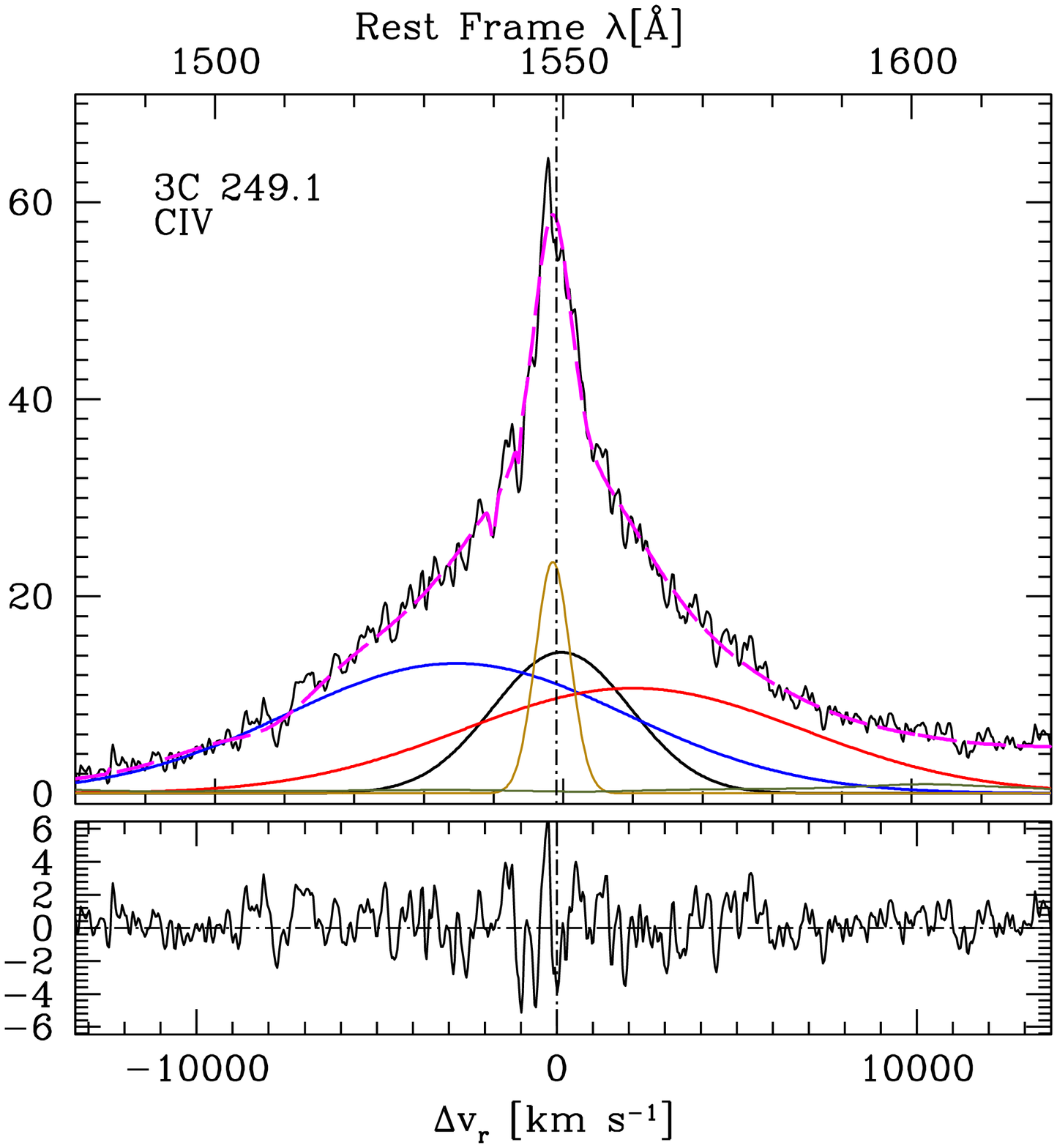}\includegraphics[scale=0.22]{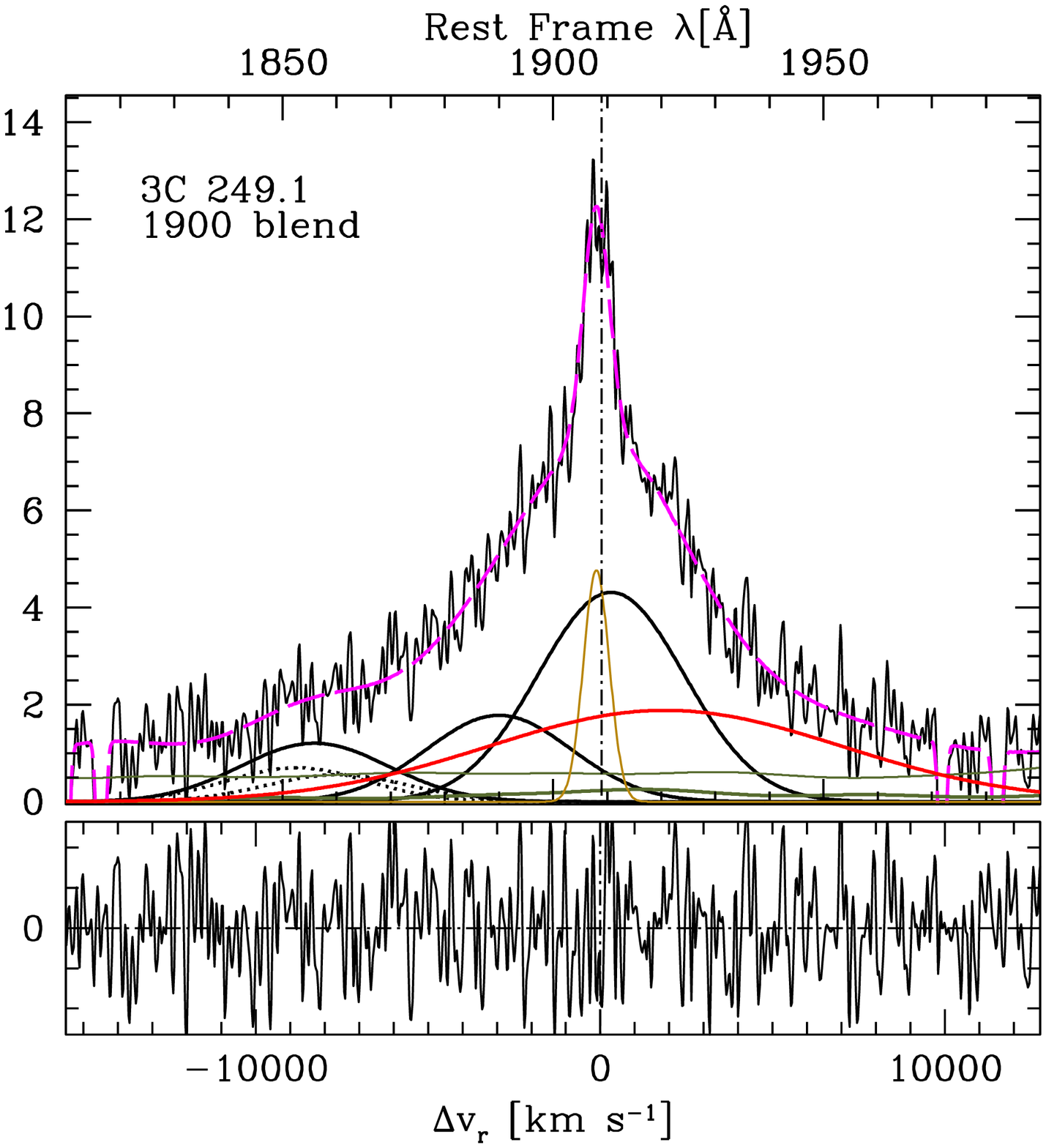}\includegraphics[scale=0.22]{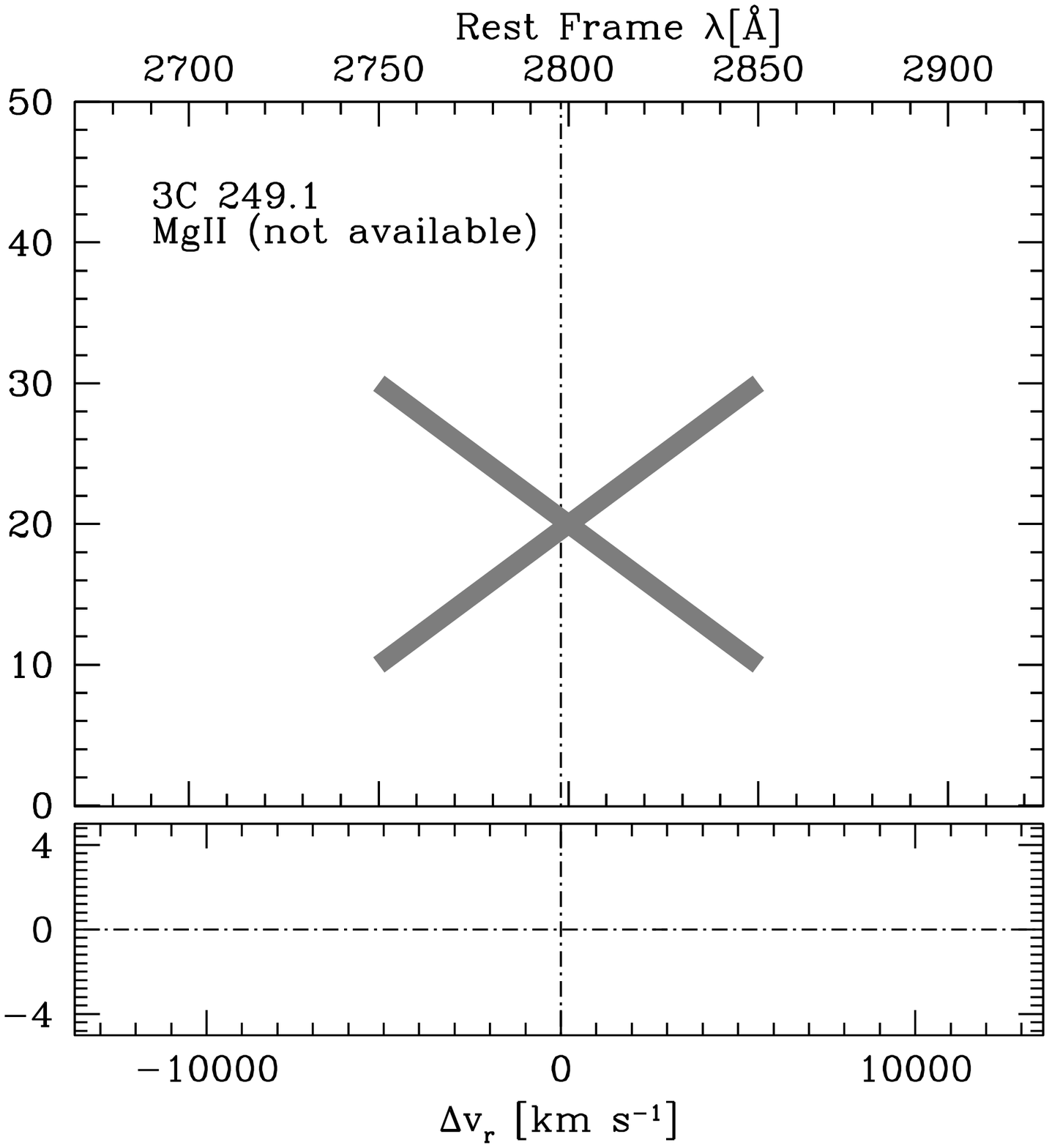}\includegraphics[scale=0.22]{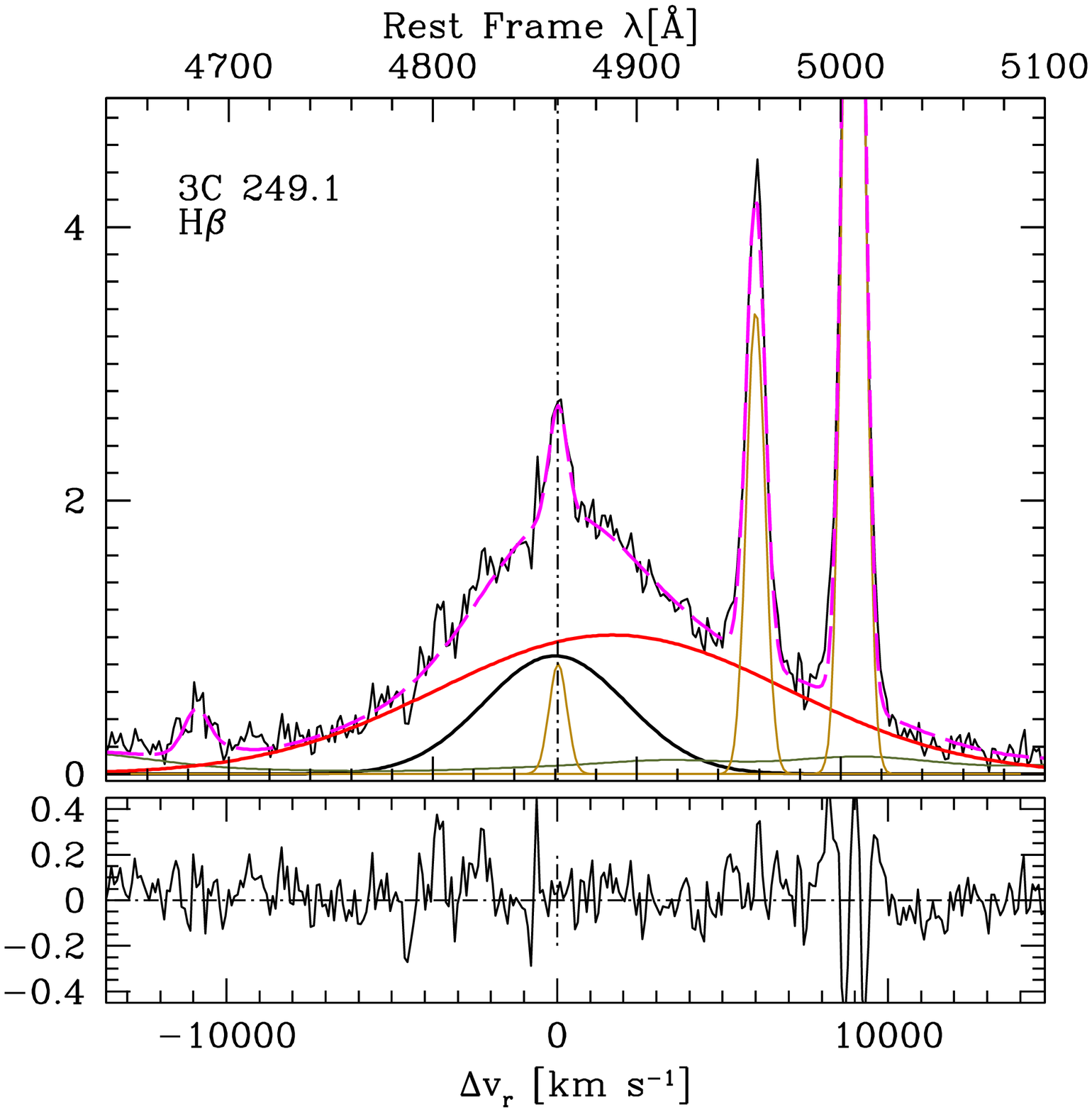}\\
\includegraphics[scale=0.22]{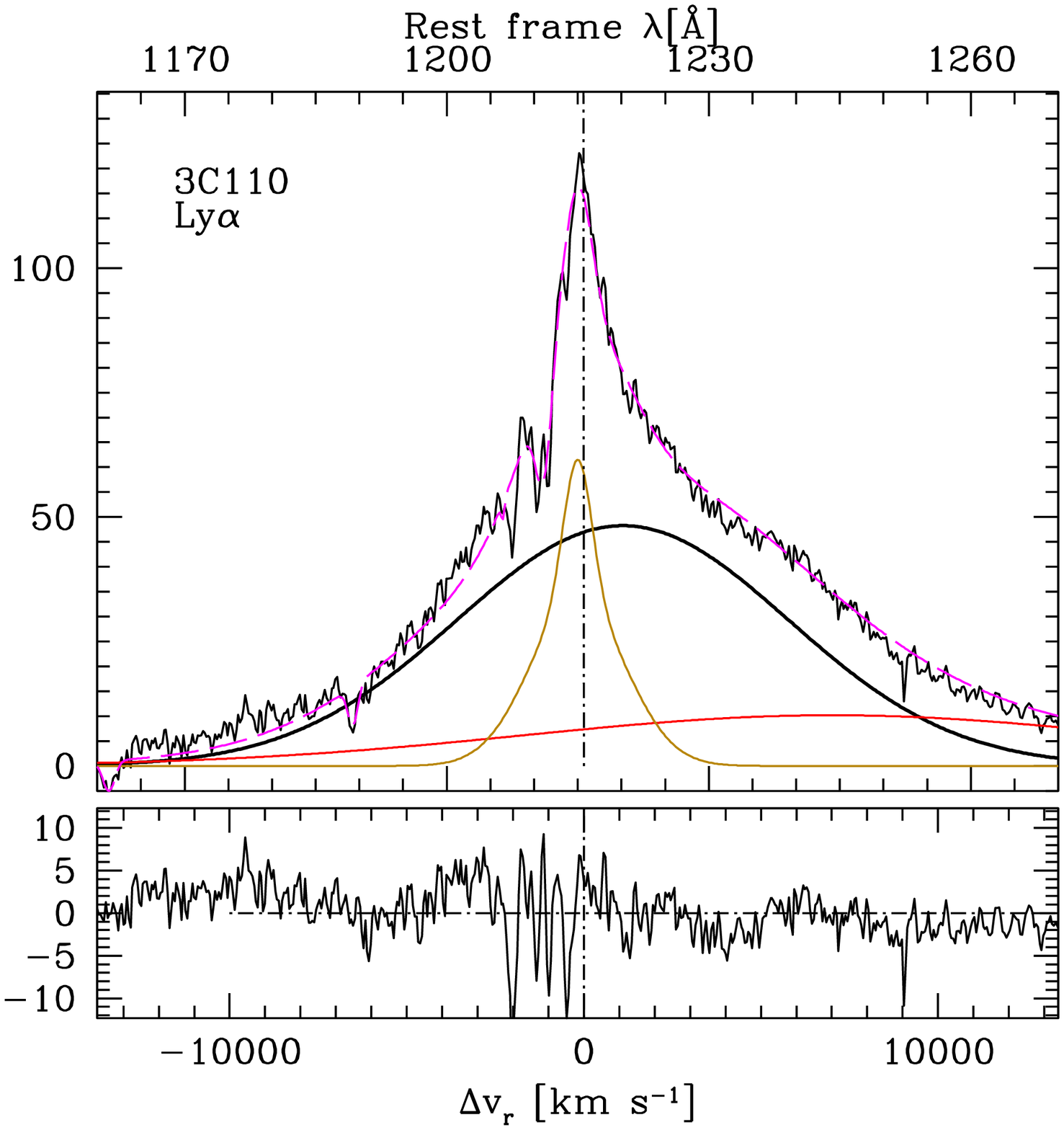}\includegraphics[scale=0.22]{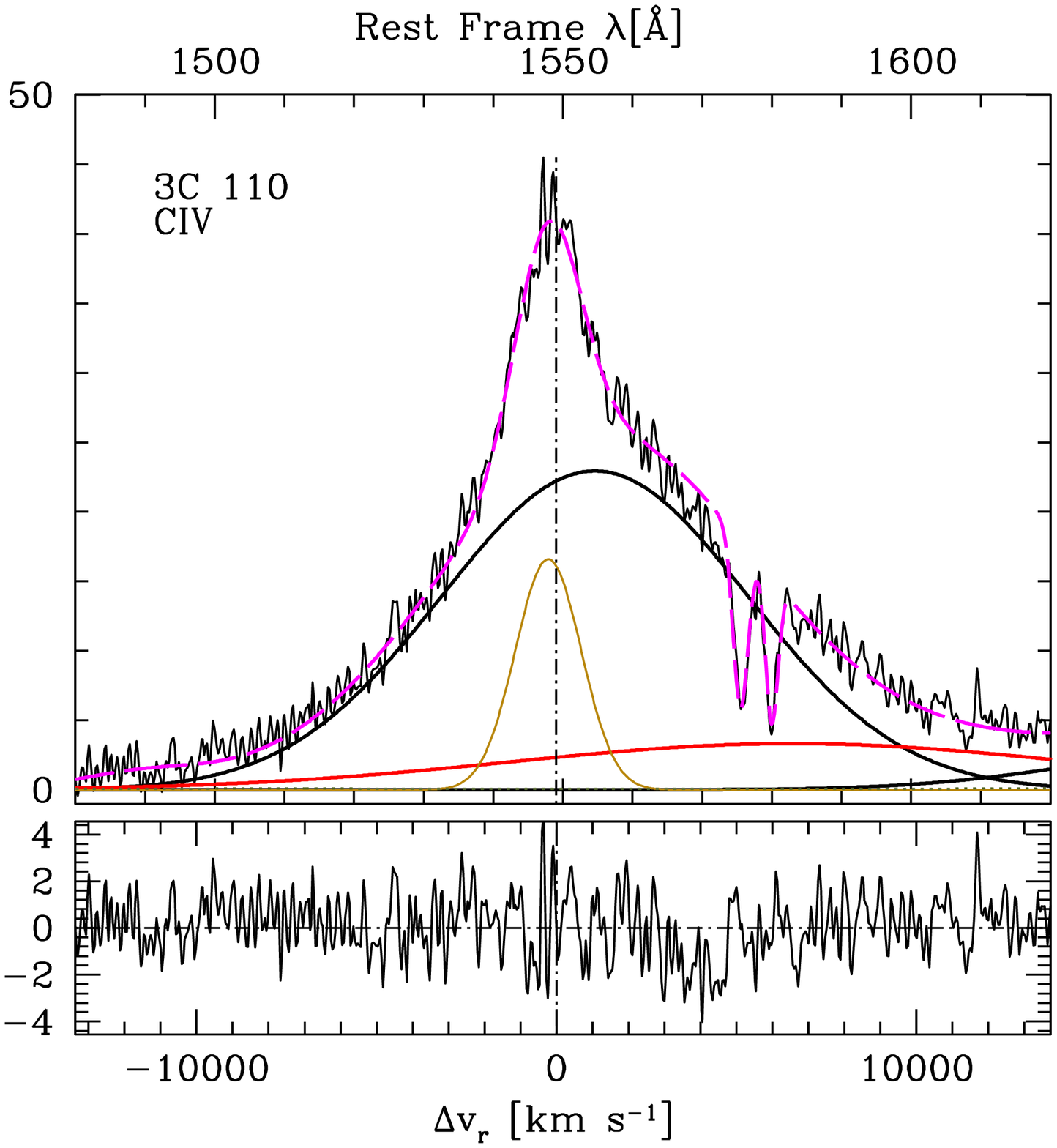}\includegraphics[scale=0.22]{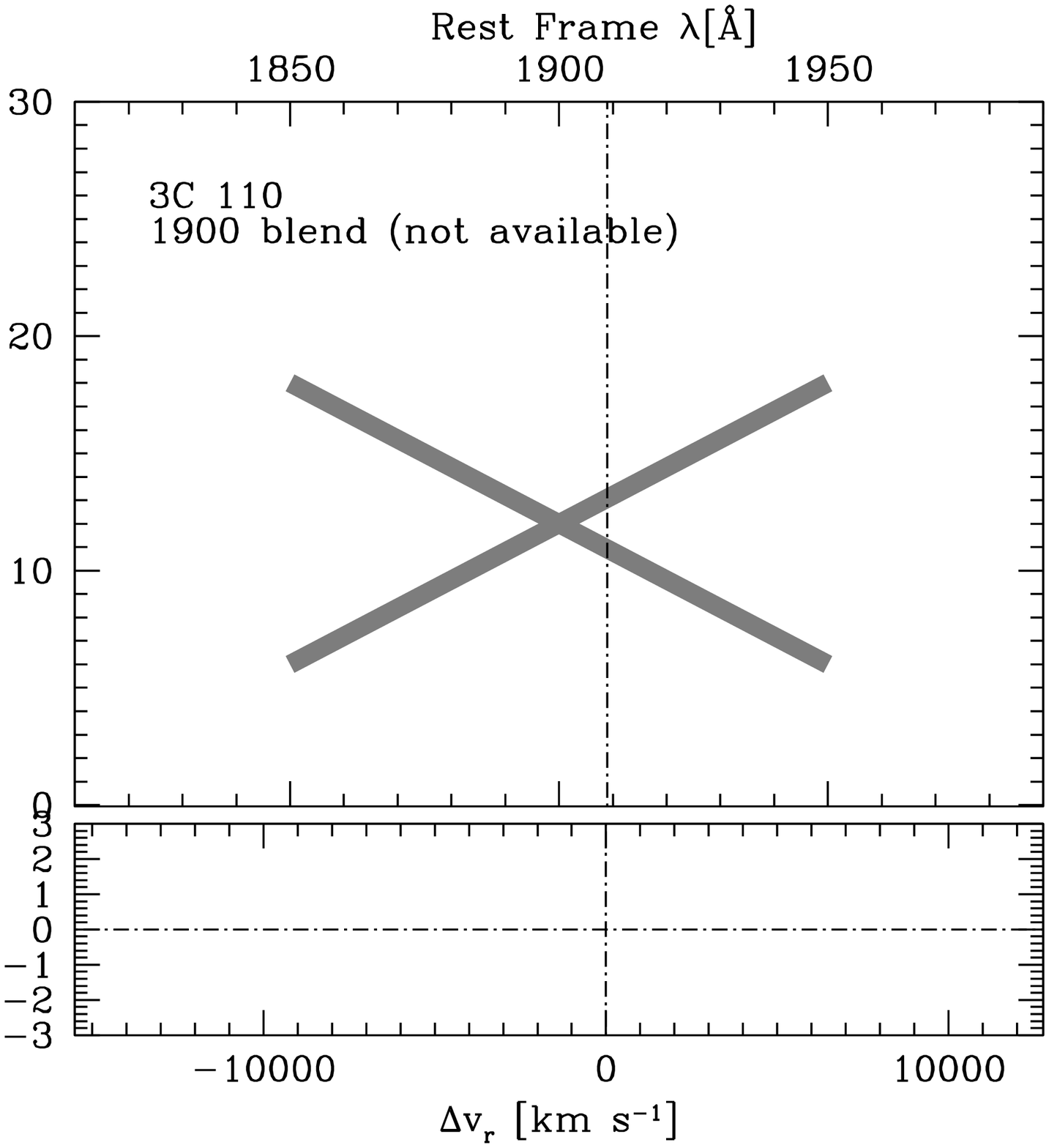}\includegraphics[scale=0.22]{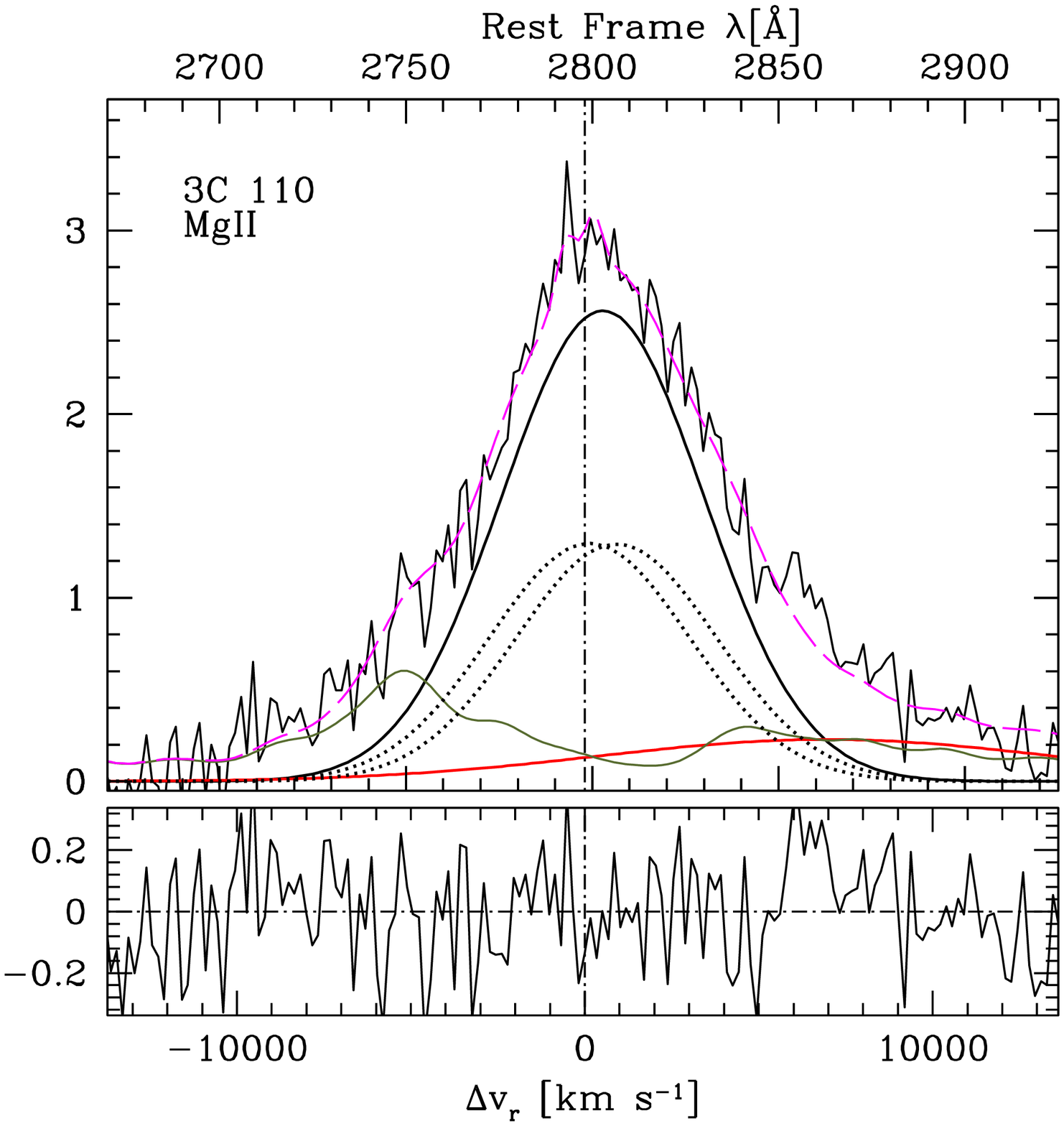}\includegraphics[scale=0.22]{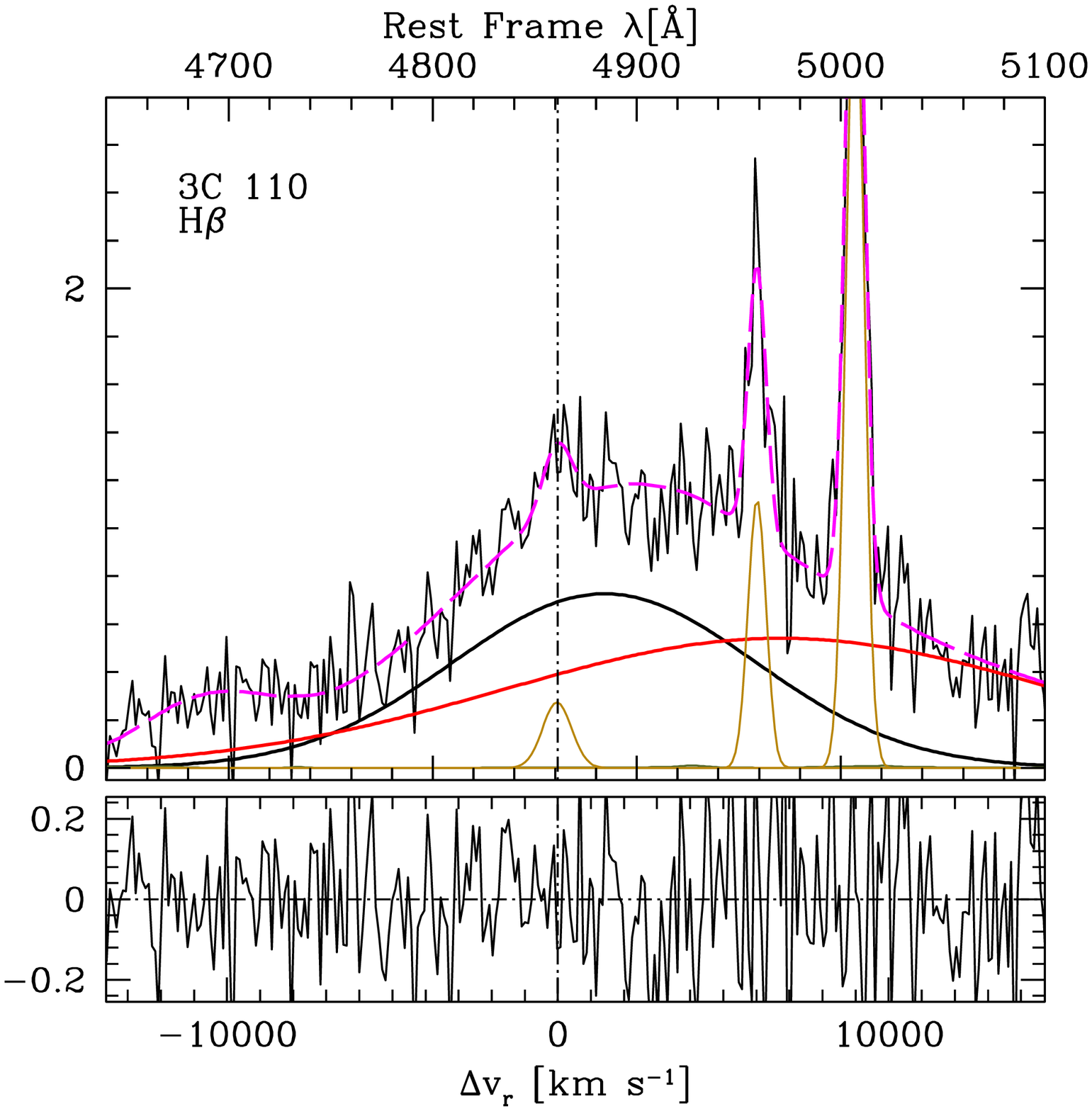}
\caption{Cont.}
\end{figure*}

\begin{figure*}
\includegraphics[scale=0.423]{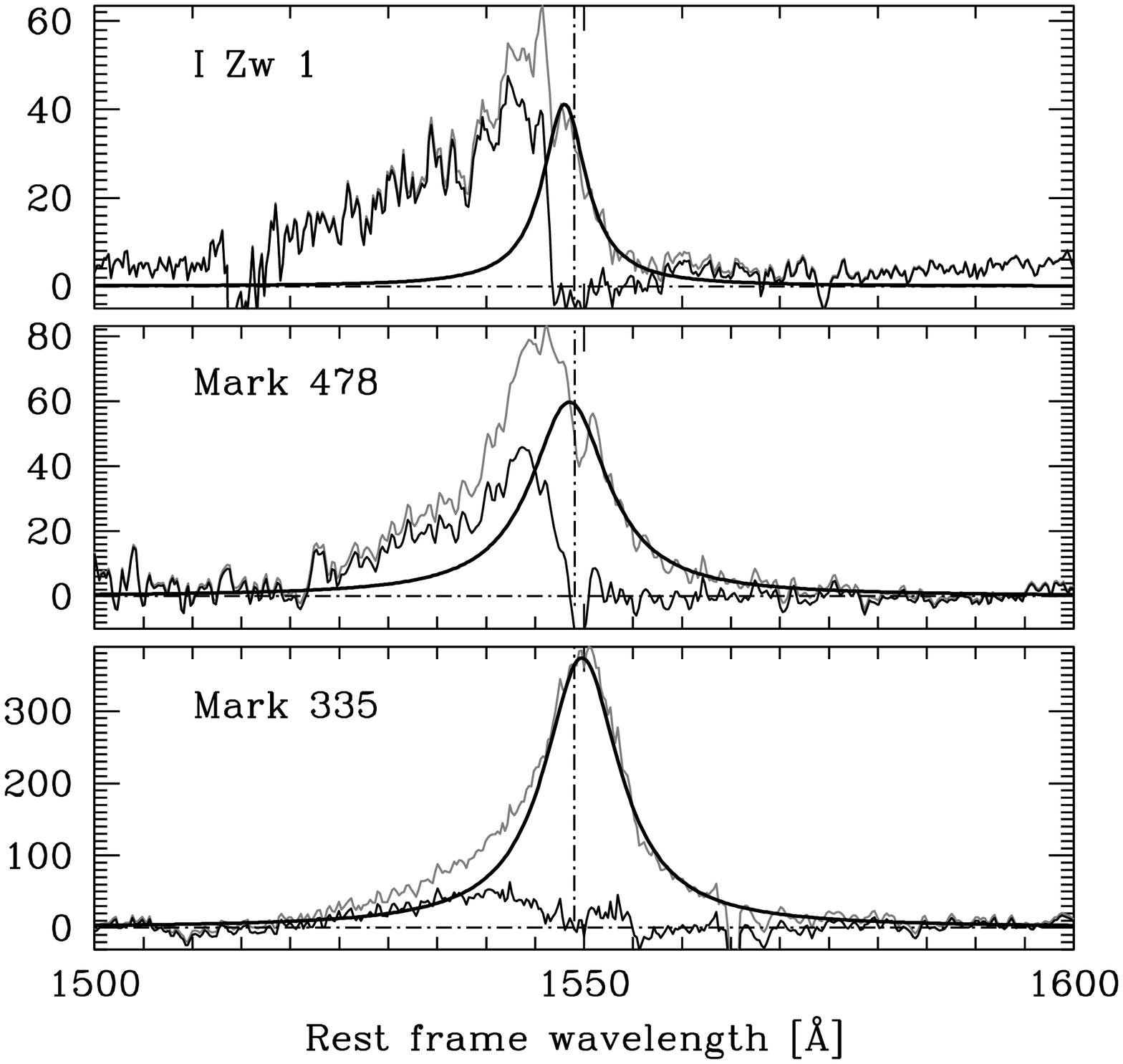}\includegraphics[scale=0.423]{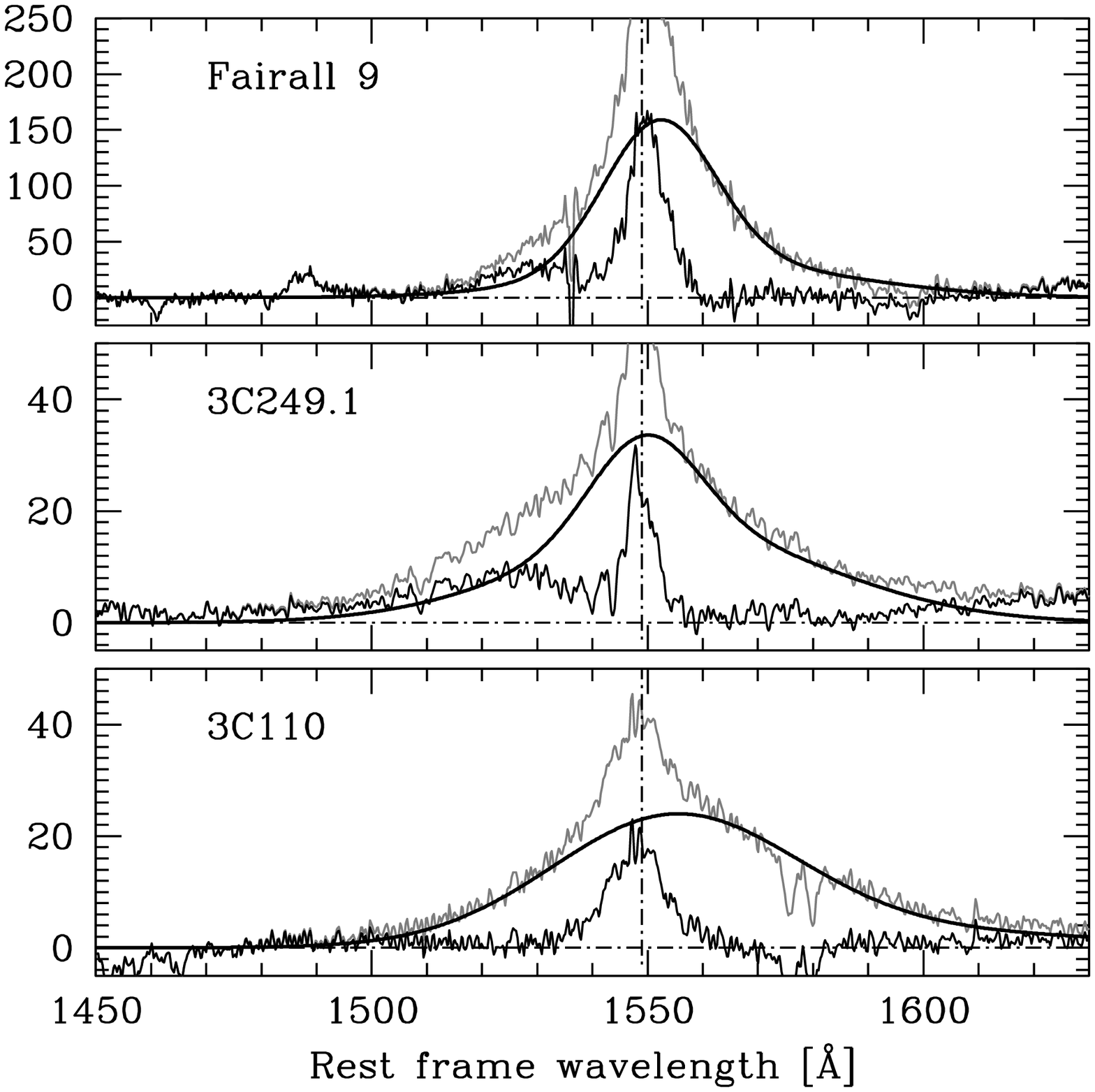}
\caption{\civ\ profile (grey line), after continuum subtraction for the six sources considered in this study. The scaled profile of \hb\ (\hbbc\ + \hbvbc) is shown as a thick line, and the residuals as a thin continuous line. For Fairall 9 and 3C249.1 it is possible to identify two major contributions to the residuals, namely one due to \civnc, and one broad and shifted to the blue with respect to the source rest frame. The scaling is meant to show a maximum possible \hb\ contribution to the \civ\ profiles; the actual \civ\ retrieved from the multicomponent fits (see text) can be somewhat different from the one shown in this Figure.\label{fig:blue}}

\end{figure*}

\rotate\rotate\rotate
\setcounter{figure}{3}
\begin{figure*}
\includegraphics[scale=0.522]{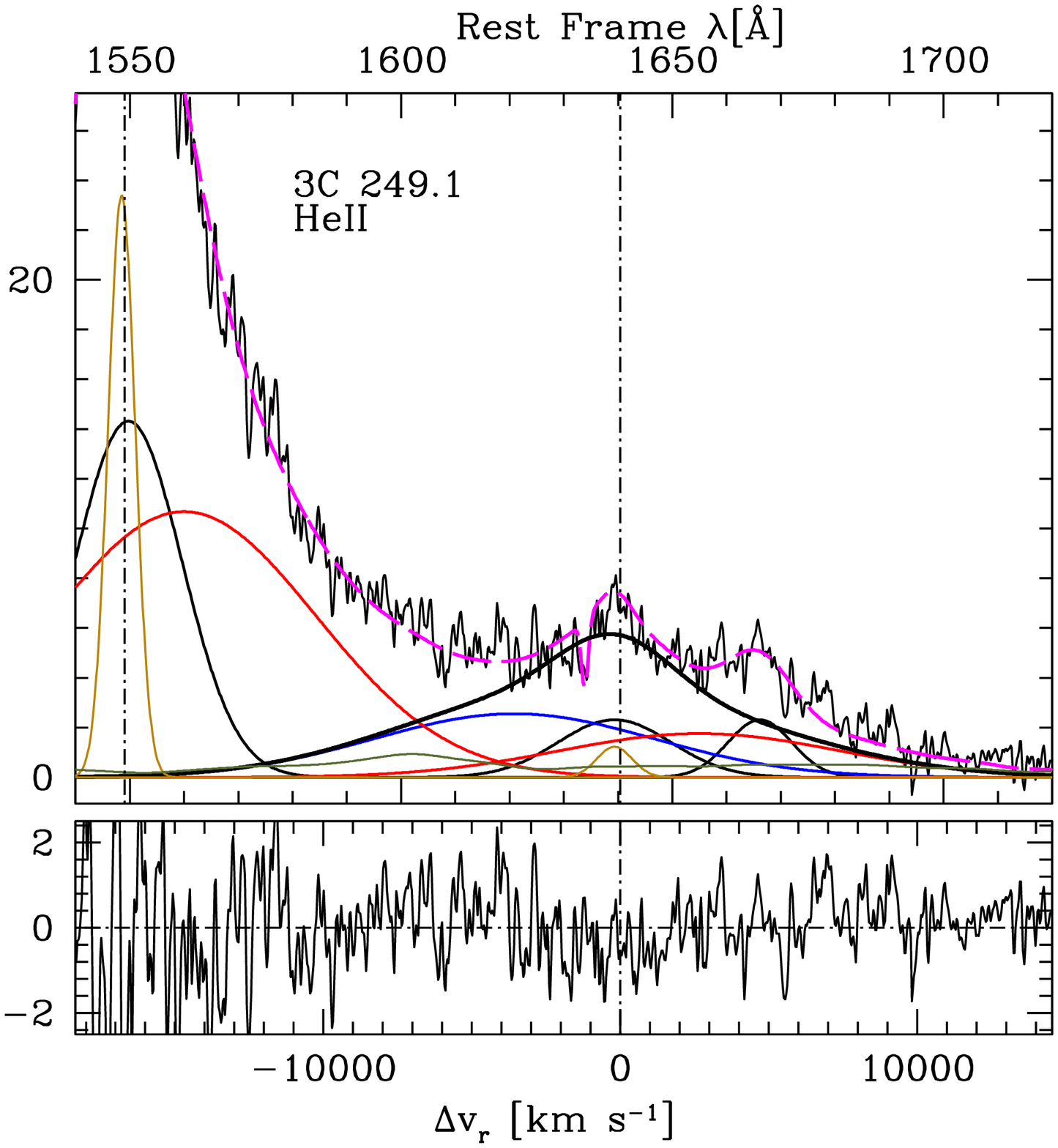}\caption{The interpretation of the \heiiuv\ profile of 3C249.1. The ``plateau'' appearance of the far red wing of \civ\ is accounted for by a blueshifted (blue) and a redshifted very broad component (red) of \heiiuv, whose shifts and widths match the ones of \civ. The resulting profile is shown as the thickest black line. Units are as for Fig. \ref{fig:profiles}. The dashed magenta lines traces the fit with all components added up. Other line styles have the same meaning of the previous Figure. \label{fig:heii}}
\end{figure*}

\begin{figure*}
\includegraphics[scale=0.522]{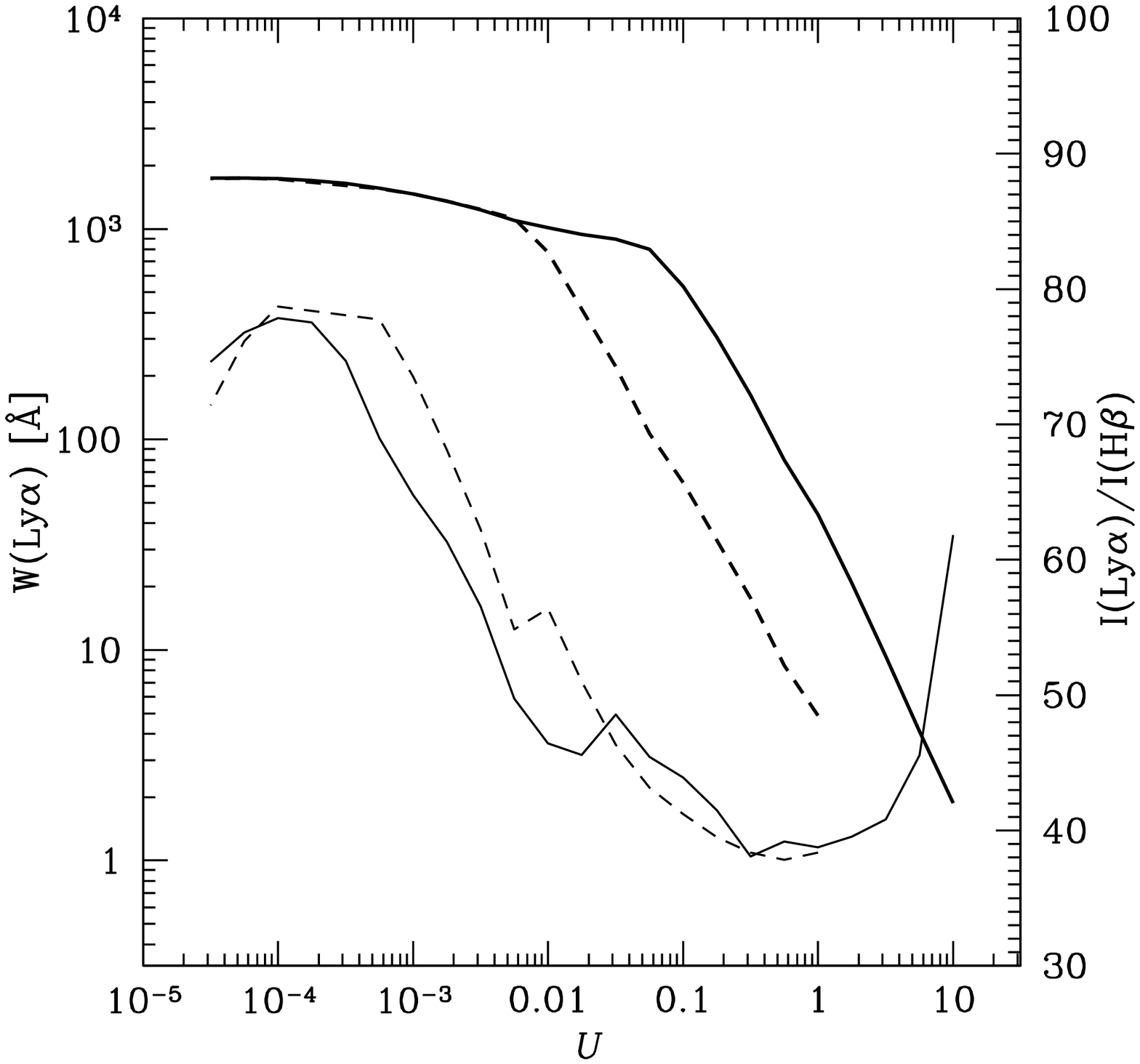}\caption{Predicted equivalent width of \lya\ in \AA\ (thick lines) and intensity ratio \lya/\hb\ (thin lines) as a function of the ionization parameter $U$.  The dashed lines refer to column density \nc $= 10^{21}$ cm$^{-2}$; the filled ones to \nc $= 10^{22}$ cm$^{-2}$.\label{fig:thin}}
\end{figure*}

\newpage\clearpage\null\vspace{-6cm}
\appendix
\section{Notes on Individual Sources}\label{individuals}
\paragraph*{I Zw 1 } is the prototypical NLSy1 source. Its UV spectrum has been analyzed in detail by many workers, notably \citet{laoretal97a}. The \mgii\ fit shows residuals that may be indicating an excess \feii\ emission around 2820 \AA.
\paragraph*{Mark 478} is spectroscopically rather similar to I Zw 1,  with less extreme properties.  The very low \civ\ BC intensity value may in part due to a narrow absorption close to the systemic velocity.
\paragraph*{Mark 335} belongs to bin A1. Its spectrum is consistent with the median spectrum of the bin, with low \rfe, weak blueshifted component. The profile of \hb\ is well described by a Lorentzian function, making this object similar to the rest of Pop. A even if \rfe\ is as low as in Pop. B.
\paragraph*{Fairall 9} is a rather typical B1 source, with a relatively modest but appreciable VBC. The \aliii/\siiii\ and \siiii/\civ\ ratios, along with \rfe $\approx$1, suggest some metal enrichment above solar. 
\paragraph*{3C 249.1} is a lobe-dominated radio source. It shares many of the properties of several radio-loud quasars at higher redshift, notably the strong \civnc, and the shape of the \lya\ and of the \civ+\heiiuv\ blend.
\paragraph*{3C 110} Some quasars show lines with strong VBC emission that can be mistaken for a BC \citep[e.g. PG1416-129]{sulenticetal00c}. The correct interpretation for 3C110 may involve a single redshifted VBC component. { The BC may be completely absent, as there is no \feii\ detection with the present data}.  This suggestion is also motivated by an analysis of 3C390.3 \citep{negreteetal10} which is a lobe-dominated RL quasar like 3C110. Physical conditions inferred for the BC of 3C 390.3 are similar to those of the VBC implying that there may be no BC in 3C 390.3.   If the BC is completely suppressed in 3C110 then the entire \civ\ profile in that source may be VBC. A fit with a single, shifted Gaussian is worse than the one shown in Fig. \ref{fig:profiles}, but not dramatically so. Then shift and FWHM values for a pure VBC \civ\  in 3C110 would be close to the canonical values (10$^3$ and 10$^4$ \kms\ respectively) for the VBC component in \hb. The spectrum of this object does lack a blueshifted component  which might be closely associated to the BC.


\label{lastpage}

\end{document}